\documentclass[12pt]{iopart}
\usepackage{amssymb}
\usepackage{amsfonts}
\usepackage{graphicx}
\usepackage{tikz,tkz-tab}
\usepackage{url}

\newtheorem{example}{Example}
\newtheorem{theo}[example]{Theorem}
\newtheorem{prop}[example]{Proposition}

\newtheorem{claim}[example]{Claim}
\newtheorem{alg}[example]{Algorithm}

\def\W{\mathtt{W}}
\def\GHZ{\mathtt{GHZ}}
\def\Swap{\mathtt{SWAP}}
\def\cZ{c-\mathtt{Z}}

\def\cX{c-\mathtt{X}}
\def\cZS{$\mathtt{cZS}$}
\def\cnot{c-\mathtt{Not}}
\def\PauliX{ Pauli-$\mathtt X$ }
\def\PauliY{ Pauli-$\mathtt Y$ }
\def\PauliZ{ Pauli-$\mathtt Z$ }
\def\<{\langle}
\def\>{\rangle}

\def\C{{\mathbb C\, }}

\def\i{{\rm i}}

\def\binom#1#2{\left(#1\atop#2\right)}

\begin{document}

\title[Quantum circuits of $\cZ$ and $\Swap$ gates]{Quantum circuits of $\cZ$ and $\Swap$ gates: optimization and entanglement}

\author{Marc Bataille}
\ead{marc.bataille@ac-rouen.fr}
\author{Jean-Gabriel Luque}
\ead{Jean-Gabriel.Luque@univ-rouen.fr}
\address{LITIS laboratory, Universit\'e Rouen-Normandie, 685 Avenue de l'Universit\'e, 76800 Saint-\'Etienne-du-Rouvray. France.}
\date{}

\begin{abstract}
We have studied the algebraic structure underlying the quantum circuits composed by $\cZ$ and $\Swap$ gates. Our results are applied to optimize the circuits and to understand the emergence of entanglement when a circuit acts on a fully factorized state.
\end{abstract}


\maketitle

\section{\label{intro}Introduction}
Back to basics: Computer Science is the science of using the laws of physics to perform calculations.
It is a multidisciplinary field in which physics, mathematics and engineering interact.
The computers we are currently handling are based on a ( {\it classical} ) mechanistic vision of the calculations: at first, the calculations were thought of as the result of a series of actions on gears or ribbons before being implemented on electronic devices that will give them their flexibility of use and the performances we know them to have today.
Since the pioneers' works, many algorithms adapted both to the representation of the information in the proposed physical devices as well as to the logic underlying them have been developed and intensively studied.
Nevertheless, many important problems (e.g. integer factorization and discrete logarithms) are known to be difficult to solve by computer. The only hope of breakthrough is to develop a computer science based on other physical laws. A fairly natural and promising way is to extend information theory to the quantum world in order to use phenomena such as state superposition and entanglement to improve performances. The idea dates back to the early 1980s
\cite{1980Benioff,1980Manin,1982Feynman}. The theories of Quantum Information and Quantum Computation have been extensively developed from this (see e.g. \cite{2011NCI}) and some spectacular algorithms have been exhibited (e.g. \cite{1992DJ,1996Grover,1997Shor}).
But building an efficient quantum computer remains one of the greatest challenge of  modern physic and one of its main difficulties is to manage with the instability of superposition. Several devices have been already experimented: optical quantum computers (e.g \cite{1995CY}), cavity
-QED technique (e.g. \cite{1997Turchette} ), trapped ions (e.g.\cite{1995CZ}), nuclear spins (e.g.\cite{1997CFH,1997GC}). One of the main drawbacks of these devices is that quantum gates can not be applied without errors. Many strategies exist to solve this problem. The first one is to find more stable devices. As an example, Topological Quantum Computers \cite{2003FKLW} are good candidates but at the present time they are only theoretical machines manipulating quasi-particles named non-abelian anyons that have not  been discovered yet. The second strategy consists in elaborating a theory of Quantum Error-Correction \cite{1995Shor}. Finally, parallel to the latter, it is possible to optimize the circuits, for instance, by minimizing the number of multiqubit gates that generate many errors. Our work is situated in this context. More specifically, we are interested in the interactions between $\Swap$ and $\cZ$ (controlled \PauliZ) gates that allow simplifications. After recalling background on quantum circuits, we'll prove that circuits generated by these two gates form a finite group. Investigating its structure, we'll find algebraic and combinatorial properties that allow us to propose a very simple algorithm of simplification. Then, we'll ask  the question of the emergence of entanglement: can we move from one entanglement level to another using only $\cZ$ and $\Swap$ gates together with $\mathtt{SLOCC}$ operations ?  

\section{\label{Qubit} Qubit systems, quantum gates and quantum circuits}The material contained in this section is rather classical in Quantum Information Theory. We  mainly recall notations and results that the reader can find in \cite{2011NCI,2018Yakimenko}.\\
In Quantum Information a qubit is a quantum state that represents the basic information storage unit. For our purpose, we consider that the states are pure, i.e. states that are described by a single ket vector in the Dirac notation $|\psi\rangle=a_{0}|0\rangle+a_{1}|1\rangle$ with $|a_{0}|^{2}+|a_{1}|^{2}=1$. The value of $|a_{i}|^{2}$  represents the probability that measurement produces the value $i$. Such a superposed state is often written as \begin{equation}|\psi\rangle=e^{i\gamma}\left(\cos\frac\theta2|0\rangle+e^{i\varphi}\sin\frac\theta2|1\rangle\right).\end{equation} The factor 
$e^{i\gamma}$ being ignored because having no observable effects, a single qubit depends only on two parameters and so defines a point on the unit three-dimensional sphere (the so called  Bloch sphere)
\begin{equation}|\psi\rangle=\cos\frac\theta2|0\rangle+e^{i\varphi}\sin\frac\theta2|1\rangle.\end{equation}
Operations on qubits must preserve the norm. So they act on qubits as $2\times 2$ unitary matrices act on two dimensional vectors. In Quantum Computation, they are represented by quantum gates. See  figure \ref{1gates} for a non exhaustive list of most used gates. 
\begin{figure}[h]
	\begin{center}

	\begin{tikzpicture}[scale=1.500000,x=1pt,y=1pt]
\filldraw[color=white] (0.000000, -7.500000) rectangle (24.000000, 7.500000);
\draw[color=black] (0.000000,0.000000) -- (24.000000,0.000000);
\draw[color=black] (0.000000,0.000000) node[left] {Hadamard\ \ \ };
\draw (45.00, 0.00) node {$\ \ \ \frac1{\sqrt2}\left[\begin{array}{cc}1&1\\1&-1\end{array}\right]$};
\begin{scope}
\draw[fill=white] (12.000000, -0.000000) +(-45.000000:8.485281pt and 8.485281pt) -- +(45.000000:8.485281pt and 8.485281pt) -- +(135.000000:8.485281pt and 8.485281pt) -- +(225.000000:8.485281pt and 8.485281pt) -- cycle;
\clip (12.000000, -0.000000) +(-45.000000:8.485281pt and 8.485281pt) -- +(45.000000:8.485281pt and 8.485281pt) -- +(135.000000:8.485281pt and 8.485281pt) -- +(225.000000:8.485281pt and 8.485281pt) -- cycle;
\draw (12.000000, -0.000000) node {$H$};
\end{scope}

\end{tikzpicture}\ \\
	\begin{tikzpicture}[scale=1.500000,x=1pt,y=1pt]
\filldraw[color=white] (0.000000, -7.500000) rectangle (24.000000, 7.500000);
\draw[color=black] (0.000000,0.000000) -- (24.000000,0.000000);
\draw[color=black] (0.000000,0.000000) node[left] {\PauliX\ \ };
\draw (45.00, 0.00) node {$\left[\begin{array}{cc}0&1\\1&0\end{array}\right]$};
\begin{scope}
\draw[fill=white] (12.000000, -0.000000) +(-45.000000:8.485281pt and 8.485281pt) -- +(45.000000:8.485281pt and 8.485281pt) -- +(135.000000:8.485281pt and 8.485281pt) -- +(225.000000:8.485281pt and 8.485281pt) -- cycle;
\clip (12.000000, -0.000000) +(-45.000000:8.485281pt and 8.485281pt) -- +(45.000000:8.485281pt and 8.485281pt) -- +(135.000000:8.485281pt and 8.485281pt) -- +(225.000000:8.485281pt and 8.485281pt) -- cycle;
\draw (12.000000, -0.000000) node {$X$};
\end{scope}

\end{tikzpicture}

	\begin{tikzpicture}[scale=1.500000,x=1pt,y=1pt]
\filldraw[color=white] (0.000000, -7.500000) rectangle (24.000000, 7.500000);
\draw[color=black] (0.000000,0.000000) -- (24.000000,0.000000);
\draw[color=black] (0.000000,0.000000) node[left] {\PauliY\ \ };
\draw (45.00, 0.00) node {$\left[\begin{array}{cc}0&-i\\i&0\end{array}\right]$};
\begin{scope}
\draw[fill=white] (12.000000, -0.000000) +(-45.000000:8.485281pt and 8.485281pt) -- +(45.000000:8.485281pt and 8.485281pt) -- +(135.000000:8.485281pt and 8.485281pt) -- +(225.000000:8.485281pt and 8.485281pt) -- cycle;
\clip (12.000000, -0.000000) +(-45.000000:8.485281pt and 8.485281pt) -- +(45.000000:8.485281pt and 8.485281pt) -- +(135.000000:8.485281pt and 8.485281pt) -- +(225.000000:8.485281pt and 8.485281pt) -- cycle;
\draw (12.000000, -0.000000) node {$Y$};
\end{scope}

\end{tikzpicture}

	\begin{tikzpicture}[scale=1.500000,x=1pt,y=1pt]
\filldraw[color=white] (0.000000, -7.500000) rectangle (24.000000, 7.500000);
\draw[color=black] (0.000000,0.000000) -- (24.000000,0.000000);
\draw[color=black] (0.000000,0.000000) node[left] {\PauliZ\ \ };
\draw (45.00, 0.00) node {$\left[\begin{array}{cc}1&0\\0&-1\end{array}\right]$};
\begin{scope}
\draw[fill=white] (12.000000, -0.000000) +(-45.000000:8.485281pt and 8.485281pt) -- +(45.000000:8.485281pt and 8.485281pt) -- +(135.000000:8.485281pt and 8.485281pt) -- +(225.000000:8.485281pt and 8.485281pt) -- cycle;
\clip (12.000000, -0.000000) +(-45.000000:8.485281pt and 8.485281pt) -- +(45.000000:8.485281pt and 8.485281pt) -- +(135.000000:8.485281pt and 8.485281pt) -- +(225.000000:8.485281pt and 8.485281pt) -- cycle;
\draw (12.000000, -0.000000) node {$Z$};
\end{scope}

\end{tikzpicture}

	\begin{tikzpicture}[scale=1.500000,x=1pt,y=1pt]
\filldraw[color=white] (0.000000, -7.500000) rectangle (24.000000, 7.500000);
\draw[color=black] (-1.000000,0.000000) -- (25.000000,0.000000);
\draw[color=black] (0.000000,0.000000) node[left] {Rotation about the $x$ axis\ \ };
\draw (65.00, 0.00) node {$\left[\begin{array}{cc}\cos\frac\theta2&-i\sin\frac\theta2\\-i\sin\frac\theta2&\cos\frac\theta2\end{array}\right]$};
\begin{scope}
\draw[fill=white] (9.000000, -0.000000) +(-45.000000:18.985281pt and 8.485281pt) -- +(45.000000:18.985281pt and 8.485281pt) -- +(135.000000:8.485281pt and 8.485281pt) -- +(225.000000:8.485281pt and 8.485281pt) -- cycle;
\clip (8.000000, -0.000000) +(-45.000000:18.985281pt and 8.485281pt) -- +(45.000000:18.985281pt and 8.485281pt) -- +(135.000000:8.485281pt and 8.485281pt) -- +(225.000000:8.485281pt and 8.485281pt) -- cycle;
\draw (12.500000, -0.000000) node {$R_{x}(\theta)$};
\end{scope}

\end{tikzpicture}

	\begin{tikzpicture}[scale=1.500000,x=1pt,y=1pt]
\filldraw[color=white] (0.000000, -7.500000) rectangle (24.000000, 7.500000);
\draw[color=black] (-1.000000,0.000000) -- (25.000000,0.000000);
\draw[color=black] (0.000000,0.000000) node[left] {Rotation about the $y$ axis\ \ };
\draw (65.00, 0.00) node {$\left[\begin{array}{cc}\cos\frac\theta2&-\sin\frac\theta2\\\sin\frac\theta2&\cos\frac\theta2\end{array}\right]$};
\begin{scope}
\draw[fill=white] (9.000000, -0.000000) +(-45.000000:18.985281pt and 8.485281pt) -- +(45.000000:18.985281pt and 8.485281pt) -- +(135.000000:8.485281pt and 8.485281pt) -- +(225.000000:8.485281pt and 8.485281pt) -- cycle;
\clip (8.000000, -0.000000) +(-45.000000:18.985281pt and 8.485281pt) -- +(45.000000:18.985281pt and 8.485281pt) -- +(135.000000:8.485281pt and 8.485281pt) -- +(225.000000:8.485281pt and 8.485281pt) -- cycle;
\draw (12.000000, -0.000000) node {$R_{y}(\theta)$};
\end{scope}

\end{tikzpicture}

	\begin{tikzpicture}[scale=1.500000,x=1pt,y=1pt]
\filldraw[color=white] (0.000000, -7.500000) rectangle (24.000000, 7.500000);
\draw[color=black] (-1.000000,0.000000) -- (25.000000,0.000000);
\draw[color=black] (0.000000,0.000000) node[left] {Rotation about the $z$ axis\ \ };
\draw (65.00, 0.00) node {$\left[\begin{array}{cc}e^{-i\frac\theta2}&0\\0&e^{i\frac\theta2}\end{array}\right]$};
\begin{scope}
\draw[fill=white] (9.000000, -0.000000) +(-45.000000:18.985281pt and 8.485281pt) -- +(45.000000:18.985281pt and 8.485281pt) -- +(135.000000:8.485281pt and 8.485281pt) -- +(225.000000:8.485281pt and 8.485281pt) -- cycle;
\clip (8.000000, -0.000000) +(-45.000000:18.985281pt and 8.485281pt) -- +(45.000000:18.985281pt and 8.485281pt) -- +(135.000000:8.485281pt and 8.485281pt) -- +(225.000000:8.485281pt and 8.485281pt) -- cycle;
\draw (12.000000, -0.000000) node {$R_{z}(\theta)$};
\end{scope}

\end{tikzpicture}

\caption{Some single qubit gates\label{1gates}}	
\end{center}
\end{figure}
The behavior of multiple qubit systems with respect to the action of its dynamic group is very rich and complicated to describe in the general case.
A $k$-qubit system is seen as a superposition:
\begin{equation}
	|\psi\rangle=\sum_{0\leq i_{1},\dots,i_{k}\leq 1}a_{i_{1}\cdots i_{k}}|i_{1}\cdots i_{k}\rangle,
\end{equation}
with $\sum |a_{i_{1}\cdots i_{k}}|^{2}=1$.
Some states have a particular interest like the Greenberger-Horne-Zeilinger state\cite{1990GHZ}
\begin{equation}
|\GHZ_{k}\rangle=\frac1{\sqrt2}\left(|\overbrace{0\cdots0}^{\times k}\rangle+\ |\overbrace{1\cdots1}^{\times k}\rangle\right){}
\end{equation}
and the $\W$-states\cite{2000DVC}
\begin{equation}
	|\W_{k}\rangle=\frac1{\sqrt k}\left(|10\cdots0\rangle+|010\cdots0\rangle+\cdots+|0\cdots01\rangle\right). 
\end{equation}
These two states represent two non-equivalent entanglements for a $k$-particles system (among a multitude of other non-equivalent ones).\\
Again operations on $k$-qubits must preserve the norm and are assimilated to $2^{k}\times 2^{k}$ matrices. For simplicity, we consider the rows and the column of the matrices encoding  operations on qubits are indexed  by integers in $\{0,\dots,2^{k}-1\}$ written in the binary representations; the index $\alpha_{k-1}\cdots \alpha_{0}$ (i.e. binary representation of $\alpha_{k-1}2^{k-1}+\cdots+\alpha_{0}2^{0}$) corresponds to the wave function $|\alpha_{k-1}\cdots \alpha_{0}\rangle$ (see  figure \ref{2Gates} for some examples of $2$-qubit gates).
\begin{figure}[h]
	\begin{tikzpicture}[scale=1.500000,x=1pt,y=1pt]
\filldraw[color=white] (0.000000, -7.500000) rectangle (36.000000, 22.500000);
\draw[color=black] (15.0,7.5) node[left] {$\Swap$};
\draw[color=black] (100.0,7.5) node[left] {$\left[\begin{array}{cccc}1&0&0&0\\0&0&1&0\\0&1&0&0\\0&0&0&1\end{array}\right]$};
\draw[color=black] (18.000000,15.000000) -- (36.000000,15.000000);
\draw[color=black] (18.000000,0.000000) -- (36.000000,0.000000);
\draw (27.000000,15.000000) -- (27.000000,0.000000);
\begin{scope}
\draw (24.878680, 12.878680) -- (29.121320, 17.121320);
\draw (24.878680, 17.121320) -- (29.121320, 12.878680);
\end{scope}
\begin{scope}
\draw (24.878680, -2.121320) -- (29.121320, 2.121320);
\draw (24.878680, 2.121320) -- (29.121320, -2.121320);
\end{scope}
\end{tikzpicture}
	\begin{tikzpicture}[scale=1.500000,x=1pt,y=1pt]
\filldraw[color=white] (0.000000, -7.500000) rectangle (36.000000, 22.500000);
\draw[color=black] (15.0,7.5) node[left] {$\cnot$};
\draw[color=black] (100.0,7.5) node[left] {$\left[\begin{array}{cccc}1&0&0&0\\0&1&0&0\\0&0&0&1\\0&0&1&0\end{array}\right]$};
\draw[color=black] (18.000000,15.000000) -- (36.000000,15.000000);
\draw[color=black] (18.000000,0.000000) -- (36.000000,0.000000);
\draw (27.000000,15.000000) -- (27.000000,0.000000);

\begin{scope}
\draw[fill=white] (27.000000, -0.000000) circle(3.000000pt);
\clip (27.000000, -0.0000000) circle(3.000000pt);
\end{scope}
\begin{scope}
\draw (24.878680, -2.121320) -- (29.121320, 2.121320);
\draw (24.878680, 2.121320) -- (29.121320, -2.121320);
\end{scope}
\filldraw (27.000000, 15.000000) circle(1.500000pt);
\end{tikzpicture}

\begin{tikzpicture}[scale=1.500000,x=1pt,y=1pt]
\filldraw[color=white] (0.000000, -7.500000) rectangle (18.000000, 22.500000);
\draw[color=black] (0.000000,15.000000) -- (18.000000,15.000000);
\draw[color=black] (0.000000,0.000000) -- (18.000000,0.000000);
\draw[color=black] (-3.000000,7.500000) node[left] {$\cZ$};
\draw (9.000000,15.000000) -- (9.000000,0.000000);
\filldraw (9.000000, 15.000000) circle(1.500000pt);
\filldraw (9.000000, 0.000000) circle(1.500000pt);
\draw[color=black] (90.0,7.5) node[left] {$\left[\begin{array}{cccc}1&0&0&0\\0&1&0&0\\0&0&1&0\\0&0&0&-1\end{array}\right]$};
\end{tikzpicture}

	\caption{Some $2$-qubit gates \label{2Gates}}
\end{figure}
 Notice that  the $\cnot$ and $\cZ$ gates are special cases of controlled gates based on \PauliX and \PauliZ matrices. Notice also that for the $\cZ$ gates the two qubits play the same role, hence the symmetrical representation of the gate. The gates can be combined in series or in parallel to form \emph{Quantum Circuits} that represent new operators.
Composing quantum gates in series allows to act on the same number of qubits and results in a multiplication of matrices read from the right to the left on the circuit\footnote{The reader must pay attention to the following fact: the circuits act to the right of the wave functions presented to their left but the associated operators act to the left of the ket, i.e. $O|\psi\rangle$.}.
Composing quantum gates in parallel makes it possible to act on larger systems and results in a Kronecker product of the matrices read from the top to the bottom on the circuit. See figure \ref{GHZCirc} for an example of quantum circuit and figure \ref{Action} for its action on a qubit system.
\begin{figure}[h]
	\begin{center}
	\begin{tikzpicture}[scale=1.500000,x=1pt,y=1pt]
\filldraw[color=white] (0.000000, -7.500000) rectangle (114.000000, 37.500000);
\draw[color=black] (0.000000,30.000000) -- (114.000000,30.000000);
\draw[color=black] (0.000000,30.000000) node[left] {$2$};
\draw[color=black] (0.000000,15.000000) -- (114.000000,15.000000);
\draw[color=black] (0.000000,15.000000) node[left] {$1$};
\draw[color=black] (0.000000,0.000000) -- (114.000000,0.000000);
\draw[color=black] (0.000000,0.000000) node[left] {$0$};
\begin{scope}
\draw[fill=white] (12.000000, 30.000000) +(-45.000000:8.485281pt and 8.485281pt) -- +(45.000000:8.485281pt and 8.485281pt) -- +(135.000000:8.485281pt and 8.485281pt) -- +(225.000000:8.485281pt and 8.485281pt) -- cycle;
\clip (12.000000, 30.000000) +(-45.000000:8.485281pt and 8.485281pt) -- +(45.000000:8.485281pt and 8.485281pt) -- +(135.000000:8.485281pt and 8.485281pt) -- +(225.000000:8.485281pt and 8.485281pt) -- cycle;
\draw (12.000000, 30.000000) node {$H$};
\end{scope}
\begin{scope}
\draw[fill=white] (12.000000, 15.000000) +(-45.000000:8.485281pt and 8.485281pt) -- +(45.000000:8.485281pt and 8.485281pt) -- +(135.000000:8.485281pt and 8.485281pt) -- +(225.000000:8.485281pt and 8.485281pt) -- cycle;
\clip (12.000000, 15.000000) +(-45.000000:8.485281pt and 8.485281pt) -- +(45.000000:8.485281pt and 8.485281pt) -- +(135.000000:8.485281pt and 8.485281pt) -- +(225.000000:8.485281pt and 8.485281pt) -- cycle;
\draw (12.000000, 15.000000) node {$H$};
\end{scope}
\begin{scope}
\draw[fill=white] (12.000000, -0.000000) +(-45.000000:8.485281pt and 8.485281pt) -- +(45.000000:8.485281pt and 8.485281pt) -- +(135.000000:8.485281pt and 8.485281pt) -- +(225.000000:8.485281pt and 8.485281pt) -- cycle;
\clip (12.000000, -0.000000) +(-45.000000:8.485281pt and 8.485281pt) -- +(45.000000:8.485281pt and 8.485281pt) -- +(135.000000:8.485281pt and 8.485281pt) -- +(225.000000:8.485281pt and 8.485281pt) -- cycle;
\draw (12.000000, -0.000000) node {$H$};
\end{scope}
\draw (33.000000,15.000000) -- (33.000000,0.000000);
\filldraw (33.000000, 15.000000) circle(1.500000pt);
\filldraw (33.000000, 0.000000) circle(1.500000pt);
\draw (54.000000,30.000000) -- (54.000000,15.000000);
\filldraw (54.000000, 30.000000) circle(1.500000pt);
\filldraw (54.000000, 15.000000) circle(1.500000pt);
\begin{scope}
\draw[fill=white] (54.000000, -0.000000) +(-45.000000:8.485281pt and 8.485281pt) -- +(45.000000:8.485281pt and 8.485281pt) -- +(135.000000:8.485281pt and 8.485281pt) -- +(225.000000:8.485281pt and 8.485281pt) -- cycle;
\clip (54.000000, -0.000000) +(-45.000000:8.485281pt and 8.485281pt) -- +(45.000000:8.485281pt and 8.485281pt) -- +(135.000000:8.485281pt and 8.485281pt) -- +(225.000000:8.485281pt and 8.485281pt) -- cycle;
\draw (54.000000, -0.000000) node {$X$};
\end{scope}
\begin{scope}
\draw[fill=white] (78.000000, 30.000000) +(-45.000000:8.485281pt and 8.485281pt) -- +(45.000000:8.485281pt and 8.485281pt) -- +(135.000000:8.485281pt and 8.485281pt) -- +(225.000000:8.485281pt and 8.485281pt) -- cycle;
\clip (78.000000, 30.000000) +(-45.000000:8.485281pt and 8.485281pt) -- +(45.000000:8.485281pt and 8.485281pt) -- +(135.000000:8.485281pt and 8.485281pt) -- +(225.000000:8.485281pt and 8.485281pt) -- cycle;
\draw (78.000000, 30.000000) node {$X$};
\end{scope}
\begin{scope}
\draw[fill=white] (78.000000, -0.000000) +(-45.000000:8.485281pt and 8.485281pt) -- +(45.000000:8.485281pt and 8.485281pt) -- +(135.000000:8.485281pt and 8.485281pt) -- +(225.000000:8.485281pt and 8.485281pt) -- cycle;
\clip (78.000000, -0.000000) +(-45.000000:8.485281pt and 8.485281pt) -- +(45.000000:8.485281pt and 8.485281pt) -- +(135.000000:8.485281pt and 8.485281pt) -- +(225.000000:8.485281pt and 8.485281pt) -- cycle;
\draw (78.000000, -0.000000) node {$H$};
\end{scope}
\begin{scope}
\draw[fill=white] (102.000000, 30.000000) +(-45.000000:8.485281pt and 8.485281pt) -- +(45.000000:8.485281pt and 8.485281pt) -- +(135.000000:8.485281pt and 8.485281pt) -- +(225.000000:8.485281pt and 8.485281pt) -- cycle;
\clip (102.000000, 30.000000) +(-45.000000:8.485281pt and 8.485281pt) -- +(45.000000:8.485281pt and 8.485281pt) -- +(135.000000:8.485281pt and 8.485281pt) -- +(225.000000:8.485281pt and 8.485281pt) -- cycle;
\draw (102.000000, 30.000000) node {$H$};
\end{scope}
\end{tikzpicture}
\end{center}
\[{}\begin{array}{rcl}
C&=&\left(H\otimes I_{4}\right)\cdot\left(X\otimes I_{2}\otimes H\right){}
\cdot\left(\cZ\otimes X\right)
\cdot\left(I_{2}\otimes \cZ\right)\cdot\left(H\otimes H\otimes H\right)\\&=&
 \frac1{\sqrt2} \left[ \begin {array}{cccccccc} 1&0&1&0&0&0&0&0\\ \noalign{\medskip}0
&-1&0&-1&0&0&0&0\\ \noalign{\medskip}0&0&0&0&0&1&0&-1
\\ \noalign{\medskip}0&0&0&0&-1&0&1&0\\ \noalign{\medskip}0&0&0&0&-1&0
&-1&0\\ \noalign{\medskip}0&0&0&0&0&1&0&1\\ \noalign{\medskip}0&-1&0&1
&0&0&0&0\\ \noalign{\medskip}1&0&-1&0&0&0&0&0\end {array} \right] 
\end{array}
\]
\caption{Example of Quantum Circuit together with its associated matrix\label{GHZCirc} ($I_{k}$ stands for the identity $k\times k$ matrix).}
\end{figure}
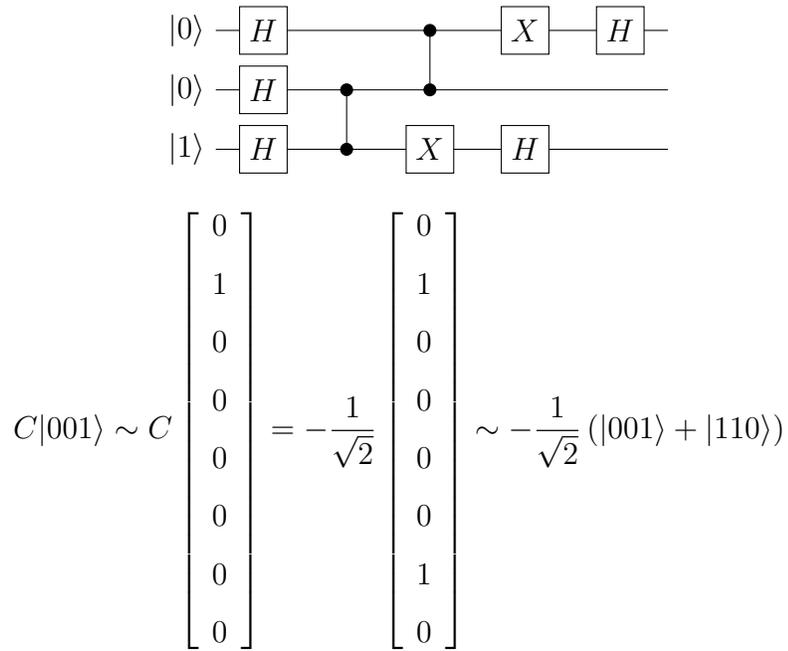
\begin{figure}[h]
	\begin{center}

\begin{tikzpicture}[scale=1.500000,x=1pt,y=1pt]
\filldraw[color=white] (0.000000, -7.500000) rectangle (114.000000, 37.500000);
\draw[color=black] (0.000000,30.000000) -- (114.000000,30.000000);
\draw[color=black] (0.000000,30.000000) node[left] {$|0\rangle$};
\draw[color=black] (0.000000,15.000000) -- (114.000000,15.000000);
\draw[color=black] (0.000000,15.000000) node[left] {$|0\rangle$};
\draw[color=black] (0.000000,0.000000) -- (114.000000,0.000000);
\draw[color=black] (0.000000,0.000000) node[left] {$|1\rangle$};
\begin{scope}
\draw[fill=white] (12.000000, 30.000000) +(-45.000000:8.485281pt and 8.485281pt) -- +(45.000000:8.485281pt and 8.485281pt) -- +(135.000000:8.485281pt and 8.485281pt) -- +(225.000000:8.485281pt and 8.485281pt) -- cycle;
\clip (12.000000, 30.000000) +(-45.000000:8.485281pt and 8.485281pt) -- +(45.000000:8.485281pt and 8.485281pt) -- +(135.000000:8.485281pt and 8.485281pt) -- +(225.000000:8.485281pt and 8.485281pt) -- cycle;
\draw (12.000000, 30.000000) node {$H$};
\end{scope}
\begin{scope}
\draw[fill=white] (12.000000, 15.000000) +(-45.000000:8.485281pt and 8.485281pt) -- +(45.000000:8.485281pt and 8.485281pt) -- +(135.000000:8.485281pt and 8.485281pt) -- +(225.000000:8.485281pt and 8.485281pt) -- cycle;
\clip (12.000000, 15.000000) +(-45.000000:8.485281pt and 8.485281pt) -- +(45.000000:8.485281pt and 8.485281pt) -- +(135.000000:8.485281pt and 8.485281pt) -- +(225.000000:8.485281pt and 8.485281pt) -- cycle;
\draw (12.000000, 15.000000) node {$H$};
\end{scope}
\begin{scope}
\draw[fill=white] (12.000000, -0.000000) +(-45.000000:8.485281pt and 8.485281pt) -- +(45.000000:8.485281pt and 8.485281pt) -- +(135.000000:8.485281pt and 8.485281pt) -- +(225.000000:8.485281pt and 8.485281pt) -- cycle;
\clip (12.000000, -0.000000) +(-45.000000:8.485281pt and 8.485281pt) -- +(45.000000:8.485281pt and 8.485281pt) -- +(135.000000:8.485281pt and 8.485281pt) -- +(225.000000:8.485281pt and 8.485281pt) -- cycle;
\draw (12.000000, -0.000000) node {$H$};
\end{scope}
\draw (33.000000,15.000000) -- (33.000000,0.000000);
\filldraw (33.000000, 15.000000) circle(1.500000pt);
\filldraw (33.000000, 0.000000) circle(1.500000pt);
\draw (54.000000,30.000000) -- (54.000000,15.000000);
\filldraw (54.000000, 30.000000) circle(1.500000pt);
\filldraw (54.000000, 15.000000) circle(1.500000pt);
\begin{scope}
\draw[fill=white] (54.000000, -0.000000) +(-45.000000:8.485281pt and 8.485281pt) -- +(45.000000:8.485281pt and 8.485281pt) -- +(135.000000:8.485281pt and 8.485281pt) -- +(225.000000:8.485281pt and 8.485281pt) -- cycle;
\clip (54.000000, -0.000000) +(-45.000000:8.485281pt and 8.485281pt) -- +(45.000000:8.485281pt and 8.485281pt) -- +(135.000000:8.485281pt and 8.485281pt) -- +(225.000000:8.485281pt and 8.485281pt) -- cycle;
\draw (54.000000, -0.000000) node {$X$};
\end{scope}
\begin{scope}
\draw[fill=white] (78.000000, 30.000000) +(-45.000000:8.485281pt and 8.485281pt) -- +(45.000000:8.485281pt and 8.485281pt) -- +(135.000000:8.485281pt and 8.485281pt) -- +(225.000000:8.485281pt and 8.485281pt) -- cycle;
\clip (78.000000, 30.000000) +(-45.000000:8.485281pt and 8.485281pt) -- +(45.000000:8.485281pt and 8.485281pt) -- +(135.000000:8.485281pt and 8.485281pt) -- +(225.000000:8.485281pt and 8.485281pt) -- cycle;
\draw (78.000000, 30.000000) node {$X$};
\end{scope}
\begin{scope}
\draw[fill=white] (78.000000, -0.000000) +(-45.000000:8.485281pt and 8.485281pt) -- +(45.000000:8.485281pt and 8.485281pt) -- +(135.000000:8.485281pt and 8.485281pt) -- +(225.000000:8.485281pt and 8.485281pt) -- cycle;
\clip (78.000000, -0.000000) +(-45.000000:8.485281pt and 8.485281pt) -- +(45.000000:8.485281pt and 8.485281pt) -- +(135.000000:8.485281pt and 8.485281pt) -- +(225.000000:8.485281pt and 8.485281pt) -- cycle;
\draw (78.000000, -0.000000) node {$H$};
\end{scope}
\begin{scope}
\draw[fill=white] (102.000000, 30.000000) +(-45.000000:8.485281pt and 8.485281pt) -- +(45.000000:8.485281pt and 8.485281pt) -- +(135.000000:8.485281pt and 8.485281pt) -- +(225.000000:8.485281pt and 8.485281pt) -- cycle;
\clip (102.000000, 30.000000) +(-45.000000:8.485281pt and 8.485281pt) -- +(45.000000:8.485281pt and 8.485281pt) -- +(135.000000:8.485281pt and 8.485281pt) -- +(225.000000:8.485281pt and 8.485281pt) -- cycle;
\draw (102.000000, 30.000000) node {$H$};
\end{scope}
\end{tikzpicture}
\end{center}
\[{}
C|001\rangle\sim C \left[ \begin {array}{c} 0\\ \noalign{\medskip}1\\ \noalign{\medskip}0
\\ \noalign{\medskip}0\\ \noalign{\medskip}0\\ \noalign{\medskip}0
\\ \noalign{\medskip}0\\ \noalign{\medskip}0\end {array} \right] = -\frac1{\sqrt2}\left[ \begin {array}{c} 0\\ \noalign{\medskip}1
\\ \noalign{\medskip}0\\ \noalign{\medskip}0\\ \noalign{\medskip}0
\\ \noalign{\medskip}0\\ \noalign{\medskip}1\\ \noalign{\medskip}0
\end {array} \right] \sim -\frac1{\sqrt2}\left(|001\rangle+|110\rangle\right)
\]
	\caption{Example of the action of a Quantum circuit on a qubit system. $C$ denotes the operator associated to the circuit of figure \ref{GHZCirc}.\label{Action}}
\end{figure}

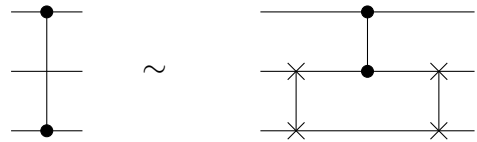
\begin{figure}[h]\begin{center}
\begin{tikzpicture}[scale=1.500000,x=1pt,y=1pt]
\filldraw[color=white] (0.000000, -7.500000) rectangle (36.000000, 37.500000);
\draw[color=black] (18.000000,30.000000) -- (36.000000,30.000000);
\draw[color=black] (18.000000,15.000000) -- (36.000000,15.000000);
\draw[color=black] (18.000000,0.000000) -- (36.000000,0.000000);
\draw (27.000000,30.000000) -- (27.000000,0.000000);
\filldraw (27.000000, 30.000000) circle(1.500000pt);
\filldraw (27.000000, 0.000000) circle(1.500000pt);
\draw (60,15)  node[left]{$\sim$};
\end{tikzpicture}
\begin{tikzpicture}[scale=1.500000,x=1pt,y=1pt]
\filldraw[color=white] (0.000000, -7.500000) rectangle (72.000000, 37.500000);
\draw[color=black] (18.000000,30.000000) -- (72.000000,30.000000);
\draw[color=black] (18.000000,15.000000) -- (72.000000,15.000000);
\draw[color=black] (18.000000,0.000000) -- (72.000000,0.000000);
\draw (27.000000,15.000000) -- (27.000000,0.000000);
\begin{scope}
\draw (24.878680, 12.878680) -- (29.121320, 17.121320);
\draw (24.878680, 17.121320) -- (29.121320, 12.878680);
\end{scope}
\begin{scope}
\draw (24.878680, -2.121320) -- (29.121320, 2.121320);
\draw (24.878680, 2.121320) -- (29.121320, -2.121320);
\end{scope}
\draw (45.000000,30.000000) -- (45.000000,15.000000);
\filldraw (45.000000, 30.000000) circle(1.500000pt);
\filldraw (45.000000, 15.000000) circle(1.500000pt);
\draw (63.000000,15.000000) -- (63.000000,0.000000);
\begin{scope}
\draw (60.878680, 12.878680) -- (65.121320, 17.121320);
\draw (60.878680, 17.121320) -- (65.121320, 12.878680);
\end{scope}
\begin{scope}
\draw (60.878680, -2.121320) -- (65.121320, 2.121320);
\draw (60.878680, 2.121320) -- (65.121320, -2.121320);
\end{scope}
\end{tikzpicture}

\end{center}
	 \caption{A $\cZ$ gate acting on two separated qubits\label{cZ13} simulated with $\Swap$ gates}
 \end{figure}

 Using $\Swap$ gates it is always possible to simulate gates acting on separate qubits on the circuits (see figure \ref{cZ13} as an example).
 From a purely algebraic point of view, the circuit compositions as well as the manipulations of the associated matrices fit in the context of the PRO theory (\emph{product categories}) \cite{1965Maclane,2004Leinster}. PRO are algebraic structures that allow to abstract behaviors of operators with several inputs and several outputs. The link between compositions of gates and manipulations of matrices fits in a representation theory of PRO \cite{2018LLMN}. This remark has no impact on the rest of the paper, but it shows that the problem takes place within a much broader framework that connect many domains in Mathematics, Physics and Computer Science.

\section{\label{AlgebraicStructure} The group generated by $\cZ$ and $\Swap$ gates}
 For a computational point of view, $2$-qubits gates are known to be universal \cite{1995Divencenzo}, i.e. any  quantum circuit admits an equivalent one composed only with single qubit  and $2$-qubits gates. More precisely, one can simulate any quantum system by using only single qubits gates together with $\cnot$ gates. In particular, we have
 \begin{equation}
	 \begin{tikzpicture}[scale=1.500000,x=1pt,y=1pt]
\filldraw[color=white] (0.000000, -7.500000) rectangle (18.000000, 22.500000);
\draw[color=black] (0.000000,15.000000) -- (18.000000,15.000000);
\draw[color=black] (0.000000,0.000000) -- (18.000000,0.000000);
\draw (9.000000,15.000000) -- (9.000000,0.000000);
\filldraw (9.000000, 0.000000) circle(1.500000pt);
\begin{scope}
\draw[fill=white] (9.000000, 15.000000) circle(3.000000pt);
\clip (9.000000, 15.000000) circle(3.000000pt);
\draw (6.000000, 15.000000) -- (12.000000, 15.000000);
\draw (9.000000, 12.000000) -- (9.000000, 18.000000);
\end{scope}
\end{tikzpicture}
 \raisebox{8mm}{$\quad \sim \quad$}
 \begin{tikzpicture}[scale=1.500000,x=1pt,y=1pt]
\filldraw[color=white] (0.000000, -7.500000) rectangle (66.000000, 22.500000);
\draw[color=black] (0.000000,15.000000) -- (66.000000,15.000000);
\draw[color=black] (0.000000,0.000000) -- (66.000000,0.000000);
\begin{scope}
\draw[fill=white] (12.000000, 15.000000) +(-45.000000:8.485281pt and 8.485281pt) -- +(45.000000:8.485281pt and 8.485281pt) -- +(135.000000:8.485281pt and 8.485281pt) -- +(225.000000:8.485281pt and 8.485281pt) -- cycle;
\clip (12.000000, 15.000000) +(-45.000000:8.485281pt and 8.485281pt) -- +(45.000000:8.485281pt and 8.485281pt) -- +(135.000000:8.485281pt and 8.485281pt) -- +(225.000000:8.485281pt and 8.485281pt) -- cycle;
\draw (12.000000, 15.000000) node {$H$};
\end{scope}
\begin{scope}
\draw[fill=white] (12.000000, -0.000000) +(-45.000000:8.485281pt and 8.485281pt) -- +(45.000000:8.485281pt and 8.485281pt) -- +(135.000000:8.485281pt and 8.485281pt) -- +(225.000000:8.485281pt and 8.485281pt) -- cycle;
\clip (12.000000, -0.000000) +(-45.000000:8.485281pt and 8.485281pt) -- +(45.000000:8.485281pt and 8.485281pt) -- +(135.000000:8.485281pt and 8.485281pt) -- +(225.000000:8.485281pt and 8.485281pt) -- cycle;
\draw (12.000000, -0.000000) node {$H$};
\end{scope}
\draw (33.000000,15.000000) -- (33.000000,0.000000);
\begin{scope}
\draw[fill=white] (33.000000, 0.000000) circle(3.000000pt);
\clip (33.000000, 0.000000) circle(3.000000pt);
\draw (30.000000, 0.000000) -- (36.000000, 0.000000);
\draw (33.000000, -3.000000) -- (33.000000, 3.000000);
\end{scope}
\filldraw (33.000000, 15.000000) circle(1.500000pt);
\begin{scope}
\draw[fill=white] (54.000000, 15.000000) +(-45.000000:8.485281pt and 8.485281pt) -- +(45.000000:8.485281pt and 8.485281pt) -- +(135.000000:8.485281pt and 8.485281pt) -- +(225.000000:8.485281pt and 8.485281pt) -- cycle;
\clip (54.000000, 15.000000) +(-45.000000:8.485281pt and 8.485281pt) -- +(45.000000:8.485281pt and 8.485281pt) -- +(135.000000:8.485281pt and 8.485281pt) -- +(225.000000:8.485281pt and 8.485281pt) -- cycle;
\draw (54.000000, 15.000000) node {H};
\end{scope}
\begin{scope}
\draw[fill=white] (54.000000, -0.000000) +(-45.000000:8.485281pt and 8.485281pt) -- +(45.000000:8.485281pt and 8.485281pt) -- +(135.000000:8.485281pt and 8.485281pt) -- +(225.000000:8.485281pt and 8.485281pt) -- cycle;
\clip (54.000000, -0.000000) +(-45.000000:8.485281pt and 8.485281pt) -- +(45.000000:8.485281pt and 8.485281pt) -- +(135.000000:8.485281pt and 8.485281pt) -- +(225.000000:8.485281pt and 8.485281pt) -- cycle;
\draw (54.000000, -0.000000) node {H};
\end{scope}
\end{tikzpicture}
 \end{equation}
 and
 \begin{equation}\label{transp}
	 \begin{tikzpicture}[scale=1.500000,x=1pt,y=1pt]
\filldraw[color=white] (0.000000, -7.500000) rectangle (36.000000, 22.500000);
\draw[color=black] (18.000000,15.000000) -- (36.000000,15.000000);
\draw[color=black] (18.000000,0.000000) -- (36.000000,0.000000);
\draw (27.000000,15.000000) -- (27.000000,0.000000);
\begin{scope}
\draw (24.878680, 12.878680) -- (29.121320, 17.121320);
\draw (24.878680, 17.121320) -- (29.121320, 12.878680);
\end{scope}
\begin{scope}
\draw (24.878680, -2.121320) -- (29.121320, 2.121320);
\draw (24.878680, 2.121320) -- (29.121320, -2.121320);
\end{scope}
\end{tikzpicture}
 \raisebox{8mm}{$\quad \sim \quad$}
 \begin{tikzpicture}[scale=1.500000,x=1pt,y=1pt]
\filldraw[color=white] (0.000000, -7.500000) rectangle (54.000000, 22.500000);
\draw[color=black] (0.000000,15.000000) -- (54.000000,15.000000);
\draw[color=black] (0.000000,0.000000) -- (54.000000,0.000000);
\draw (9.000000,15.000000) -- (9.000000,0.000000);
\begin{scope}
\draw[fill=white] (9.000000, 15.000000) circle(3.000000pt);
\clip (9.000000, 15.000000) circle(3.000000pt);
\draw (6.000000, 15.000000) -- (12.000000, 15.000000);
\draw (9.000000, 12.000000) -- (9.000000, 18.000000);
\end{scope}
\filldraw (9.000000, 0.000000) circle(1.500000pt);
\draw (27.000000,15.000000) -- (27.000000,0.000000);
\filldraw (27.000000, 15.000000) circle(1.500000pt);
\begin{scope}
\draw[fill=white] (27.000000, 0.000000) circle(3.000000pt);
\clip (27.000000, 0.000000) circle(3.000000pt);
\draw (24.000000, 0.000000) -- (30.000000, 0.000000);
\draw (27.000000, -3.000000) -- (27.000000, 3.000000);
\end{scope}
\draw (45.000000,15.000000) -- (45.000000,0.000000);
\begin{scope}
\draw[fill=white] (45.000000, 15.000000) circle(3.000000pt);
\clip (45.000000, 15.000000) circle(3.000000pt);
\draw (42.000000, 15.000000) -- (48.000000, 15.000000);
\draw (45.000000, 12.000000) -- (45.000000, 18.000000);
\end{scope}
\filldraw (45.000000, 0.000000) circle(1.500000pt);
\end{tikzpicture} \raisebox{4mm}{$\quad. \quad$}
 \end{equation}
We notice also that the $\cZ$ has the same property of universality as $\cnot$ because we have
\begin{equation}
	\begin{tikzpicture}[scale=1.500000,x=1pt,y=1pt]
\filldraw[color=white] (0.000000, -7.500000) rectangle (18.000000, 22.500000);
\draw[color=black] (0.000000,15.000000) -- (18.000000,15.000000);
\draw[color=black] (0.000000,0.000000) -- (18.000000,0.000000);
\draw (9.000000,15.000000) -- (9.000000,0.000000);
\filldraw (9.000000, 15.000000) circle(1.500000pt);
\begin{scope}
\draw[fill=white] (9.000000, 0.000000) circle(3.000000pt);
\clip (9.000000, 0.000000) circle(3.000000pt);
\draw (6.000000, 0.000000) -- (12.000000, 0.000000);
\draw (9.000000, -3.000000) -- (9.000000, 3.000000);
\end{scope}
\end{tikzpicture}
\raisebox{8mm}{$\quad \sim \quad$}
\begin{tikzpicture}[scale=1.500000,x=1pt,y=1pt]
\filldraw[color=white] (0.000000, -7.500000) rectangle (66.000000, 22.500000);
\draw[color=black] (0.000000,15.000000) -- (66.000000,15.000000);
\draw[color=black] (0.000000,0.000000) -- (66.000000,0.000000);
\begin{scope}
\draw[fill=white] (12.000000, -0.000000) +(-45.000000:8.485281pt and 8.485281pt) -- +(45.000000:8.485281pt and 8.485281pt) -- +(135.000000:8.485281pt and 8.485281pt) -- +(225.000000:8.485281pt and 8.485281pt) -- cycle;
\clip (12.000000, -0.000000) +(-45.000000:8.485281pt and 8.485281pt) -- +(45.000000:8.485281pt and 8.485281pt) -- +(135.000000:8.485281pt and 8.485281pt) -- +(225.000000:8.485281pt and 8.485281pt) -- cycle;
\draw (12.000000, -0.000000) node {$H$};
\end{scope}
\draw (33.000000,15.000000) -- (33.000000,0.000000);
\filldraw (33.000000, 15.000000) circle(1.500000pt);
\filldraw (33.000000, 0.000000) circle(1.500000pt);
\begin{scope}
\draw[fill=white] (54.000000, -0.000000) +(-45.000000:8.485281pt and 8.485281pt) -- +(45.000000:8.485281pt and 8.485281pt) -- +(135.000000:8.485281pt and 8.485281pt) -- +(225.000000:8.485281pt and 8.485281pt) -- cycle;
\clip (54.000000, -0.000000) +(-45.000000:8.485281pt and 8.485281pt) -- +(45.000000:8.485281pt and 8.485281pt) -- +(135.000000:8.485281pt and 8.485281pt) -- +(225.000000:8.485281pt and 8.485281pt) -- cycle;
\draw (54.000000, -0.000000) node {$H$};
\end{scope}
\end{tikzpicture}\raisebox{4mm}{$\quad.\quad$}
\end{equation}
In general, two qubits quantum gates implementations are unreliable and may cause many execution errors (see appendix \ref{appExperiment}). It is therefore of crucial importance to study the algebraic nature of the circuits in order to know how to use as few  two-qubits gates as possible.
In that context, we study the group \cZS$_{k}$\  generated by the $\cZ$ and $\Swap$ gates acting on $k$-qubits.
 The motivation for studying such a (toy) model  is that experimentally the order of the group is $k!2^{\binom k2}$, which suggests that it has interesting algebraic and combinatorial structures that can be exploited to simplify circuits.
 
 
 \subsection{The group \cZS$_{k}$ as a semi-direct product}
 Let us prove the formula for the order of  \cZS$_{k}$ and, at the same time, we exhibit the algebraic structure of this group. 
  The  \cZS$_{k}$\  group is not the only interesting finite subgroup of the unitary transform of $k$-qubit systems. Let us denote by $S_{i}$ the $\Swap$ gate acting simultaneously on the qubits $i$ and $i+1$ of the system. The group  $\mathcal S_{k}$ generated by the $S_{i}$'s is straightforwardly isomorphic to the symmetric group $\mathfrak S_{k}$, i.e. $\mathcal S_{k}$  is a faithful (but non irreducible) representation of  $\mathfrak S_{k}$. In  literature, a permutation is usually a bijection of $\{1,\dots,k\}$ but to make it  compatible with our notations, we instead consider that a permutation acts on $\{0,\dots,k-1\}$. 
   This  changes nothing to the theory of symmetric group (except a shift $-1$ for the notations).  A $\Swap$ gate is nothing but a transposition. We have seen that there is a one to one correspondence between the permutations of $\mathfrak S_{k}$ and the matrices of $\mathcal S_{k}$. To each permutation $\sigma$, we associate its corresponding matrix $S_{\sigma}$. 
  
  Similarly, denote by $Z_{ij}$ the matrix corresponding to the $\cZ$ gates acting on the qubits  $i$ and  $j$. We notice that 
  \begin{equation}
	  Z_{ij}=Z_{ji},\ 
	  Z_{ij}^{-1}=Z_{ij}, \mbox{and } Z_{ij}Z_{i'j'}=Z_{i'j'}Z_{ij}.
  \end{equation}
  Indeed, $Z_{ij}$ is a diagonal matrix where the entry $(\alpha,\alpha)$ equals $-1$ if both the bit $i$  of $\alpha$ and the  bit $j$ of $\alpha$ equal $1$, and $1$ otherwise.
  We deduce that the group $\mathcal P_{k}$ generated by the $Z_{ij}$'s is isomorphic to the group $\mathfrak P_{k}$ whose elements are the subsets of $\{\{i,j\}\mid 0\leq i,j\leq k-1\}$ and the product is the symmetric difference $\oplus$. For any $E\subset \{\{i,j\}\mid 0\leq i,j\leq k-1\}$, we denote by $Z_{E}$ the preimage of $E$ by this isomorphism.
  Each matrix $Z_{E}$ is  diagonal with only entries $1$ and $-1$ on the diagonal. More precisely, the set $E$ is completely encoded in the diagonal of $Z_{E}$ since the entry of coordinates $(\alpha_{k-1}\cdots \alpha_{0},\alpha_{k-1}\cdots \alpha_{0})$ equals 
  $(-1)^{\mathtt{card}\{\{i,j\}\in E\mid \alpha_{i}=\alpha_{j}=1\}}.$ 
  As an example,  \begin{equation}Z_{\{\{0,1\},\{0,2\}\}}=\mathtt{diag}(1,1,1,-1,1,-1,1,1),\end{equation} where $\mathtt{diag}(e_{1},\dots,e_{k})$ stands for the diagonal matrix with diagonal entries $e_{1},\dots,e_{k}$.
   So, the product is easy to describe as
  \begin{equation}\label{oplus}
	  Z_{E}Z_{E'}=Z_{E\oplus E'}.
  \end{equation}
  For instance, consider the following elements of $\mathcal P_{3}$: 
  \begin{equation} 
  Z_{\{\{0,1\},\{0,2\}\}}=Z_{01}Z_{02}= \mathtt{diag}(1,1,1,-1,1,-1,1,1),
  \end{equation}
  \begin{equation}
	  Z_{\{\{0,1\},\{1,2\}\}}=\mathtt{diag}(1,1,1,-1,1,1,-1,1)
,
  \end{equation}
and
\begin{equation}\begin{array}{rcl}
	 Z_{\{\{0,1\},\{0,2\}\}} Z_{\{\{0,1\},\{1,2\}\}}&= &\mathtt{diag}(1,1,1,1,1,-1,-1,1)\\
&=&Z_{\{\{0,2\},\{1,2\}\}}
.\end{array}
\end{equation}

The group $\mathcal P_{k}$ is abelian with order $2^{\binom k2}$. The group \cZS$_{k}$ is the smallest group containing both $\mathcal S_{k}$ and $\mathcal P_{k}$ as subgroups. The conjectured order suggests that the underlying set of  \cZS$_{k}$  is in bijection with the cartesian product $\mathfrak S_{k}\times \mathfrak P_{k}$. We remark also that the orbit of $Z_{01}$ for conjugation by the elements of $\mathcal S_{k}$ is the set $\{Z_{ij}\mid 0\leq i,j\leq k-1\}$ (see figure \ref{conj} for an example with $k=6$). To be more precise, we have
\begin{equation}
	S_{\sigma}Z_{ij}S_{\sigma}^{-1}=Z_{\sigma(i),\sigma(j)}.
\end{equation}
This extends to any element of $\mathcal P_{k}$ by
\begin{equation}\label{conj}
	S_{\sigma}Z_{E}S_{\sigma}^{-1}=S_{\sigma}\prod_{\{i,j\}\in E}Z_{ij}S_{\sigma}^{-1}=
	\prod_{\{i,j\}\in E}S_{\sigma}Z_{ij}S_{\sigma}^{-1}=
	Z_{\sigma(E)}.
\end{equation}
From equality (\ref{conj}), we deduce that any $G\in$\cZS$_{k}$ admits a unique decomposition $G=PS$ with $P\in\mathcal P_{k}$ and $S\in\mathcal S$. Indeed, such a decomposition exists since if $Z_{E}, Z_{E'}\in\mathcal P_{k}$ and $S_{\sigma}, S_{\sigma'}\in \mathcal S_{k}$ we have
\begin{equation}\label{ZSform}
	Z_{E}S_{\sigma}Z_{E'}S_{\sigma'}=Z_{E}S_{\sigma}Z_{E'}S_{\sigma}^{-1}S_{\sigma}S_{\sigma'}=
	Z_{E\oplus \sigma(E')}S_{\sigma\sigma'}.
\end{equation}
The decomposition is unique because if $PS=P'S'$ with $P,P'\in\mathcal P_{k}$ and $S,S'\in\mathcal S_{k}$ then $SS'^{-1}=P^{-1}P\in \mathcal P_{k}\cap\mathcal S_{k}=\{I_{2^{k}}\}$, and so $S=S'$ and $P=P'$.

The results of this section are summarized in the following statement.
\begin{theo}\label{TSemiDirect}
	The group \cZS$_{k}$ is a finite  group of order $k!2^{\binom k2}$. It is isomorphic to the semi-direct product $\mathfrak P_{k}\rtimes \mathfrak S_{k}$. We recall that the underlying set of  $\mathfrak P_{k}\rtimes \mathfrak S_{k}$ is the cartesian product  $\mathfrak P_{k}\times \mathfrak S_{k}$ and its product is defined by $(E,\sigma)(E',\sigma')=(E\oplus\sigma(E'),\sigma\sigma')$.
\end{theo} 

\subsection{\label{Quotient} The group \cZS$_{k}$ as the quotient of a Coxeter group}
\def\LU{\mathtt{LU}}
\def\LOCC{\mathtt{LOCC}}
\def\SLOCC{\mathtt{SLOCC}}
We give a few expressions of the group \cZS$_{k}$ as the quotient of  some Coxeter groups.
We recall that a Coxeter group (see eg \cite{2005BB}) is generated by a set of elements $g_{0},g_{1},\dots$ satisfying $(g_{i}g_{j})^{m_{ij}}=1$ where 
$m_{ij}\in \mathbb N\cup\{\infty\}\setminus\{0,1\}$ and $m_{ij}=1$ if and only if $i=j$; the condition $m_{ij}=\infty$ means that there is no relation of the form $(g_{i}g_{j})^{m}=1$. The relations are encoded in a Coxeter matrix $M=(m_{ij})_{ij}$ or, equivalently, in a Coxeter-Dynkin diagram which is the graph of the matrix $M$ where the edges $\{i,j\}$ where $m_{ij}\leq 2$ are removed and the edges where $m_{ij}=3$ are unlabeled. 
\begin{theo}\label{TFirstCox}
Let us denote by $\mathcal W_{k}$ the Coxeter group generated by $2(k-1)$ elements $g_{0},g_{1},\dots, g_{2(k-2)+1}$ submitted to the relations given by the Coxeter matrix \begin{equation}
M_{k}=\left[\begin{array}{ccccc}D&A&B&\cdots&B\\A&\ddots&\ddots&\ddots&\vdots\\B&\ddots&\ddots&\ddots&B\\\vdots&\ddots&\ddots&\ddots&A\\
B&\cdots&B&A&D \end{array}\right]\end{equation}
where $A=\left[\begin{array}{cc}2&4\\4&3 \end{array}\right]$, $B=\left[\begin{array}{cc}2&2\\2&2\end{array}\right]$, and $D=\left[\begin{array}{cc}1&2\\2&1\end{array}\right]$.
\\ The group \cZS$_{k}$ is isomorphic to the quotient $\mathcal W_{k}/_{\mathcal R_{k}}$ of $\mathcal W_{k}$ 
	by the relations $\mathcal R_{k}:=\{g_{2i+1}g_{2i+3}g_{2i}g_{2i+3}g_{2i+1}g_{2i+2}=1\mid 0\leq i\leq k-3\}$. The explicit isomorphism sends $Z_{i}$ to $g_{2i}$ and $S_{i}$ to $g_{2i+1}$.
\end{theo}
Although the proof is not very difficult, it is relatively long and technical. In order not to distract	 the reader, it has been relegated in   \ref{ProofTh}. 
\begin{example}
	\rm{}
	For $k=5$, the elements of the group \cZS$_{5}$  submitted to the relations $S_{i}^{2}=Z_{i}^{2}=1$, $(S_{0}S_{1})^{3}=(S_{1}S_{2})^{3}=(S_{2}S_{3})^{3}=1$, $(S_{0}S_{2})^{2}=(S_{0}S_{3})^{2}=(S_{1}S_{3})^{2}=1$, $(Z_{0}S_{1})^{4}=(Z_{1}S_{0})^{4}=(Z_{1}S_{2})^{4}=(Z_{2}S_{1})^{4}=(Z_{2}S_{3})^{4}=1$, $(Z_{0}S_{0})^{2}=(Z_{0}S_{2})^{2}=(Z_{0}S_{3})^{2}=(Z_{1}S_{1})^{2}=(Z_{1}S_{3})^{2}=(Z_{2}S_{0})^{2}=(Z_{2}S_{2})^{2}=(Z_{3}S_{0})^{2}=(Z_{3}S_{1})^{2}=(Z_{3}S_{3})^{2}$, and $S_{0}S_{1}Z_{0}S_{1}S_{0}Z_{1}=S_{1}S_{2}Z_{1}S_{2}S_{1}Z_{2}=S_{2}S_{3}Z_{2}S_{3}S_{2}Z_{3}=1$.  The group \cZS$_{5}$ is the quotient of the Coxeter group $\mathcal W_{5}$ with Coxeter diagram
	\begin{center} \begin{tikzpicture}
\draw[fill=black] 
(0,0)                         
      circle [radius=.1] node [below] {$Z_{0}$} --
(1,1) 
      circle [radius=.1] node [above] {$S_{1}$}
      node [midway,below] {$4$}--
(2,0) 
      circle [radius=.1] node [below] {$Z_{2}$}
      node [midway,below] {$4$}      --
(3,1) 
      circle [radius=.1] node [above] {$S_{3}$}
      node [midway,below] {$4$}      
      ;
\draw[fill=black]  (0,1)                         
      circle [radius=.1] node [below] {$S_{0}$} --
(1,2) 
      circle [radius=.1] node [above] {$Z_{1}$}
      node [midway,above] {$4$} -- node [midway,above] {$4$}
  (2,1)      circle [radius=.1] node [below] {$S_{2}$} --
  (3,2) 
      circle [radius=.1] node [above] {$Z_{3}$}
      node [midway,above] {$4$}
      ;
\draw (0,1)--(1,1)--(2,1)--(3,1);
\end{tikzpicture}\end{center}
by the relations $S_{0}S_{1}Z_{0}S_{1}S_{0}Z_{1}=S_{1}S_{2}Z_{1}S_{2}S_{1}Z_{2}=S_{2}S_{3}Z_{2}S_{3}S_{2}Z_{3}=1$.
\end{example}
There exists other ways to write \cZS$_{k}$ as the quotient of a Coxeter group. As an example, consider the following result that proves that it is isomorphic to the quotient of a Coxeter group generated by $k$ elements by a single relation.
\begin{theo}\label{ThPres}
	Let us denote by $\mathcal C_{k}$ the Coxeter group generated by $k$ elements $g_{0},g_{1},\dots, g_{k-1}$ submitted to the relations encoded in the Coxeter matrix
	\begin{equation}
		N_{k}=\left[\begin{array}{cccccc}1&2&4&2&\cdots&2\\2&1&3&2&&\vdots\\4&3&1&3&\ddots&\vdots\\
		2&2&3&1&\ddots&2\\\vdots&&\ddots&\ddots&\ddots&3\\2&\cdots&\cdots&2&3&1\end{array}\right].
	\end{equation}
	The group \cZS$_{k}$ is isomorphic to the quotient $\mathcal C_{k}/_{(g_{0}g_{2}g_{3}g_{1}g_{2})^{4}}$. The explicit isomorphism sends $Z_{0}$ to $g_{0}$ and each $S_{i}$ to $g_{i+1}$
\end{theo}
\begin{example}\rm
	The group \cZS$_{5}$ is isomorphic to the quotient of the Coxeter group $\mathcal C_{5}$ with Coxeter diagram
	\begin{center} \begin{tikzpicture}
\draw[fill=black] 
(0,0)                         
      circle [radius=.1] node [below] {$Z_{0}$} --
(1,1) 
      circle [radius=.1] node [above] {$S_{1}$}
      node [midway,below] {$4$}--
(2,1) 
      circle [radius=.1] node [above] {$S_{2}$}
       --
(3,1) 
      circle [radius=.1] node [above] {$S_{3}$}
         
      ;
      \draw[fill=black] 
(0,1)                         
      circle [radius=.1] node [above] {$S_{0}$} -- (1,1);
\end{tikzpicture}
\end{center}
by the relation $(Z_{0}S_{1}S_{2}S_{0}S_{1})^{4}$.
\end{example}
\section{\label{Optimization} Optimization of circuits of $\cZ$ and $\Swap$ gates}
We apply the result of the previous section in order to exhibit algorithms for simplifying circuits.
When we manipulate more than $2$ qubits, the network structure of the qubits  must be taken into account.  Indeed, if some connections are missing, then it is necessary to simulate some gates from the others and this can  increase dramatically the size of the circuit. For our purpose, we consider only two cases: the complete graph topology and the line topology.
\subsection{Optimization in circuits of $\Swap$ gates}
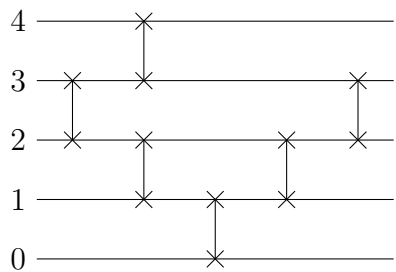
\begin{figure}[h]
	  \[\mathtt{Rothe}((0,3)(2,4))= \left[ \begin {array}{ccccc} 0&1&2&\ &\ 
\\ \noalign{\medskip}1&\ &\ &\ &\ \\
 \noalign{\medskip}2&\ &3&\ &\ 
\\ \noalign{\medskip}\ &\ &\ &\ &\ \\ \noalign{\medskip}\ 
&\ &\ &\ &\ \end {array}
 \right] \]
 \[(0,3)(2,4)=S_{2}S_{1}S_{0}S_{1}S_{3}S_{2}\]
 \begin{center}
 \begin{tikzpicture}[scale=1.500000,x=1pt,y=1pt]

\filldraw[color=white] (0.000000, -7.500000) rectangle (90.000000, 67.500000);
\draw (0.0,60) node[left]{$4$};
\draw (0.0,45) node[left]{$3$};
\draw (0.0,30) node[left]{$2$};
\draw (0.0,15) node[left]{$1$};
\draw (0.0,0) node[left]{$0$};

\draw[color=black] (0.000000,60.000000) -- (90.000000,60.000000);
\draw[color=black] (0.000000,45.000000) -- (90.000000,45.000000);
\draw[color=black] (0.000000,30.000000) -- (90.000000,30.000000);
\draw[color=black] (0.000000,15.000000) -- (90.000000,15.000000);
\draw[color=black] (0.000000,0.000000) -- (90.000000,0.000000);
\draw (9.000000,45.000000) -- (9.000000,30.000000);
\begin{scope}
\draw (6.878680, 27.878680) -- (11.121320, 32.121320);
\draw (6.878680, 32.121320) -- (11.121320, 27.878680);
\end{scope}
\begin{scope}
\draw (6.878680, 42.878680) -- (11.121320, 47.121320);
\draw (6.878680, 47.121320) -- (11.121320, 42.878680);
\end{scope}
\draw (27.000000,30.000000) -- (27.000000,15.000000);
\begin{scope}
\draw (24.878680, 12.878680) -- (29.121320, 17.121320);
\draw (24.878680, 17.121320) -- (29.121320, 12.878680);
\end{scope}
\begin{scope}
\draw (24.878680, 27.878680) -- (29.121320, 32.121320);
\draw (24.878680, 32.121320) -- (29.121320, 27.878680);
\end{scope}
\draw (27.000000,60.000000) -- (27.000000,45.000000);
\begin{scope}
\draw (24.878680, 42.878680) -- (29.121320, 47.121320);
\draw (24.878680, 47.121320) -- (29.121320, 42.878680);
\end{scope}
\begin{scope}
\draw (24.878680, 57.878680) -- (29.121320, 62.121320);
\draw (24.878680, 62.121320) -- (29.121320, 57.878680);
\end{scope}
\draw (45.000000,15.000000) -- (45.000000,0.000000);
\begin{scope}
\draw (42.878680, -2.121320) -- (47.121320, 2.121320);
\draw (42.878680, 2.121320) -- (47.121320, -2.121320);
\end{scope}
\begin{scope}
\draw (42.878680, 12.878680) -- (47.121320, 17.121320);
\draw (42.878680, 17.121320) -- (47.121320, 12.878680);
\end{scope}
\draw (63.000000,30.000000) -- (63.000000,15.000000);
\begin{scope}
\draw (60.878680, 12.878680) -- (65.121320, 17.121320);
\draw (60.878680, 17.121320) -- (65.121320, 12.878680);
\end{scope}
\begin{scope}
\draw (60.878680, 27.878680) -- (65.121320, 32.121320);
\draw (60.878680, 32.121320) -- (65.121320, 27.878680);
\end{scope}
\draw (81.000000,45.000000) -- (81.000000,30.000000);
\begin{scope}
\draw (78.878680, 27.878680) -- (83.121320, 32.121320);
\draw (78.878680, 32.121320) -- (83.121320, 27.878680);
\end{scope}
\begin{scope}
\draw (78.878680, 42.878680) -- (83.121320, 47.121320);
\draw (78.878680, 47.121320) -- (83.121320, 42.878680);
\end{scope}
\end{tikzpicture}
\end{center}
\caption{A permutation, its Rothe diagram, a reduced decomposition and the associated quantum circuit\label{PermEx}. In this figure, a permutation (acting on the set $\{0,\dots,k-1\}$) is represented by its decomposition into cycles. The symbol $S_{i}$ denotes the elementary transposition $(i,i+1)$.}
  \end{figure}

In order to illustrate the fact that the algebraic structure allows us to find efficient algorithm, let us investigate the simplest examples of circuits: those constituted only of $\Swap$ gates.  Indeed,  the group $\mathcal S_{k}$, generated by the gates $S_{i}$, is isomorphic to the symmetric group $\S_{k}$ and so is the simplest example of a finite subgroup of \cZS$_{k}$ for which the mechanism of simplification can be completely described. 
 In the case of the complete graph topology, the process to find a minimal decomposition of a permutation into transpositions is well known. It suffices to first decompose the permutation into cycles and hence decompose each cycle $(i_{1},\dots,i_{\ell})$ of length $\ell$ into $\ell-1$ transpositions
   \begin{equation}\label{transpo}
	   (i_{1},\dots,i_{\ell})=(i_{1}i_{2})\cdot(i_{2},i_{3})\cdots (i_{\ell-1},i_{\ell}).
   \end{equation}
In the case of the line topology, the algorithm is a bit more subtle but also is well known.
The minimal number of $\Swap$ gates necessary to obtain a given permutation $\sigma\in\mathfrak S_{k}$ is known to be the length $\ell(\sigma)$ of $\sigma$ and such a decomposition of $\sigma$ is called \emph{reduced}. One can compute a reduced decomposition of $\sigma$ by constructing its \emph{Rothe diagram} (see e.g. \cite{1973Knuth} pp.14-15). The construction being very classical, we will recall it briefly. The Rothe diagram of a permutation $\sigma\in\mathfrak S_{k}$ is a $k\times k$  square matrix, whose rows and columns are indexed by integers in $\{0,\dots,k-1\}$,  such that the only non-empty  entries have  coordinates $(r,c)$ ($r$ stands for the row number and $c$ for the column number) such that $(r,\sigma^{-1}(c))$ is an inversion in $\sigma$. The non-empty entries in the same column in the Rothe diagram are labeled with increasing successive integers from the top to the bottom; the highest entry in a column being labeled with the column number. A reduced decomposition is found by reading the entries from the right to the left and the top to the bottom (see figure \ref{PermEx} for an example).
\subsection{Optimization of circuits in \cZS$_{k}$ for the complete graph topology}

The main application of  Theorem \ref{TSemiDirect} is that it allows us to exhibit an algorithm for simplifying circuits constituted of $\Swap$ and $\cZ$ gates. The principle is very simple: first we write the circuit as a product of many elements $PS$ with $P\in\mathcal P_{k}$ and $S\in\mathcal S_{k}$, then we use successively many
times formula (\ref{ZSform}) in order to get only one element $PS$ and finally we use formula (\ref{transpo}) in order to write the permutation $S$ using $\Swap$.	{}
More precisely, as a direct consequence of  Theorem \ref{TSemiDirect} and  formula (\ref{ZSform}), the following algorithm allows us to give a reduced expression of an element of \cZS$_{k}$ in terms of the generators $Z_{\{i,j\}}$ and $S_{\{i,j\}}$ ($0\leq i<j\leq k-1$).
\begin{alg} {\rm \texttt{CtoZS}}\\
{\bf Input}: A circuit described as a sequence of gates $C=Z_{E_{0}}(S_{\sigma_{1}}Z_{E_{1}})\cdots (S_{\sigma_{\ell-1}}Z_{E_{\ell-1}})S_{\sigma_{\ell}}$ with $E_{0},\dots, E_{\ell-1}\subset \{\{i,j\}|0\leq i<j\leq k-1\}$ and $\sigma_{1},\dots,\sigma_{\ell}\in\mathfrak S_{k}$.\\
{\bf Ouput}: An equivalent description of the circuits under the form $Z_{E}S_{\sigma}$.
\begin{enumerate}
	\item Compute $\sigma'_{i}=\sigma_{1}\cdots \sigma_{i}$, for $i=1\dots \ell$.{}
	\item Compute $E'_{i}=E_{0}\oplus\sigma'_{1}(E_{1})\oplus\cdots \sigma'_{i}(E_{i})$, for $i=0\dots \ell-1$.{}
	\item Return $Z_{E'_{\ell-1}}S_{\sigma'_{\ell}}.$
\end{enumerate}
\end{alg}
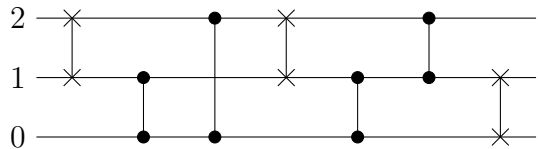
\begin{figure}[h]
	\begin{center}
\begin{tikzpicture}[scale=1.500000,x=1pt,y=1pt]
\filldraw[color=white] (0.000000, -7.500000) rectangle (126.000000, 37.500000);
\draw[color=black] (0.000000,30.000000) -- (126.000000,30.000000);
\draw(0,30) node[left] {$2$};
\draw[color=black] (0.000000,15.000000) -- (126.000000,15.000000);

\draw (0,15) node[left] {$1$};
\draw[color=black] (0.000000,0.000000) -- (126.000000,0.000000);
\draw (0,0) node[left] {$0$};
\draw (9.000000,30.000000) -- (9.000000,15.000000);
\begin{scope}
\draw (6.878680, 12.878680) -- (11.121320, 17.121320);
\draw (6.878680, 17.121320) -- (11.121320, 12.878680);
\end{scope}
\begin{scope}
\draw (6.878680, 27.878680) -- (11.121320, 32.121320);
\draw (6.878680, 32.121320) -- (11.121320, 27.878680);
\end{scope}
\draw (27.000000,15.000000) -- (27.000000,0.000000);
\filldraw (27.000000, 0.000000) circle(1.500000pt);
\filldraw (27.000000, 15.000000) circle(1.500000pt);
\draw (45.000000,30.000000) -- (45.000000,0.000000);
\filldraw (45.000000, 0.000000) circle(1.500000pt);
\filldraw (45.000000, 30.000000) circle(1.500000pt);
\draw (63.000000,30.000000) -- (63.000000,15.000000);
\begin{scope}
\draw (60.878680, 12.878680) -- (65.121320, 17.121320);
\draw (60.878680, 17.121320) -- (65.121320, 12.878680);
\end{scope}
\begin{scope}
\draw (60.878680, 27.878680) -- (65.121320, 32.121320);
\draw (60.878680, 32.121320) -- (65.121320, 27.878680);
\end{scope}
\draw (81.000000,15.000000) -- (81.000000,0.000000);
\filldraw (81.000000, 0.000000) circle(1.500000pt);
\filldraw (81.000000, 15.000000) circle(1.500000pt);
\draw (99.000000,30.000000) -- (99.000000,15.000000);
\filldraw (99.000000, 15.000000) circle(1.500000pt);
\filldraw (99.000000, 30.000000) circle(1.500000pt);
\draw (117.000000,15.000000) -- (117.000000,0.000000);
\begin{scope}
\draw (114.878680, -2.121320) -- (119.121320, 2.121320);
\draw (114.878680, 2.121320) -- (119.121320, -2.121320);
\end{scope}
\begin{scope}
\draw (114.878680, 12.878680) -- (119.121320, 17.121320);
\draw (114.878680, 17.121320) -- (119.121320, 12.878680);
\end{scope}
\end{tikzpicture}
\end{center}
\caption{Circuits corresponding to the operator $A:=S_{0}Z_{12}Z_{01}S_{1}Z_{02}Z_{01}S_{1}$\label{bigcirc}}
\end{figure}

\begin{figure}[h]\begin{center}
\begin{tikzpicture}[scale=1.500000,x=1pt,y=1pt]
\filldraw[color=white] (0.000000, -7.500000) rectangle (54.000000, 37.500000);
\draw[color=black] (0.000000,30.000000) -- (54.000000,30.000000);
\draw(0,30) node[left] {$2$};
\draw[color=black] (0.000000,15.000000) -- (54.000000,15.000000);
\draw(0,15) node[left] {$1$};
\draw[color=black] (0.000000,0.000000) -- (54.000000,0.000000);
\draw(0,0) node[left] {$0$};
\draw (9.000000,15.000000) -- (9.000000,0.000000);
\begin{scope}
\draw (6.878680, -2.121320) -- (11.121320, 2.121320);
\draw (6.878680, 2.121320) -- (11.121320, -2.121320);
\end{scope}
\begin{scope}
\draw (6.878680, 12.878680) -- (11.121320, 17.121320);
\draw (6.878680, 17.121320) -- (11.121320, 12.878680);
\end{scope}
\draw (27.000000,30.000000) -- (27.000000,15.000000);
\filldraw (27.000000, 15.000000) circle(1.500000pt);
\filldraw (27.000000, 30.000000) circle(1.500000pt);
\draw (45.000000,30.000000) -- (45.000000,0.000000);
\filldraw (45.000000, 0.000000) circle(1.500000pt);
\filldraw (45.000000, 30.000000) circle(1.500000pt);
\end{tikzpicture}
\end{center}

\caption{Simplification of the circuit of figure \ref{bigcirc}.\label{Ared}}
\end{figure}
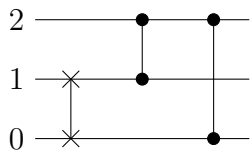

As an example, consider the circuit given in figure \ref{bigcirc}. We apply our algorithm on the associated operator,
\begin{equation}
	\begin{array}{rcl}
		A&=&Z_{\emptyset}S_{(0,1)}\cdot Z_{\{\{0,1\},\{1,2\},\}}S_{(1,2)}\cdot Z_{\{0,1\},\{0,2\}\}}S_{(1,2)}\\
		&=&Z_{\{\{0,1\},\{0,2\}\}}S_{(0,1,2)}\cdot Z_{\{0,1\},\{0,2\}\}}S_{(1,2)}\\
		&=& Z_{\{\{0,1\},\{0,2\}\}\oplus \{\{0,1\},\{1,2\}\}}S_{(0,1,2)}S_{(1,2)}\\
		&=& Z_{02}Z_{12}S_{(0,1)},
	\end{array}
\end{equation}
and we obtain the reduced circuit drawn in figure \ref{Ared}.

\subsection{Simplification of circuits in \cZS$_{k}$ for the line topology}Algorithm \texttt{CtoZS} is therefore suitable for  complete graphs. For technical reasons, current machines impose more restrictive conditions on the qubits network. Let us consider machines in which only gates acting on two adjacent qubits are allowed.  More precisely, we address the following problem: given a $cZS_k$ circuit written only with $Z_i$ and $S_i$ gates for $0\leq i\leq k-2$, find an efficient algorithm (i.e. having a reasonable polynomial complexity)  optimizing this circuit or, if it is not possible to obtain a minimal equivalent circuit in a polynomial time, at least giving a way to improve it.\\
First of all, it should be noted that there is a fairly obvious algorithm for finding the reduced decomposition. It consists in constructing the Cayley graph of the group according to the generators $Z_{i}$ and $S_{i}$ and hence deducing a reduced form by appliying a shorted path algorithm.
Of course, such an algorithm has  exponential time and space complexities with respect to the number of qubits, and is not practicable as soon as we exceed $6$ or $7$ qubits.\\
Another strategy consists in using the presentation of the group in order to reduce expressions. In that context, one of the main tools is the Dehn algorithm \cite{1989Lysenok}. Let us recall the principle. The starting point is a finite presentation of the group $G\sim\langle \mathcal S|\mathcal R\rangle$. Denote by $\tilde\mathcal R$ the  closure of ${\mathcal R}$ 
 under cyclic permutation of the symbols and inverse. We consider a reduced word $w$ in the free group $\mathbb F_{\mathcal S}$ generated by $\mathcal S$. The Dehn algorithm allows us to construct a sequence of word $w_{0}=w$, $w_{1}, w_{2},\dots,$ through the following process. It stops if $w_{i}$ is the empty word. Otherwise, if it exists a factor $u$ in $w_{i}$ which is the prefix of a word $r=uv$ in $\tilde\mathcal R$ with $|u|>|v|$, then the factor $u$ in $w_{i}$ is replaced by $v^{-1}$ and $w_{i+1}$ is the reduced word (in the free group) of this word. If such a word does not exist the algorithm stops. Obviously, we observe that the length of the words is strictly decreasing as the algorithm goes along. Hence, the Dehn algorithm allows us to compute a reduced (but not minimal) expression in a finite number of steps. In our special case, the Coxeter structure helps to improve the method. It suffices to apply successively many times the computation of a (minimal) reduced words in the Coxeter group $\mathcal W_{k}$ (see e.g. \cite{2005BB}) followed by the Dehn algorithm applied to the remaining relations. 
\begin{example}\rm
For instance, consider the circuit that implements the transform $C=Z_{0}Z_{3}S_{1}S_{0}Z_{1}Z_{3}S_{0}$. This circuit is not minimal in $\mathcal W_{5}$, since by using successively the relations $(Z_{3}Z_{1})^{2}$, $(Z_{3}S_{0})^{2}$, $(Z_{3}S_{1})^{2}$, and $Z_{3}^{2}$ it reduces to $C=Z_{0}S_{1}S_{0}Z_{1}S_{0}$. Hence, we use the Dehn algorithm, remarking that $Z_{0}S_{1}S_{0}Z_{1}S_{0}$ is a prefix of the additional relation $Z_{0}S_{1}S_{0}Z_{1}S_{0}S_{1}$  (obtained from  $S_{0}S_{1}Z_{0}S_{1}S_{0}Z_{1}$ by applying a cyclic permutation), and we deduce that $C$ reduces to $S_{1}$.
\end{example}
Nevertheless, this is often not sufficient to obtain a minimal circuit as shown by the following example.
\begin{example}\rm{}
  The circuit describing the composition $C=S_{3}S_{2}Z_{1}S_{2}S_{3}S_{2}Z_{1}S_{2}$ is minimal in $\mathcal W_{5}$. The minimality is checked by using classical algorithms of reduction for Coxeter groups  (See e.g. \cite{2005BB}).

  The only relations that could be used in the Dehn algorithm are $(S_{2}S_{3})^{3}$, $(Z_{1}S_{2})^{4}$, $(S_{2}Z_{1})^{4}$ and $S_{1}S_{2}Z_{1}S_{2}S_{1}Z_{2}$. But no subword of $S_{3}S_{2}Z_{1}S_{2}S_{3}S_{2}Z_{1}S_{2}$ is a prefix of a relation $r=uv$  with $|u|>|v|$, so the circuit cannot be reduced using Dehn algorithm and the strategy which consists to apply successively a reduction in the Coxeter group and the Dehn algorithm fails to compute a shorter equivalent circuit. Nevertheless, from $S_{2}Z_{1}S_{2}=S_{1}Z_{2}S_{1}$, one obtains $C=S_{2}S_{1}Z_{2}S_{1}S_{3}S_{1}Z_{2}S_{1}$ and it reduces to  $C=S_{2}S_{1}Z_{2}S_{3}Z_{2}S_{1}$  from $(S_{1}S_{3})^{2}$ and $S_{1}^{2}$.
\end{example}
We will continue to investigate technics of reduction in future works.
\section{\label{Entanglement} $\cZ$ gates and  entanglement}
Local unitary operations ($\LU=U(2)^{\otimes k}$) can be implemented on quantum circuits. They transform  states into others without changing their entanglement properties. The notion of $\LU$-equivalence appears to be the finest allowing to distinguish two states but it does not take into account more subtle communication protocols. The notion of entanglement is usually defined through the group of stochastic local operations assisted by classical communication ($\SLOCC$). This group of operations allows to locally change the amplitude of a state for instance, by applying  unitary operations on  bigger Hilbert spaces obtained by adding ancillary particles.  Mathematically, two states $|\psi\rangle$ and $|\phi\rangle$ are $\SLOCC$-equivalent if there exist $k$ operators $A_{1}, \dots, A_{k}$ such that $A_{1}\otimes\cdots\otimes A_{k}|\psi\rangle=\lambda |\phi\rangle$ for some complex number $\lambda$. In other words, $|\psi\rangle$ and $|\phi\rangle$ are $\SLOCC$-equivalent if they are in the same  orbit of $GL(2)^{\otimes k}$ acting on the Hilbert space $\C^{2}\otimes\cdots\otimes\C^{2}$. 
Since the relevant states belong to the unit sphere, one has only to consider the action of $SL(2)^{\otimes k}$ on the projective space   $\mathbb P(\C^{2}\otimes\cdots\otimes\C^{2})$.
Each orbit is a set of states which are entangled in the same way. Let us conclude this introductory paragraph by noting that the naive definition of entanglement as it can be naturally generalized from $2$-qubit systems (the measurement of one component of the system determines the measurement of the other components) cannot be applied to systems of $3$ qubits and more. For instance, the state $|\GHZ_{3}\rangle$ is $\SLOCC$-equivalent  (and also $\LU$-equivalent) to the state $|\GHZ'_{3}\rangle:=\frac12(|000\rangle + |011\rangle+|101\rangle +|110\rangle)$ through the map $|i\rangle\rightarrow \frac1{\sqrt2}(|0\rangle+(-1)^{i}|1\rangle)$, i.e. $|\GHZ'_{3}\rangle=H^{\otimes 3}|\GHZ_{3}\rangle$. In $|\GHZ'_{3}\rangle$, the value of the first qubit does not determine the values of the others but the property of entanglement can be seen when the state is rewritten as $|\GHZ'_{3}\rangle=\frac1{\sqrt2}\left(\left(|0\rangle+|1\rangle\over\sqrt2\right)^{\otimes 3}+\left(|0\rangle-|1\rangle\over\sqrt2\right)^{\otimes 3}\right)$. In fact, entanglement is not a property that depends only on one of the observables but on the whole space of observables.
Choosing an observable is equivalent to modifying the base of the Hilbert space by acting with an element of the Lie group associated to the Lie algebra of observables. Hence, entanglement is a $\SLOCC$ (or $\LU$, depending on the problem you're considering) invariant property.
\subsection{\label{Max Entanglement} $\LU$-equivalence to $|\GHZ_{k}\rangle$}
Although the group \cZS$_{k}$ does not contain all quantum gates, it is still powerful enough to generate a state equivalent to $|\GHZ_{k}\rangle$. Remark that $\Swap$ gates can be avoided as they do not generate any entanglement.
More precisely, let us show the following result.
\begin{prop}\label{F2GHZ} The state 
\begin{equation}\label{ZZ}
	\frac1{\sqrt{2^{k}}}Z_{\{\{0,1\},\{0,2\},\cdots,\{0,k-1\}\}}\left(|0\rangle+|1\rangle\right)^{\otimes k}
\end{equation}
is $\LU$-equivalent to $|\GHZ_{k}\rangle$.
\end{prop}
A fast computation shows
\begin{equation}
|\Psi_{E}\rangle:=\frac1{\sqrt{2^{k}}}Z_{E}\left(|0\rangle+|1\rangle\right)^{\otimes k}=
\displaystyle\frac1{\sqrt{2^{k}}}\sum_{0\leq i_{0},\dots,i_{k-1}\leq 1}(-1)^{\sum_{\{\alpha,\beta\}\in E}i_{\alpha}i_{\beta}}|i_{k-1}\cdots i_{0}\rangle.
\end{equation}
In particular
\begin{equation}\begin{array}{l}
	|\Psi_{\{\{0,1\},\{0,2\},\cdots,\{0,k-1\}\}}\rangle=\displaystyle\frac1{\sqrt{2^{k}}}\sum_{0\leq i_{0},\dots,i_{k-1}\leq 1}(-1)^{i_{0}(i_{1}+\cdots+i_{k-1})}|i_{k-1}\cdots i_{0}\rangle.\end{array}
\end{equation}
Since it is not factorizing, it is entangled. However, it is not completely obvious to see that such a state is $\LU$-equivalent to $|\GHZ_{k}\rangle$. 
Acting by $H$ on the qubit $\ell>0$, one obtains
\begin{equation}
	\begin{array}{l}
	\left(I_{2^{k-\ell-1}}\otimes H\otimes I_{2^{\ell}}\right)|\Psi_{\{\{0,1\},\{0,2\},\cdots,\{0,k-1\}\}}\rangle\\=\displaystyle\frac1{\sqrt{2^{k-1}}}\sum_{0\leq i_{1},\dots,i_{\ell-1},i_{\ell+1}\dots i_{k-1}\leq 1}\left(|i_{k-1}\cdots i_{\ell-1}0i_{\ell+1}\cdots i_{1}0\rangle\right.\\\left.	+(-1)^{i_{1}+\cdots+i_{\ell-1}+i_{\ell+1}+\cdots+i_{k-1}	}|i_{k-1}\cdots i_{\ell-1}1i_{\ell+1}\cdots i_{1} 1\rangle\right).
	\end{array}
\end{equation}
By iterating on all the qubits but the qubit $0$, one finds
\begin{equation}\label{enGHZ}
	\left(H^{\otimes k-1}\otimes I_{2}\right)|\Psi_{\{\{0,1\},\{0,2\},\cdots,\{0,k-1\}\}}\rangle=|\mathtt{GHZ}_{k}\rangle.
\end{equation}
Since the action on a single qubit does not change the entanglement properties, equation (\ref{enGHZ}) shows that the state (\ref{ZZ}) is $\LU$-equivalent to $|\GHZ_{k}\rangle$. Figure \ref{F2GHZ} contains an example of such a circuit for $k=5$.
\begin{figure}[h]\begin{center}
\begin{tikzpicture}[scale=1.200000,x=1pt,y=1pt]
\filldraw[color=white] (0.000000, -7.500000) rectangle (120.000000, 67.500000);
\draw[color=black] (0.000000,60.000000) -- (120.000000,60.000000);
\draw (0,60) node[left]{$|0\rangle$};
\draw (0,45) node[left]{$|0\rangle$};
\draw (0,30) node[left]{$|0\rangle$};
\draw (0,15) node[left]{$|0\rangle$};
\draw (0,0) node[left]{$|0\rangle$};
\draw (120,30) node[right]{\Huge$|\mathtt{GHZ}_{5}\rangle$};
\draw[color=black] (0.000000,45.000000) -- (120.000000,45.000000);
\draw[color=black] (0.000000,30.000000) -- (120.000000,30.000000);
\draw[color=black] (0.000000,15.000000) -- (120.000000,15.000000);
\draw[color=black] (0.000000,0.000000) -- (120.000000,0.000000);
\begin{scope}
\draw[fill=white] (12.000000, 60.000000) +(-45.000000:8.485281pt and 8.485281pt) -- +(45.000000:8.485281pt and 8.485281pt) -- +(135.000000:8.485281pt and 8.485281pt) -- +(225.000000:8.485281pt and 8.485281pt) -- cycle;
\clip (12.000000, 60.000000) +(-45.000000:8.485281pt and 8.485281pt) -- +(45.000000:8.485281pt and 8.485281pt) -- +(135.000000:8.485281pt and 8.485281pt) -- +(225.000000:8.485281pt and 8.485281pt) -- cycle;
\draw (12.000000, 60.000000) node {$H$};
\end{scope}
\begin{scope}
\draw[fill=white] (12.000000, 45.000000) +(-45.000000:8.485281pt and 8.485281pt) -- +(45.000000:8.485281pt and 8.485281pt) -- +(135.000000:8.485281pt and 8.485281pt) -- +(225.000000:8.485281pt and 8.485281pt) -- cycle;
\clip (12.000000, 45.000000) +(-45.000000:8.485281pt and 8.485281pt) -- +(45.000000:8.485281pt and 8.485281pt) -- +(135.000000:8.485281pt and 8.485281pt) -- +(225.000000:8.485281pt and 8.485281pt) -- cycle;
\draw (12.000000, 45.000000) node {$H$};
\end{scope}
\begin{scope}
\draw[fill=white] (12.000000, 30.000000) +(-45.000000:8.485281pt and 8.485281pt) -- +(45.000000:8.485281pt and 8.485281pt) -- +(135.000000:8.485281pt and 8.485281pt) -- +(225.000000:8.485281pt and 8.485281pt) -- cycle;
\clip (12.000000, 30.000000) +(-45.000000:8.485281pt and 8.485281pt) -- +(45.000000:8.485281pt and 8.485281pt) -- +(135.000000:8.485281pt and 8.485281pt) -- +(225.000000:8.485281pt and 8.485281pt) -- cycle;
\draw (12.000000, 30.000000) node {$H$};
\end{scope}
\begin{scope}
\draw[fill=white] (12.000000, 15.000000) +(-45.000000:8.485281pt and 8.485281pt) -- +(45.000000:8.485281pt and 8.485281pt) -- +(135.000000:8.485281pt and 8.485281pt) -- +(225.000000:8.485281pt and 8.485281pt) -- cycle;
\clip (12.000000, 15.000000) +(-45.000000:8.485281pt and 8.485281pt) -- +(45.000000:8.485281pt and 8.485281pt) -- +(135.000000:8.485281pt and 8.485281pt) -- +(225.000000:8.485281pt and 8.485281pt) -- cycle;
\draw (12.000000, 15.000000) node {$H$};
\end{scope}
\begin{scope}
\draw[fill=white] (12.000000, -0.000000) +(-45.000000:8.485281pt and 8.485281pt) -- +(45.000000:8.485281pt and 8.485281pt) -- +(135.000000:8.485281pt and 8.485281pt) -- +(225.000000:8.485281pt and 8.485281pt) -- cycle;
\clip (12.000000, -0.000000) +(-45.000000:8.485281pt and 8.485281pt) -- +(45.000000:8.485281pt and 8.485281pt) -- +(135.000000:8.485281pt and 8.485281pt) -- +(225.000000:8.485281pt and 8.485281pt) -- cycle;
\draw (12.000000, -0.000000) node {$H$};
\end{scope}
\draw (33.000000,15.000000) -- (33.000000,0.000000);
\filldraw (33.000000, 15.000000) circle(1.500000pt);
\filldraw (33.000000, 0.000000) circle(1.500000pt);
\draw (51.000000,30.000000) -- (51.000000,0.000000);
\filldraw (51.000000, 30.000000) circle(1.500000pt);
\filldraw (51.000000, 0.000000) circle(1.500000pt);
\draw (69.000000,45.000000) -- (69.000000,0.000000);
\filldraw (69.000000, 45.000000) circle(1.500000pt);
\filldraw (69.000000, 0.000000) circle(1.500000pt);
\draw (87.000000,60.000000) -- (87.000000,0.000000);
\filldraw (87.000000, 60.000000) circle(1.500000pt);
\filldraw (87.000000, 0.000000) circle(1.500000pt);
\begin{scope}
\draw[fill=white] (108.000000, 15.000000) +(-45.000000:8.485281pt and 8.485281pt) -- +(45.000000:8.485281pt and 8.485281pt) -- +(135.000000:8.485281pt and 8.485281pt) -- +(225.000000:8.485281pt and 8.485281pt) -- cycle;
\clip (108.000000, 15.000000) +(-45.000000:8.485281pt and 8.485281pt) -- +(45.000000:8.485281pt and 8.485281pt) -- +(135.000000:8.485281pt and 8.485281pt) -- +(225.000000:8.485281pt and 8.485281pt) -- cycle;
\draw (108.000000, 15.000000) node {$H$};
\end{scope}
\begin{scope}
\draw[fill=white] (108.000000, 30.000000) +(-45.000000:8.485281pt and 8.485281pt) -- +(45.000000:8.485281pt and 8.485281pt) -- +(135.000000:8.485281pt and 8.485281pt) -- +(225.000000:8.485281pt and 8.485281pt) -- cycle;
\clip (108.000000, 30.000000) +(-45.000000:8.485281pt and 8.485281pt) -- +(45.000000:8.485281pt and 8.485281pt) -- +(135.000000:8.485281pt and 8.485281pt) -- +(225.000000:8.485281pt and 8.485281pt) -- cycle;
\draw (108.000000, 30.000000) node {$H$};
\end{scope}
\begin{scope}
\draw[fill=white] (108.000000, 45.000000) +(-45.000000:8.485281pt and 8.485281pt) -- +(45.000000:8.485281pt and 8.485281pt) -- +(135.000000:8.485281pt and 8.485281pt) -- +(225.000000:8.485281pt and 8.485281pt) -- cycle;
\clip (108.000000, 45.000000) +(-45.000000:8.485281pt and 8.485281pt) -- +(45.000000:8.485281pt and 8.485281pt) -- +(135.000000:8.485281pt and 8.485281pt) -- +(225.000000:8.485281pt and 8.485281pt) -- cycle;
\draw (108.000000, 45.000000) node {$H$};
\end{scope}
\begin{scope}
\draw[fill=white] (108.000000, 60.000000) +(-45.000000:8.485281pt and 8.485281pt) -- +(45.000000:8.485281pt and 8.485281pt) -- +(135.000000:8.485281pt and 8.485281pt) -- +(225.000000:8.485281pt and 8.485281pt) -- cycle;
\clip (108.000000, 60.000000) +(-45.000000:8.485281pt and 8.485281pt) -- +(45.000000:8.485281pt and 8.485281pt) -- +(135.000000:8.485281pt and 8.485281pt) -- +(225.000000:8.485281pt and 8.485281pt) -- cycle;
\draw (108.000000, 60.000000) node {$H$};
\end{scope}
\end{tikzpicture}
\end{center}
\caption{The  entangled state $|\mathtt{GHZ}_{5}\rangle$ created from the completely factorized state $|00000\rangle$. The first five $H$ gates are used to create superposition and generate the state $\frac1{4\sqrt2}(|0\rangle+|1\rangle)^{5}$ from $|00000\rangle$.\label{F2GHZ}}
\end{figure}
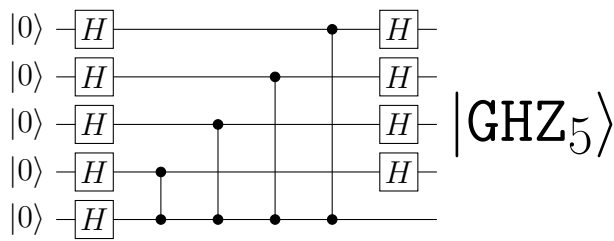
Notice that $|\GHZ_{k}\rangle$ is a generic entangled state in the sense of Miyake \cite{2003Miyake} only for $k\leq 3$ (see appendix \ref{appGHZ}). In the rest of the section, we prove that the group  \cZS$_{k}$ does not generate all entanglement types from a completely factorized state and in general, it is not possible to compute a generic entangled state by using only these operations.
\subsection{$\SLOCC$-equivalence to $|\mathtt{W}_3\rangle$}
We shall prove that the  group \cZS$_{k}$  is not powerful enough to generate any entanglement type from a completely factorized state. More precisely, we shall find  a counter-example for $k=3$ qubits: it is not possible to generates a state which is $\SLOCC$-equivalent to $|\mathtt W_{3}\rangle$. We use the method pioneered by Klyachko  in \cite{2002Klyachko}  wherein he promoted the use of Algebraic Theory of Invariant. The states of a $\SLOCC$-orbit are characterized by their values on covariant polynomials. Let $|\psi\rangle=\sum_{i,j,k}\alpha_{ijk}|ijk\rangle$. 
For our purpose we consider only two polynomials: 
\begin{equation}\begin{array}{rcl}
	\Delta(|\psi\rangle)&=& \left(\alpha_{000} \alpha_{111} - \alpha_{001} \alpha_{110} - \alpha_{010} \alpha_{101} + \alpha_{011} \alpha_{100}\right)^{2}\\&&-4\left( \alpha_{000} \alpha_{011} -  \alpha_{001} \alpha_{010}\right) \left(\alpha_{100} \alpha_{111} - \alpha_{101} \alpha_{110}\right)\end{array}
\end{equation}
and the catalecticant
\begin{equation}
	C(x_0,x_1,y_0,y_1,z_0,z_1)=\left[\begin{array}{cc}
  \frac{\partial A}{\partial x_0}&\frac{\partial A}{\partial x_1}\\
  \frac{\partial B_x}{\partial x_0}&\frac{\partial B_x}{\partial x_1}\\
\end{array}\right],
\end{equation}
with $A=\sum_{ijk}\alpha_{ijk}x_{i}y_{j}z_{k}$ and
\begin{equation}
	B_x(x_0,x_1)=\left[\begin{array}{cc}
  \frac{\partial^2A}{\partial y_0\partial z_0}&\frac{\partial^2A}{\partial y_0\partial z_1}\\
  \frac{\partial^2A}{\partial y_1\partial z_0}&\frac{\partial^2A}{\partial y_1\partial z_1}\\
\end{array}\right].
\end{equation}
The states which are $\SLOCC$-equivalent to $|\GHZ_{3}\rangle$ are characterized by $\Delta\neq 0$ while the states which are $\SLOCC$-equivalent to $|W_{3}\rangle$ are characterized by $C\neq 0$ and $\Delta=0$ \cite{2012HLT}. For the other states, we have $\Delta=C=0$.\\
We remark that 
\begin{equation}
	Z_{E}=\mathtt{diag}(1,1,1,\epsilon_{\{0,1\}},1,\epsilon_{\{0,2\}},\epsilon_{\{1,2\}},\epsilon_{\{0,1\}}\epsilon_{\{0,2\}}\epsilon_{\{1,2\}})
\end{equation}
where $\epsilon_{\{0,1\}}, \epsilon_{\{0,2\}}$, and  $\epsilon_{\{1,2\}}\in\{-1,1\}$ and we consider the states
\begin{equation}
	|\phi_{E}\rangle:=Z_{E}\sum_{ijk}a_{i}b_{j}c_{k}|ijk\rangle=Z_{E}(a_{0}|0\rangle+a_{1}|1\rangle){}
	(b_{0}|0\rangle+b_{1}|1\rangle)(c_{0}|0\rangle+c_{1}|1\rangle).
\end{equation}
We have
\begin{equation}
	C(|\phi_{E}\rangle)=a_{0}b_{0}c_{0}a_{1}b_{1}c_{1}(2-\epsilon_{\{0,1\}}-\epsilon_{\{0,2\}}-\epsilon_{\{1,2\}}+\epsilon_{\{0,1\}}\epsilon_{\{0,2\}}\epsilon_{\{1,2\}})P
\end{equation}
where $P=\displaystyle\sum_{i_{0}i_{1}i_{2}}a_{i_{2}}b_{i_{1}}c_{i_{0}}(-1)^{\mathrm{card}\{j\mid i_{j}=1\}}\left(\prod_{\{j,k\}\mid i_{j}+i_{k}>0}\epsilon_{\{j,k\}}\right)x_{i_{2}}y_{i_{1}}z_{i_{0}}$ is a non zero trilinear form,
and
\begin{equation}
\Delta(|\phi_{E}\rangle)=4(a_{0}b_{0}c_{0}a_{1}b_{1}c_{1})\epsilon_{01}\epsilon_{02}\epsilon_{12}(2-\epsilon_{01}-\epsilon_{02}-\epsilon_{12}+\epsilon_{01}\epsilon_{02}\epsilon_{12}).\end{equation}
So if $\Delta$ vanishes then $C$ also vanishes. This implies the following result
\begin{prop}
	The state $|\phi_{E}\rangle$ is not $\SLOCC$-equivalent to $|W_{3}\rangle$.
\end{prop}
However, it is interesting to note that it is possible to join a state of the $\LU$-orbit of $|\GHZ_{3}\rangle$ to a state of the $\SLOCC$-orbit of $|\mathtt W_{3}\rangle$. Consider the state
\[{}
|\phi_{1}\rangle:=\left(R_{y}(\frac\pi4) HX\otimes R_{y}(\frac\pi4)\otimes HX\right)Z_{01}Z_{12}(|0\rangle+|1\rangle)^{3}
\]
The state $|\phi_{1}\rangle$ is in the $\LU$-orbit of $|GHZ\rangle$ since 
\begin{equation}
	\left(R_{y}(\frac\pi4)^{-1} \otimes R_{y}(\frac\pi4)^{-1}\otimes I\right)|\phi_{1}\rangle=|\GHZ_{3}\rangle.
\end{equation}
A fast computation shows 
\begin{equation}
	\Delta\left(Z_{12}|\phi_{1}\rangle\right)=0\mbox{ and } C\left(Z_{12}|\phi_{1}\rangle\right)\neq 0,
\end{equation}
equivalently $Z_{12}|\phi_{1}\rangle$ is in the $\SLOCC$-orbit of $|\mathtt W_{3}\rangle$.{}
\subsection{Four qubits systems\label{four}}
The situation of 4-qubits systems is more complex than for 3-qubits systems. Nevertheless, there still is a classification of entanglement as well as related mathematical tools. We  refer to the classification of Verstraete et al \cite{2002VDDV} which assigns any $4$-qubit state to one of $9$ families. Any state in the more general situation (in the sense of Zarisky topology) is $\SLOCC$-equivalent to $192$ Verstraete states of the family
\begin{equation}\begin{array}{rcl}
	G_{abcd}&=&{a+d\over 2}\left(|0000\rangle + |1111\rangle\right)+{a-d\over 2}\left(|0011\rangle+|1100\rangle\right){}
	\\&&+{b+c\over 2}\left(|0101\rangle+|1010\rangle\right)+{b-c\over 2}\left(|0110\rangle+|1001\rangle\right).\end{array}
\end{equation}
for  independent parameters $a,b, c,$ and $d$ \cite{2002VDDV,2014HLT,2017HLT}.
To determine the Verstraete family to which a state belongs, we use an algorithm  described in a previous paper \cite{2017HLT}. This algorithm is based on the evaluation of some covariants. Recall that the algebra of (relative) $\SLOCC$-invariant is freely generated by the four following polynomials \cite{2003LT}:
\begin{itemize}
	\item The smallest degree invariant
	\begin{equation}{}
	B:=\sum_{0\leq i_{1},i_{2},i_{3}\leq 1}(-1)^{i_{1}+i_{2}+i_{3}}\alpha_{0i_{1}i_{2}i_{3}}\alpha_{1(1-i_{1})(1-i_{2})(1-i_{3})},
	\end{equation}
	\item Two polynomials of degree $4$
	\begin{equation}
		L:=\left|\begin{array}{cccc}\alpha_{0000}&\alpha_{0010}&\alpha_{0001}&\alpha_{0011}\\
		\alpha_{1000}&\alpha_{1010}&\alpha_{1001}&\alpha_{1011}\\
		\alpha_{0100}&\alpha_{0110}&\alpha_{0101}&\alpha_{0111}\\
		\alpha_{1100}&\alpha_{1110}&\alpha_{1101}&\alpha_{1111}\end{array}
		\right|\end{equation}  and 
		\begin{equation} 
		M:=\left|\begin{array}{cccc} \alpha_{0000}&\alpha_{0001}&\alpha_{0100}&\alpha_{0101}\\
		 \alpha_{1000}&\alpha_{1001}&\alpha_{1100}&\alpha_{1101}\\
		  \alpha_{0010}&\alpha_{0011}&\alpha_{0110}&\alpha_{0111}\\
		   \alpha_{1010}&\alpha_{1011}&\alpha_{1110}&\alpha_{1111}\end{array}\right|.
	\end{equation}
	\item and a polynomial of degree $6$ defined by $D_{xy}=-\det(B_{xy})$ where $B_{xy}$ is the $3\times 3$ matrix satisfying
	\begin{equation}
		\left[x_{0}^{2},x_{0}x_{1},x_{1}^{2}\right]B_{xy}\left[\begin{array}{c} y_{0}^{2}\\y_{0}y_{1}\\y_{1}^{2}\end{array}\right]=\det\left({\partial^{2}\over\partial z_{i}\partial t_{j}}A\right)
	\end{equation}
\noindent with $A=\sum_{ijkl}\alpha_{ijkl}x_{i}y_{j}z_{k}t_{l}$.
\end{itemize}
For our purpose we define also \begin{equation}
N=-L-M=\left|\begin{array}{cccc} \alpha_{0000}&\alpha_{1000}&\alpha_{0001}&\alpha_{1001}\\
		 \alpha_{0100}&\alpha_{1100}&\alpha_{0101}&\alpha_{1101}\\
		  \alpha_{0010}&\alpha_{1010}&\alpha_{0011}&\alpha_{1011}\\
		   \alpha_{0110}&\alpha_{1110}&\alpha_{0111}&\alpha_{1111}\end{array}\right|.
\end{equation}
We need also the covariant polynomials $\overline{\mathcal G}$, $\mathcal G$, $\mathcal H$, $\mathcal K_{3}$, and $\mathcal L$ defined in \cite{2017HLT} and those complete definition is relegated to appendix.
We recall the principle of the algorithm as described in \cite{2017HLT}. A first coarser classification is obtained by investigating the roots of the three quartics
\begin{equation}
	Q_{1}=x^{4}-2Bx^{3}y+(B^{2}+2L+4M)x^{2}y^{2}+4(D_{xy}-B(M+\frac12L))xy^{3}+L^{2}y^{4},
\end{equation}
\begin{equation}
	Q_{2}=x^{4}-2Bx^{3}y+(B^{2}-4L-2M)x^{2}y^{2}+(4D_{xy}-2MB)xy^{3}+M^{2}y^{4},
\end{equation}
and
\begin{equation}
	Q_{3}=x^{4}-2Bx^{3y}+(B^{2}+2L-2M)x^{2}y^{2}-(2(L+M)B-4D_{xy})xy^{3}+N^{2}y^{4}.
\end{equation}
We determine the roots configuration of a quartic $Q=\alpha x^{4}-4\beta x^{3}y+6\gamma x^{2}y^{2}-4\delta xy^{3}+\omega y^{4}$ by examining the vanishing of the five covariants 
\begin{equation}I_{2}=\alpha\omega-4\beta\delta+3\gamma^{2},\end{equation}
	\begin{equation}
I_{3}=\alpha\gamma\omega-\alpha\delta^{2}-\omega\beta^{2}-\gamma^{3}+2\beta\gamma\delta,
\end{equation}
\begin{equation}\Delta=I_{2}^{3}-27I_{3}^{2},
\end{equation}
\begin{equation}
	Hess=\left|\begin{array}{cc}{\partial^{2}\over\partial x^{2}}Q&{\partial^{2}\over\partial x\partial y}Q\\
	{\partial^{2}\over\partial x\partial y}Q&{\partial^{2}\over\partial y^{2}}Q\end{array}\right|,
\end{equation}
and
\begin{equation}
	T=\left|\begin{array}{cc}{\partial\over\partial x}Q&{\partial\over\partial y}Q\\
	{\partial\over\partial x}Hess(Q)&{\partial\over\partial y}Hess(Q)\end{array}\right|.
\end{equation}
The interpretation of the values of the covariants in terms of roots is summarized in table \ref{rootq} (see eg \cite{1999Olver}).
\begin{table}[h]
	\[{}
	\begin{array}{|c|c|}
		\hline covariants&Interpretation\\\hline
		\Delta\neq 0&\mbox{Four distinct roots}\\
		\Delta=0\mbox{ and } T\neq 0&\mbox{ Exactly one double root}\\
		T=0\mbox{ and }I_{2}\neq 0&\mbox{Two distinct double roots}\\
		I_{2}=I_{3}=0\mbox{ and }Hess\neq 0&\mbox{ A triple root}\\
		Hess=0&\mbox{ a quadruple root}\\\hline
	\end{array}
	\]\caption{Roots of a quartic\label{rootq}}
\end{table}
Notice that the values of the invariant polynomials $I_{2}$, $I_{3}$ and $\Delta$ are the same for the three quartics. These invariants are also invariant polynomials of the binary quadrilinear form $A$. Furthermore, $\Delta$ is nothing but the hyperdeterminant, in the sense of Gelfand et al. \cite{1992GKL}, of $A$ \cite{2017HLT}. 
Remark also that 
\begin{equation}
	Q_{1}(G_{abcd})=(x-a^{2})(x-b^{2})(x-c^{2})(x-d^{2}).
\end{equation}
Once the configuration of the roots has been identified, we can refine our result by looking at the values of the other covariants and refer to  the classification described  in \cite{2017HLT} p32.

We have to investigate the $64$ possible values of $E$.
\begin{enumerate}
	\item\label{un} If $E=\emptyset$ then $|\Phi_{E}\rangle$ is completely factorized.
	\item\label{deux} If $E=\{\{i,j\}\}$ for some $i, j=0,\dots, 3$, $i\neq j$ ($6$ cases), then $|\Phi_{E}\rangle$ belongs to the nilpotent cone and so each quartic equals $x^{4}$. The state $|\Phi_{E}\rangle$  is partially factorized as a state which is $\SLOCC$-equivalent to an EPR pair on the qubits $i,j$ together with two independent particles.
	\item\label{trois} If  $E=\{\{i,j\},\{i,k\}\}$ for some $i,j,k$ distinct ($12$ cases), then each quartic equals $x^{4}$. The state $|\Phi_{E}\rangle$ factorizes as a state which is $\SLOCC$-equivalent to $|\GHZ_{3}\rangle$ on the qubits $i,j, k$ together with an  independent qubit.
	\item\label{quatre} If $E=\{\{i,j\},\{k,l\}\}$ with  $\{i,j\}\cap\{k,l\}=\emptyset$ ($3$ cases) then one of the quartic equals $x^{3}(x-4\exp\{i\diamond\} y)$ and the two others equal $(x-\frac14\exp\{i\diamond\} y)^{4}$ with $a_{0}b_{0}c_{0}d_{0}a_{1}b_{1}c_{1}d_{1}=\frac1{16}\exp\{i\diamond\}$. For generic values of the parameters we have $\diamond\neq 0$ and this implies that $|\Phi_{E}\rangle$  factorizes as a two $2$-qubits state which are $\SLOCC$-equivalent to two EPR pairs. Let us examine only the case where $E=\{\{1,2\},\{3,4\}\}$, the other cases are obtained symmetrically. In this case, we have $Q_{1}=x^{3}(x-4\exp\{i\diamond\} y)$ and $\mathcal C=\mathcal D=\mathcal K_{5}=\mathcal L=0$. From \cite{2017HLT}, this implies that it is in the same orbit as $G_{a000}$ with $a=\frac12\exp\{\frac12i\diamond\}$.
	\item\label{cinq} If $E=\{\{i,j\},\{j,k\},\{i,k\}\}$ with $i, j, k$ distinct ($4$ cases) then each quartic equals $x^{4}$. The state $|\Phi_{E}\rangle$ factorizes as a state which is $\SLOCC$-equivalent to $|\GHZ_{3}\rangle$ on the qubits $i,j, k$ together with a single independent qubit. 
	\item\label{six} If  $E=\{\{i,j\},\{j,k\},\{k,l\}\}$ with $\{i,j,k,l\}=\{0,1,2,3\}$ ($12$ cases) then one of the quartics equals 
	$x^2(x^2+\frac14\exp\{2i\diamond\}y^2)$ and the two others equal $(x^{2}-\frac1{16}\exp\{2i\diamond\})^2$. For generic values of the parameters, one quartic has a double zero root together with two simple roots and the two other quartics have two double roots. Let us examine only the case $E=\{\{1,2\},\{2,3\},\{3,4\}\}$ for which $Q_{1}=x^2(x^2+\frac14\exp\{2i\diamond\}y^2)$.  Following the algorithm described in \cite{2017HLT}, we have to compute the values of the $\mathcal K_{3}$ and $\mathcal L$. The two covariants being zero, we deduce that  $|\Phi_{E}\rangle$ is in the $\SLOCC$-orbit of a degenerated $G_{abcd}$. More precisely, following the value of $E$, $|\Phi_{E}\rangle$  is equivalent to $G_{ab00}$  with $a=\frac1{\sqrt{2}}\exp\{\frac12(\diamond +\frac\pi2)\}$ and $b=\frac1{\sqrt{2}}\exp\{\frac12(\diamond -\frac\pi2)\}$. 
	\item\label{sept} If $E=\{\{i,j\},\{i,k\},\{i,l\}\}$ with  $\{i,j,k,l\}=\{0,1,2,3\}$ ($4$ cases) then the three quartics are equal to $x^2(x+\frac12\exp\{i\diamond\})^2$. Following \cite{2017HLT}, the Verstraete type of $|\Phi_{E}\rangle$  is determined by evaluating $\mathcal K_{3}$ and $\mathcal L$. Since the two covariants vanish, we deduce that  $|\Phi_{E}\rangle$  is in the $\SLOCC$-orbit of $G_{aa00}$ with  $a=\frac1{\sqrt{2}}\exp\{\frac12i(\diamond+\frac\pi2)\}$. We are in the case where $L=M=B=0$. From \cite{2012HLT}, there is only one dense $\SLOCC$-orbit in that variety.
	\item\label{huit} If $E=\{\{i,j\},\{i,k\},\{i,l\},\{j,k\}\}$   with  $\{i,j,k,l\}=\{0,1,2,3\}$ ($12$ cases) then one of the quartics equals $x^{2}(x-\frac12y\diamond)(x+\frac12y\diamond)$ and the two others equal $(x^{2}+\frac1{16}y^{2}\diamond^{2})^{2}$. We only investigate the case $E=\{\{1,2\},\{1,3\},\{1,4\},\{3,4\}\}$, since one can easily deduce the others by symmetry. From \cite{2017HLT}, one has to compute the values of $\mathcal K_{3}$ and $\mathcal L$.	Both covariants vanish and we deduce that $|\Phi_{E}\rangle$ is $\SLOCC$-equivalent to $G_{ab00}$ with $a=\frac12\exp\{\frac12i\diamond\}$ et $b=\frac12\exp\{\frac12i(\diamond+\pi)\}$. 
	Furthermore, it belongs to the variety defined by $L=0$.
	\item\label{neuf} If $E=\{\{i,j\},\{j,k\},\{k,l\},\{l,i\}\}$ with  $\{i,j,k,l\}=\{0,1,2,3\}$  ($3$ cases) then we find that one of the quartics equals $x^{2}(x^{2}+\frac1{16}y^{2}\exp\{2i\diamond\})$ and the others equal $(x^{2}-\frac1{16}y^{2}\diamond^{2})^{2}$. Let us only examine  the case $E=\{\{1,3\},\{2,3\},\{2,4\},\{1,4\}\}$. From \cite{2017HLT}, one has to compute the values of $\mathcal K_{3}$ and $\mathcal L$. Both covariants vanish and we deduce that $|\Phi_{E}\rangle$ is $\SLOCC$-equivalent to $G_{ab00}$ with $a=\frac12\exp\{\frac12i(\diamond+\frac\pi2)\}$ et $b=\frac12\exp\{\frac12i(\diamond-\frac\pi2)\}$. 
	Furthermore, it belongs to the variety defined by $L=0$.
	\item\label{dix} If $E=\{\{i,j\}\mid 0\leq i<j\leq 3\}\setminus \{k,l\}\}$ for some $k\neq l$ ($6$ cases) then one of the quartics equal $x^{2}(x^{2}-\frac14\exp\{2i\diamond\} y^{2})$ while the two others equal $(x^2+\frac1{16}\exp\{2i\diamond\}y^2)^2$. Without loss of generalities one supposes that $E=\{\{1,2\},\{1,3\},\{1,4\},\{2,3\},\{2,4\}\}$; the other cases being obtained by symmetry. We have $Q_{1}=x^{2}(x^{2}-\frac14\exp\{2i\diamond\} y^{2})$. Following \cite{2017HLT}, we have computed $\mathcal K_{3}$ and $\mathcal L$. Since the two invariants vanish, we have deduced that $|\Phi_{E}\rangle$ is equivalent to a degenerated $G_{abcd}$. More precisely, following the value of $E$, it is equivalent to $G_{ab00}$  with
	 $a=\sqrt{2}\exp\{\frac12i\diamond\}$ and
	$b=\sqrt{2}\exp\{\frac12i(\diamond-\pi)\}$.
	\item\label{onze} If $E=\{\{i,j\}\mid 0\leq i<j\leq 3\}$ then the two quartics are equal to $x^2(x-\frac12\exp\{i\diamond\}y)^2$. Following \cite{2017HLT}, we have computed the values of the four covariants $\overline{\mathcal G}$, $\mathcal G$, $\mathcal H$, and $\mathcal L$. Since they all vanish, we have deduced that  $|\Phi_{E}\rangle$ is equivalent to a degenerated $G_{abcd}$. More precisely, $|\Phi_{E}\rangle$ is $\SLOCC$-equivalent to $G_{aa00}$ with $a=\frac1{\sqrt{2}}\exp\{\frac12i\diamond\}$.
\end{enumerate}
Viewing $E$ as the set of the edges of a $4$ vertices graph, the cases (\ref{un}) to (\ref{cinq}) above correspond to disconnected graphs and factorized states. The other ones correspond to connected graphs and degenerated $G_{abcd}$ states. Notice that some degenerated $G_{abcd}$  factorize.  

Miyake \cite{2003Miyake} has shown that the more generic entanglement holds for $\Delta\neq 0$. So we have,
\begin{prop}
	The states $|\Phi_{E}\rangle$ are not generically entangled.
\end{prop}

We can be more precise by noticing that the invariant polynomial  $LMN$ vanishes for any state $|\Phi_{E}\rangle$. In terms of (projective) geometry, this means that $|\Phi_{E}\rangle$ corresponds to a point of the  secant variety of one of the three Segre embedding $\sigma(Seg_{ij}(\mathbb P^{3}\times \mathbb P^{3}))$ (see \cite{2017HLT} p18-21), $\{i,j\}\in\{\{0,1\}, \{0,2\}, \{0,3\}\}$. 
The three Segre varieties $Seg_{ij}(\mathbb P^{3}\times \mathbb P^{3})$ are isomorphic and are the image of one of the three  Segre bilinear map $Seg_{ij}:\mathbb P^{3}\times \mathbb P^{3}\rightarrow \mathbb P^{15}=\mathbb P(\mathbb C^{2}\otimes\mathbb C^{2}\otimes \mathbb C^{2}\otimes\mathbb C^{2})$, defined by
\begin{eqnarray}
Seg_{01}([v_{0}\otimes v_{1}],[w_{0}\otimes w_{1}])=[v_{0}\otimes v_{1}\otimes w_{0}\otimes w_{1}],\\
Seg_{02}([v_{0}\otimes v_{1}],[w_{0}\otimes w_{1}])=[v_{0}\otimes w_{0}\otimes v_{1}\otimes w_{1}], \mbox{ and }\\
Seg_{03}([v_{0}\otimes v_{1}],[w_{0}\otimes w_{1}])=[v_{0}\otimes w_{0}\otimes w_{1}\otimes v_{1}].
\end{eqnarray}
Alternatively, the Segre varieties  are the zero locus of the $2\times 2$-minors of one of the matrices involved in the definition of $L$, $M$ and $N$. 
The secant variety of a projective variety $Y$ is the algebraic closure of the union of secant lines $\mathbb P^{1}_{xy}$, $x,y\in Y$. The variety $\sigma(Seg_{ij}(\mathbb P^{3}\times \mathbb P^{3}))$ corresponds to the zero locus the $3\times 3$-minors of  one of the matrices involved in the definition of $L$, $M$, and $N$.\\
From \cite{2017HLT} and the vanishing of $LMN$, we deduce that all the $|\Phi_{E}\rangle$ belong to one of the third secant variety 
\begin{equation}
	\sigma_{3}(Seq_{ij}(\mathbb P^{3}\times \mathbb P^{3})):=\overline{\displaystyle\bigcup_{x_{1},x_{2},x_{3}}\mathbb P^2_{x_{1}x_{2}x_{3}}},
\end{equation}
where $\overline Y$ denotes the algebraic closure of $Y$ and $P^2_{x_{1}x_{2}x_{3}}$ is the only projective plane containing the point $x_{1}$, $x_{2}$, and $x_{3}$.
The computation of corresponding Verstraete forms allows us to refine this result by exhibiting for each state a strictly included variety to which it belongs. 
\begin{table}[h]
\begin{center}
$
	\begin{array}{|c|c|}
		\hline Varieties & Cases\\
		\hline{}
		\sigma(Seg_{ij}(\mathbb P^{3}\times \mathbb P^{3}))&(\ref{un}), (\ref{deux}), (\ref{quatre}), (\ref{six}), (\ref{sept}), (\ref{huit}), (\ref{neuf}), (\ref{dix}), (\ref{onze})\\\hline
		\displaystyle\sigma(\mathbb P^{1}\times \mathbb P^{1}\times\mathbb P^{1}\times\mathbb P^{1})=\bigcap_{ij}\sigma(Seg_{ij}(\mathbb P^{3}\times \mathbb P^{3})&(\ref{un}), (\ref{six}), (\ref{onze})\\\hline
		Seg_{ij}(\mathbb P^{3}\times \mathbb P^{3})&(\ref{un}), (\ref{deux}), (\ref{quatre})\\\hline
		Seg_{ij}(\mathbb P^{1}\times \mathbb P^{1}\times \mathbb P^{3})&(\ref{un}), (\ref{deux})
		\\\hline
		\mathbb P^{1}\times \mathbb P^{1}\times \mathbb P^{1}\times\mathbb P^{1}&(\ref{un}) \\\hline 
	\end{array}
$
\end{center}
\caption{Varieties associated to cases (\ref{un}), (\ref{deux}), (\ref{quatre}), (\ref{six}), (\ref{sept}), (\ref{huit}), (\ref{neuf}), (\ref{dix}), (\ref{onze})\label{tablecase}}
\end{table}
For all the cases but (\ref{trois}) and (\ref{cinq}), the corresponding varieties are summarized in table \ref{tablecase}. The $16$ remaining cases, corresponding to (\ref{trois}) and (\ref{cinq}), belong to one of the four Segre varieties $\mathbb P^{1}\times\mathbb P^{7}\rightarrow \mathbb P^{15}$.




Notice also that $|\GHZ_{4}\rangle$ is $\SLOCC$-equivalent to a $G_{aa00}$ state \cite{2017HLT}  as case (\ref{sept}) above.
Moreover when $a_{i}=b_{i}=c_{i}=d_{i}=\frac1{\sqrt 2}$ for $i=0,1$, it is possible to create a state which is $\LU$-equivalent to  $|\GHZ_{4}\rangle$ (see proposition \ref{F2GHZ}).\\
The state $|\W_{4}\rangle=|1000\rangle+|0100\rangle+|0010\rangle+|0001\rangle$ belongs to the null cone (ie, all the invariant polynomials vanish). Since all the $|\Phi_{E}\rangle$ belonging to the null cone factorize and the factorization properties is a $\SLOCC$-invariant property, we deduce that no $|\Phi_{E}\rangle$ is $\SLOCC$-equivalent to  $|\W_{4}\rangle$, as in the case of $3$-qubit systems.

\subsection{Five qubits and beyond}
 For more five qubits and more, the number of tools is much smaller. Indeed, the description of the algebra of covariant polynomials is out of reach and even some important invariant polynomial, such that the hyperdeterminant, are too huge to be computed in a suitable form. We recall that the importance of the hyperdeterminant is due to the fact that it vanishes when the system is not generically entangled \cite{2003Miyake}. Although this polynomial is very difficult to calculate, its nullity can be tested thanks to its interpretation in terms of solution to a system of equations, e.g. \cite{1992GKL} p445. For instance , if $A=\displaystyle\sum_{0\leq i,j,k,l,n\leq 1}\alpha_{ijkln}x_{i}y_{j}z_{k}t_{l}s_{n}$ is the ground form associated to the five qubits state $|\phi\rangle=\displaystyle\sum_{0\leq i,j,k,l,n\leq 1}\alpha_{ijkln}|ijkln\rangle$, the condition $\Delta(|\phi\rangle)=0$ means that the system
 \begin{equation}
	 S_{\phi}:=\{A={d\over dx_{0}}A={d\over dx_{1}}A={d\over dy_{0}}A={d\over dy_{1}}A=\cdots={d\over ds_{0}}A={d\over ds_{1}}A=0\} \end{equation}
 has a solution  $\hat x_{0},\hat x_{1},\hat y_{0},\hat y_{1},\dots,\hat s_{0},\hat s_{1}$ in the variables $x_{0},x_{1},y_{0},y_{1},\dots,s_{0},s_{1}$ such that $(\hat x_{0},\hat x_{1}),(\hat y_{0},\hat y_{1}),\dots,(\hat s_{0},\hat s_{1})\neq (0,0)$. Such a solution is called non trivial.
 We process exhaustively by exhibiting a non trivial solution for each of the $1024$ systems $S_{\Phi_{E}}$ where
\begin{equation}
	|\Phi_{E}\rangle=Z_{E}(a_{0}|0\rangle+a_{1}|0\rangle)(b_{0}|0\rangle+b_{1}|0\rangle)
	(c_{0}|0\rangle+c_{1}|0\rangle)(d_{0}|0\rangle+d_{1}|0\rangle)(e_{0}|0\rangle+e_{1}|0\rangle).
\end{equation}
Since permutations of the qubits let the value of the hyperdeterminant $\Delta$ unchanged, the set of systems $S_{\Phi_{E}}$ splits into $34$ classes  corresponding to undirected unlabeled graphs. For each of these classes, we find a non trivial solution with $x_{1}=y_{1}=z_{1}=t_{1}=s_{1}=1$ for a given representative element. The solutions are summarized in tables \ref{NTS<5}, \ref{NTS=5} and \ref{NTS>5} with the notations $\triangle_{1}:=\displaystyle\sum_{i,j,k}(-1)^{j(i+1)+ik}a_{i}b_{j}c_{k}$, 
$\triangle_{2}:=\displaystyle\sum_{i,j,k}(-1)^{i(j+k)}a_{i}b_{j}c_{k}$, $\triangle_{3}=\displaystyle\sum_{i,j,k}(-1)^{k(1+j)+ij}a_{i}b_{j}c_{k}$,
$\triangle_{4}:=\displaystyle\sum_{i,j,k}(-1)^{j(i+k)}a_{i}b_{j}c_{k}$, 
$\triangle_{5}:=\displaystyle\sum_{ijk}(-1)^{k+j(k+i)}a_{i}b_{j}c_{k}$, and $\triangle_{6}:=\displaystyle\sum_{ijk}(-1)^{j(i+k)}a_{i}b_{j}c_{k}$.\\
\begin{table}[h]
\begin {center}
	$
\begin{array}{|c|c|c|c|}
\hline
	Classes& Representative& Cardinals& Solutions\\
	&elements&&[x_{0},y_{0},z_{0},t_{0},s_{0}]\\\hline
	\{\}&\{\}&1&[1,-{b_{1}\over b_{0}},-{c_{1}\over c_{0}},1,1]\\\hline
	\{\{i,j\}\}&\{\{0,1\}\}&10&[1,1,1,-{d_{1}\over d_{0}},-{e_{1}\over e_{0}}]\\\hline
	\{\{i,j\},\{i,k\}\}&\{\{0,1\},\{0,2\}\}&30&[1,1,1,-{d_{1}\over d_{0}},-{e_{1}\over e_{0}}]\\
	\{\{i,j\},\{k,l\}\}&\{\{0,1\},\{2,3\}\}&15&[1,1,1,-{d_{1}(c_{0}-c_{1}\over d_{0}(c_{0}+c_{1})},-{e_{1}\over e_{0}}]\\\hline
	\{\{i,j\},\{i,k\},\{i,l\}\}&\{\{0,1\},\{0,2\},\{0,3\}\}&20&[1,{b_{1}\over b_{0}},-{c_{1}\over c_{0}},1,-{e_{1}\over e_{0}}]\\
	\{\{i,j\},\{j,k\},\{i,k\}\}&\{\{0,1\},\{1,2\},\{0,2\}\}&10&[1,1,1,-{d_{1}\over d_{0}},-{e_{1}\over e_{0}}]\\
	\{\{i,j\},\{i,k\},\{j,l\}\}&\{\{0,1\},\{0,2\},\{1,3\}\}&60&[1,1,1,-{d_{1}\triangle_{1}\over d_{0}\triangle_{2}},-{e_{1}\over e_{0}}]\\
	\{\{i,j\},\{i,k\},\{l,n\}\}&\{\{0,1\},\{0,2\},\{3,4\}\}&30&[1,1,1,-{d_{1}\over d_{0}},-{e_{1}\over e_{0}}]\\\hline
	\begin{array}{r}\{\{i,j\},\{i,k\},\{i,l\},\\\{i,n\}\}\end{array}&\begin{array}{r}\{\{0,1\},\{0,2\},\{0,3\},\\\{0,4\}\}\end{array}&5&[1,{b_{1}\over b_{0}},-{c_{1}\over c_{0}},1,-{e_{1}\over e_{0}}]\\
	\begin{array}{r}\{\{i,j\},\{i,k\},\{i,l\},\\\{j,n\}\}\end{array}&\begin{array}{r}\{\{0,1\},\{0,2\},\{0,3\},\\\{1,4\}\}\end{array}&60&[1,{b_{1}\over b_{0}},-{c_{1}\over c_{0}},1,-{e_{1}\over e_{0}}]\\
	\begin{array}{r}\{\{i,j\},\{j,k\},\{i,k\},\\\{i,l\}\}\end{array}&\begin{array}{r}\{\{0,1\},\{1,2\},\{0,2\},\\\{0,3\}\}\end{array}&60&[1,-{(i-1)b_{1}\over (i+1)b_{0}},(i+1){c_{1}\over c_{0}},1,-{e_{1}\over e_{0}}]\\
	\begin{array}{r}\{\{i,j\},\{j,k\},\{k,l\},\\\{i,l\}\}\end{array}&\begin{array}{r}\{\{0,1\},\{1,2\},\{2,3\},\\\{0,3\}\}\end{array}&15&[{a_{1}\over a_{0}},{b_{1}\over b_{0}},-{c_{1}\over c_{0}},{d_{1}\over d_{0}},1]\\
	\begin{array}{r}\{\{i,j\},\{k,l\},\{l,n\},\\\{k,n\}\}\end{array}&\begin{array}{r}\{\{0,1\},\{2,3\},\{3,4\},\\\{2,4\}\}\end{array}&10&[1,1,1,-{d_{1}(c_{0}-c_{1})\over d_{1}(c_{0}+c_{1})},-{e_{1}\over e_{0}}]\\
	\begin{array}{r}\{\{i,j\},\{j,k\},\{k,l\},\\\{l,n\}\}\end{array}&\begin{array}{r}\{\{0,1\},\{1,2\},\{2,3\},\\\{3,4\}\}\end{array}&60&[1,1,1,-{d_{1}\triangle_{3}\over d_{0}\triangle_{4}},{e_{1}\over e_{0}}]\\\hline
\end{array}
$\end{center}
\caption{Non trivial solutions of $S_{\Phi_{E}}$ for $\mathtt{card}(E)<5$ \label{NTS<5}}
\end{table}
\begin{table}[h]
\begin {center}
	$
\begin{array}{|c|c|c|c|}
\hline
	Classes& Representative& Cardinals& Solutions\\
	&elements&&[x_{0},y_{0},z_{0},t_{0},s_{0}]\\\hline
	\begin{array}{r}\{\{i,j\},\{j,k\},\{k,l\},\\\{l,n\},\{i,n\}\}\end{array}&\begin{array}{r}\{\{0,1\},\{1,2\},\{2,3\},\\\{3,4\},\{0,4\}\}\end{array}&12&[1,1,1,{-d_{1}\triangle_{5}\over d_{0}
	\triangle_{6}},-{e_{1}\over e_{0}}]\\
	\begin{array}{r}\{\{i,j\},\{i,k\},\{i,l\},\\\{i,n\},\{j,k\}\}\end{array}&\begin{array}{r}\{\{0,1\},\{0,2\},\{0,3\},\\\{0,4\},\{1,2\}\}\end{array}&30&[1,-{b_{1}(i-1)\over b_{0}(i+1)},(i+1){c_{1}\over c_{0}},1,-{e_{1}\over e_{0}}]\\
	\begin{array}{r}\{\{i,j\},\{i,k\},\{i,l\},\\\{j,n\},\{j,k\}\}\end{array}&\begin{array}{r}\{\{0,1\},\{0,2\},\{0,3\},\\\{1,4\},\{1,2\}\}\end{array}&60&[1,-{b_{1}(i-1)\over b_{0}(i+1)},(i+1){c_{1}\over c_{0}},1,-{e_{1}\over e_{0}}]\\
	\begin{array}{r}\{\{i,j\},\{i,k\},\{i,l\},\\\{j,k\},\{k,l\}\}\end{array}&\begin{array}{r}\{\{0,1\},\{0,2\},\{0,3\},\\\{1,2\},\{1,3\}\}\end{array}&30&[-i{a_{1}\over a_{0}},i{b_{1}(i-1)\over b_{0}(i+1)},{c_{1}\over c_{0}},{d_{1}\over d_{0}},1]\\
	\begin{array}{r}\{\{i,j\},\{j,k\},\{k,l\},\\\{i,l\},\{j,n\}\}\end{array}&\begin{array}{r}\{\{0,1\},\{1,2\},\{2,3\},\\\{0,3\},\{1,4\}\}\end{array}&60&[{a_{1}(i-1)(b_{0}-b_{1})\over a_{0}(i+1)(b_{0}+b_{1)}},1,{c_{1}\over c_{0}}i,1,1]\\
	\begin{array}{r}\{\{i,j\},\{j,k\},\{i,k\},\\\{i,l\},\{l,n\}\}\end{array}&\begin{array}{r}\{\{0,1\},\{1,2\},\{0,2\},\\\{0,3\},\{3,4\}\}\end{array}&60&[1,-{b_{1}(i-1)\over b_{0}(i+1)},i{c_{1}\over c_{0}},1,-{e_{1}\over e_{0}}]\\
	\hline
\end{array}
$\end{center}
\caption{Non trivial solutions of $S_{\Phi_{E}}$ for $\mathtt{card}(E)=5$ \label{NTS=5}}
\end{table}
\begin{table}[h]
\begin {center}
	$
\begin{array}{|c|c|c|c|}
\hline
	Classes& Representative& Cardinals& Solutions\\
\{\{i,j\}\mid 0\leq i<j\leq 4\}\setminus E'	&elements&&[x_{0},y_{0},z_{0},t_{0},s_{0}]\\
with\mbox{ }E'=&E'=&&\\\hline
	\{\}&\{\}&1&[{a_{1}\over a_{0}},-{b_{1}\over b_{0}},-{c_{1}\over c_{0}},{d_{1}\over d_{0}},1]\\\hline
	\{\{i,j\}\}&\{\{0,1\}\}&10&[{a_{1}\over a_{0}},-{b_{1}\over b_{0}},-{c_{1}\over c_{0}},{d_{1}\over d_{0}},1]\\\hline
	\{\{i,j\},\{i,k\}\}&\{\{0,1\},\{0,2\}\}&30&[1,-{b_{1}(i-1)\over b_{0}(i+1)},i{c_{1}\over c_{0}},1,-{e_{1}\over e_{0}}]\\
	\{\{i,j\},\{k,l\}\}&\{\{0,1\},\{2,3\}\}&15&[{a_{1}\over a_{0}},-{b_{1}(i-1)\over b_{0}(i+1)},i{c_{1}\over c_{0}},1,-{e_{1}\over e_{0}}]\\\hline
	\{\{i,j\},\{i,k\},\{i,l\}\}&\{\{0,1\},\{0,2\},\{0,3\}\}&20&[-{a_{1}\over a_{0}},-{b_{1}(i-1)\over b_{0}(i+1)},i{c_{1}\over c_{0}},1,1]\\
	\{\{i,j\},\{j,k\},\{i,k\}\}&\{\{0,1\},\{1,2\},\{0,2\}\}&10&[1,{b_{1}\over b_{0}},-{c_{0}\over c_{1}},1,-{e_{1}\over e_{0}}]\\
	\{\{i,j\},\{i,k\},\{j,l\}\}&\{\{0,1\},\{0,2\},\{1,3\}\}&60&[{a_{1}\over a_{0}},1,-{b_{1}(c_{0}-c_{1})\over b_{0}(c_{0}+c_{1})},1,1,-{e_{1}\over e_{0}}]\\
	\{\{i,j\},\{i,k\},\{l,n\}\}&\{\{0,1\},\{0,2\},\{3,4\}\}&30&[1,-{b_{1}(i-1)\over b_{0}(i+1)},i{c_{1}\over c_{0}},1,-{e_{1}\over e_{0}}]\\\hline
	\begin{array}{r}\{\{i,j\},\{i,k\},\{i,l\},\\\{i,n\}\}\end{array}&\begin{array}{r}\{\{0,1\},\{0,2\},\{0,3\},\\\{0,4\}\}\end{array}&5&[-{a_{1}\over a_{0}},-{b_{1}(i-1)\over b_{0}(i+1)},i{c_{1}\over c_{0}},1,1]\\
	\begin{array}{r}\{\{i,j\},\{i,k\},\{i,l\},\\\{j,n\}\}\end{array}&\begin{array}{r}\{\{0,1\},\{0,2\},\{0,3\},\\\{1,4\}\}\end{array}&60&[-{a_{1}\over a_{0}},-{b_{1}(i-1)\over b_{0}(i+1)},i{c_{1}\over c_{0}},1,1]\\
	\begin{array}{r}\{\{i,j\},\{j,k\},\{i,k\},\\\{i,l\}\}\end{array}&\begin{array}{r}\{\{0,1\},\{1,2\},\{0,2\},\\\{0,3\}\}\end{array}&60&[-{a_{1}\over a_{0}},{b_{1}\over b_{0}},-{c_{1}\over c_{0}},1,1]\\
	\begin{array}{r}\{\{i,j\},\{j,k\},\{k,l\},\\\{i,l\}\}\end{array}&\begin{array}{r}\{\{0,1\},\{1,2\},\{2,3\},\\\{0,3\}\}\end{array}&15&[-{a_{1}(c_{0}-c_{1})\over a_{0}(c_{0}+c_{1})},1,1,-{d_{1}(b_{0}-b_{1})\over d_{0}(b_{0}+b_{1})},1]\\
	\begin{array}{r}\{\{i,j\},\{k,l\},\{l,n\},\\\{k,n\}\}\end{array}&\begin{array}{r}\{\{0,1\},\{2,3\},\{3,4\},\\\{2,4\}\}\end{array}&10&[{a_{1}\over a_{0}},-{b_{1}\over b_{0}},-{c_{1}\over c_{0}},{d_{1}\over d_{1}},1]\\
	\begin{array}{r}\{\{i,j\},\{j,k\},\{k,l\},\\\{l,n\}\}\end{array}&\begin{array}{r}\{\{0,1\},\{1,2\},\{2,3\},\\\{3,4\}\}\end{array}&60&[{a_{0}(c_{0}-c_{1})\over a_{1}(c_{0}+c_{1})},-{b_{1}\over b_{0}},1,1,-{e_{1}\over e_{0}}]\\\hline
\end{array}
$\end{center}
\caption{Non trivial solutions of $S_{\Phi_{E}}$ for $\mathtt{card}(E)>5$ \label{NTS>5}}
\end{table}
So we deduce
\begin{prop}
	For five qubit systems, the states $|\Phi_{E}\rangle$ are not generically entangled.	
\end{prop}
In principle the same strategy can be applied for more than five qubits and we conjecture that the property is still true.
\section{Conclusion and perspectives}
We have investigated some properties of the controlled-$Z$ gates. In particular, we have described combinatorially and algebraically the group generated by controlled-$Z$ and swap gates and we have studied its action with respect to the entanglement ($\SLOCC$-equivalence). In terms of algebra, this group is isomorphic to the semi-direct product of two well known groups and this property allows us to propose algorithms for simplifying circuits. About entanglement, we have shown that the group is powerful enough to generate the states $|\GHZ_{k}\rangle$ from a completely factorized state but not to generate a representative element for every $\SLOCC$-classes. In particular, we have shown that for four and five qubits, it is not possible to produce a generically entangled state (in the sense of Miyake \cite{2003Miyake}). Furthermore, by adding unitary single qubit operations, all the unitary operations can be encoded and the associated circuits can be implemented on actual quantum machines without too many adjustments. So we have here an interesting toy model for the study of quantum circuits with many connections with algebra, combinatorics and geometry. Nevertheless, there are still many interesting questions to explore. Let us list some of them.
\subsection{On the network structure of the qubits} 
It would be interesting to design circuit simplification algorithms that work regardless of the configuration of the network of the qubits. Since the controlled-$Z$ operations are symmetrical, we have only to investigate networks which are undirected graphs.
If the network is organized as a line (the bit $0$ is connected to the bit $1$, the bit $1$ is connected to the bit $2$ etc.) then one has to manage with generators which are elementary transpositions $S_{i}$, $i=0\dots k-2$, that permute the values of $i$ and $i+1$, together with the controlled-$Z$: $Z_{i}:=Z_{i,i+1}$. Minimizing the number of elementary transposition in a permutation is a well known exercise in combinatorics. This is  important for quantum computing, because transpositions are often implemented from $3$ quantum $2$-qubit gates, see equality (\ref{transp}), and so are particularly unreliable. Finding an optimal algorithm for reduction requires a deeper knowledge of the algebraic and combinatorial structure of the group \cZS$_{k}$ and will be the topic of future works.
\subsection{Algebraic structure}
Beyond applications in quantum information, the groups \cZS$_{k}$ deserve to be studied for themselves.
First, as for symmetric group, we have a tower of groups \cZS$_{2}\subset$\cZS$_{3}\subset\cdots\subset$\cZS$_{k}\subset\cdots$. This suggests connections with combinatorial Hopf algebras. In particular, the conjugacy classes and the representation theory of these groups must be studied in details. One of the underlying question is:
Is there a polynomial representation whose base would be indexed by combinatorial objects  and which could be provided with a co-product giving it a Hopf algebra structure?\\
Representations as matrices are also highly relevant in our context. Indeed, any circuit composed of controlled-$Z$ and $\Swap$ gates is nothing but the image of an element of the abstract group \cZS$_{k}$ through a certain representation. This representation has the particularity that it is also a linear representation of a free PRO \cite{2018LLMN}. So it is not irreducible and it would be interesting to understand its decomposition into irreducible representations.

\subsection{Generalizations of \cZS$_{k}$ and entanglement}
Even if we   investigated a few properties of \cZS$_{k}$ with respect to entanglement, a detailed and complete study remains to be done for any number of qubits. In particular, we focused on $\SLOCC$-equivalence but $\LU$-equivalence is also highly relevant in that context. G{\"u}hne at al. \cite{2014GCSMRDKM} investigated $\LU$-equivalence with respect to more general operations indexed by hypergraphs (instead of simple graphs in our paper). They proved that some generically entangled states, like
\begin{equation}\begin{array}{rcl}
V_{3}&:=&\frac1{\sqrt8}\left(|0011\rangle+|0101\rangle+|1001\rangle+|0110\rangle +|1010\rangle+|1100\rangle\right.\\&&\left.+|0000\rangle -|1111\rangle\right),\\
V_{9}&:=&\frac12\left(|0000\rangle -|1111\rangle\right)+\frac14\left(|0100\rangle+|0101\rangle{}
-0110\rangle+|0111\rangle\right.\\&&\left.+|1000\rangle-|1001\rangle+|1010\rangle+|1011\rangle\right)\\
V_{14}&=&\frac1{\sqrt8}\left(|0011\rangle+|0101\rangle+|1001\rangle+|0110\rangle +|1010\rangle+|1100\rangle\right.\\&&\left.+|0001\rangle -|1110\rangle\right).\end{array}
\end{equation}
can be obtained from $|0000\rangle$.
Remark that even if these states are generically entangled in the sense of \cite{2003Miyake}, ie. $\Delta\neq 0$, they have , however, some specificities. For instance $V_{3}$ and $V_{14}$ belongs to the third secant  variety $\sigma_{3}(\mathbb P^{1}\times\mathbb P^{1}\times \mathbb P^{1}\times \mathbb P^{1})$, ie. $L=M=N=0$ \cite{2017HLT}.
 For the state $V_{14}$, we have $Q_{1}(V_{14})=Q_{2}(V_{14})=Q_{3}(V_{14})=-x(x-\frac14y)(x^{2}-\frac34xy+\frac1{16}x^{2})$. 
 The state $V_{9}$ is in a certain sense more general than $V_{3}$ and $V_{14}$ because only the invariant $L$ vanishes but this means that it belongs to the  third secant variety $\sigma_{3}(\mathbb P^{3}\times \mathbb P^{3})$.
It would be interesting to know if one can generate more general entanglement types such that $L,M,N\neq 0$.\\

\noindent{\bf Acknowledgement}\ 
We acknowledge Irina Yakimenko and Fr\'ed\'eric Holweck the fruitful discussions.
We thank Ammar Husain for suggesting us the possibility of using Dehn's algorithm for simplifying quantum circuits.

\noindent This work is partially supported by the projects MOUSTIC (ERDF/GRR) and  ARTIQ (ERDF/RIN).
\\ \\
\bibliographystyle{unsrt}
\bibliography{cZ}
\appendix
\section{\label{appExperiment}Quantum circuits on actual quantum computers}
\begin{figure}[h]
 \begin{center}
 	\includegraphics[scale=0.3]{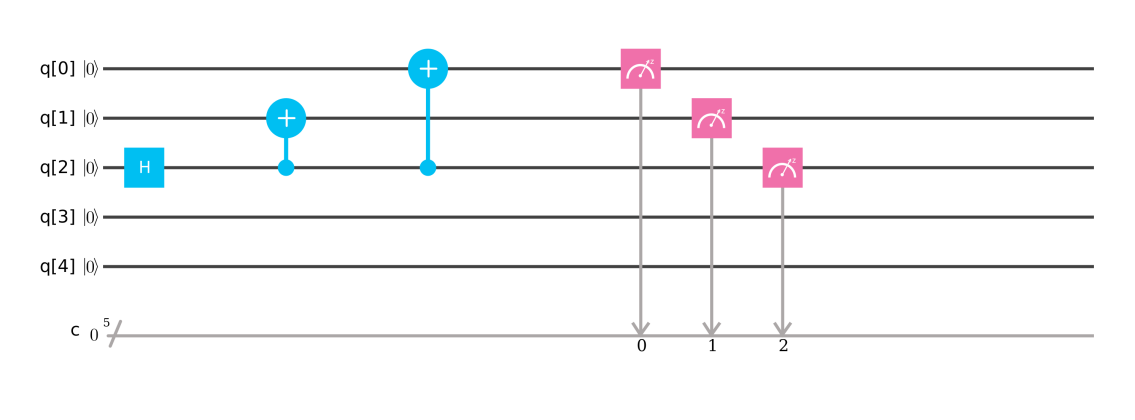}
 \end{center}
 \caption{Circuit producing a $|\GHZ_{3}\rangle$ state on the IBM Q 5 Tenerife machine\label{GHZcircIBM}}
 \end{figure}

Although the model we have investigated is a toy model, the calculations can be performed on actual quantum machines and illustrate the importance of using the less of 2-qubit gates as possible.
To perform our computations, we use the \texttt{IBM} Q experience.
On its website \url{https://quantumexperience.ng.bluemix.net/qx/experience}, \texttt{IBM} offers a free
online access to three quantum computers. Two of these computers have 5 qubits and the third has 16 qubits. We used them to test our quantum circuits. The $5$ qubits computers are not organized as a complete graph since there are only $6$ connections implemented. There are several differences between the circuits that can be realized on these machines and those presented in the paper:
\begin{itemize}
	\item The network is not a complete graph.
	\item The only $2$-qubit gates that can be used are $\cX$. The others must be obtained by combining gates.
	\item The uses of the gates induce  a  probability of error, with a very significant probability of error for $2$-qubit gates.
\end{itemize}
The $\cZ$ gates are implementable since we have
\begin{equation}
	\begin{tikzpicture}[scale=1.500000,x=1pt,y=1pt]
\filldraw[color=white] (0.000000, -7.500000) rectangle (18.000000, 22.500000);
\draw[color=black] (0.000000,15.000000) -- (18.000000,15.000000);
\draw[color=black] (0.000000,0.000000) -- (18.000000,0.000000);
\draw (9.000000,15.000000) -- (9.000000,0.000000);
\filldraw (9.000000, 15.000000) circle(1.500000pt);
\begin{scope}
\filldraw (9.000000, 0.000000) circle(1.500000pt);
\clip (9.000000, 0.000000) circle(3.000000pt);
\draw (6.000000, 0.000000) -- (12.000000, 0.000000);
\draw (9.000000, 0.000000) -- (9.000000, 3.000000);
\end{scope}
\end{tikzpicture}
\raisebox{8mm}{$\quad \sim \quad$}
\begin{tikzpicture}[scale=1.500000,x=1pt,y=1pt]
\filldraw[color=white] (0.000000, -7.500000) rectangle (66.000000, 22.500000);
\filldraw (33.000000, 15.000000) circle(1.500000pt);
\draw[fill=white] (33.000000, 0.000000) circle(3.000000pt);
\draw[color=black] (0.000000,15.000000) -- (66.000000,15.000000);
\draw[color=black] (0.000000,0.000000) -- (66.000000,0.000000);
\begin{scope}
\draw[fill=white] (12.000000, -0.000000) +(-45.000000:8.485281pt and 8.485281pt) -- +(45.000000:8.485281pt and 8.485281pt) -- +(135.000000:8.485281pt and 8.485281pt) -- +(225.000000:8.485281pt and 8.485281pt) -- cycle;
\clip (12.000000, -0.000000) +(-45.000000:8.485281pt and 8.485281pt) -- +(45.000000:8.485281pt and 8.485281pt) -- +(135.000000:8.485281pt and 8.485281pt) -- +(225.000000:8.485281pt and 8.485281pt) -- cycle;
\draw (12.000000, -0.000000) node {$H$};
\end{scope}
\draw (33.000000,15.000000) -- (33.000000,-3.00000);
\begin{scope}
\draw[fill=white] (54.000000, -0.000000) +(-45.000000:8.485281pt and 8.485281pt) -- +(45.000000:8.485281pt and 8.485281pt) -- +(135.000000:8.485281pt and 8.485281pt) -- +(225.000000:8.485281pt and 8.485281pt) -- cycle;
\clip (54.000000, -0.000000) +(-45.000000:8.485281pt and 8.485281pt) -- +(45.000000:8.485281pt and 8.485281pt) -- +(135.000000:8.485281pt and 8.485281pt) -- +(225.000000:8.485281pt and 8.485281pt) -- cycle;
\draw (54.000000, -0.000000) node {$H$};
\end{scope}
\end{tikzpicture}\raisebox{4mm}{$\quad.\quad$}
\end{equation}
Elementary transpositions are obtained by using (\ref{transp}). Nevertheless, they used three $2$-qubit gates and then are very unreliable.
Despite all this, we can test some of our circuits in real situations and check some of their properties. For instance, consider the  circuit of figure \ref{GHZcircIBM} computed on the IBM Q 5 Tenerife machine.
The result of the experiment is in figure \ref{GHZexpIBM}. It is interesting to note that although theoretically only the states $|00000\rangle$ and $|00111\rangle$ can be reached, the other states have a low but not zero probability of being obtained after the measure.
 \begin{figure}[h]
 	\begin{center}
 		\includegraphics[scale=0.5]{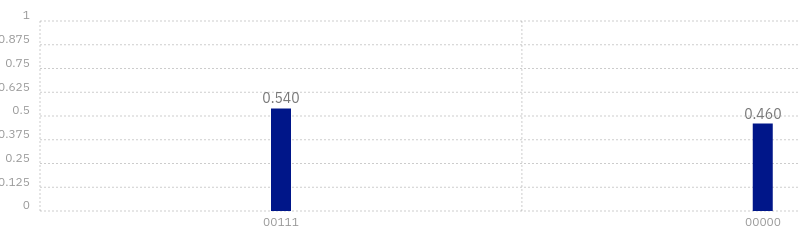}\\
 		\includegraphics[scale=0.5]{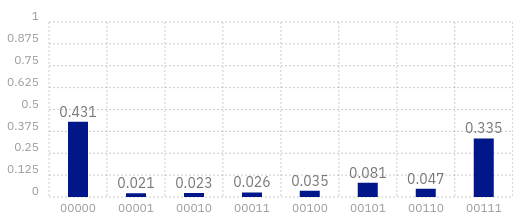}
 	\end{center}
 	\caption{Execution of the circuit of figure \ref{GHZcircIBM}. 
 	The top graphic is a simulation while the  graphic on the bottom is obtained after $1024$ executions on the IBM Q 5 Tenerife  machine. Both graphics have been produced through the IBM website.\label{GHZexpIBM}}
 \end{figure}
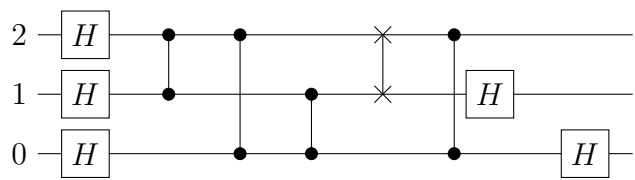
\begin{figure}[h]
	\begin{center}
		\begin{tikzpicture}[scale=1.500000,x=1pt,y=1pt]
\filldraw[color=white] (0.000000, -7.500000) rectangle (150.000000, 37.500000);
\draw (0.0,0) node[left]{$0$};
\draw (0.0,15) node[left]{$1$};
\draw (0.0,30) node[left]{$2$};
\draw[color=black] (0.000000,30.000000) -- (150.000000,30.000000);
\draw[color=black] (0.000000,15.000000) -- (150.000000,15.000000);
\draw[color=black] (0.000000,0.000000) -- (150.000000,0.000000);
\begin{scope}
\draw[fill=white] (12.000000, -0.000000) +(-45.000000:8.485281pt and 8.485281pt) -- +(45.000000:8.485281pt and 8.485281pt) -- +(135.000000:8.485281pt and 8.485281pt) -- +(225.000000:8.485281pt and 8.485281pt) -- cycle;
\clip (12.000000, -0.000000) +(-45.000000:8.485281pt and 8.485281pt) -- +(45.000000:8.485281pt and 8.485281pt) -- +(135.000000:8.485281pt and 8.485281pt) -- +(225.000000:8.485281pt and 8.485281pt) -- cycle;
\draw (12.000000, -0.000000) node {$H$};
\end{scope}
\begin{scope}
\draw[fill=white] (12.000000, 15.000000) +(-45.000000:8.485281pt and 8.485281pt) -- +(45.000000:8.485281pt and 8.485281pt) -- +(135.000000:8.485281pt and 8.485281pt) -- +(225.000000:8.485281pt and 8.485281pt) -- cycle;
\clip (12.000000, 15.000000) +(-45.000000:8.485281pt and 8.485281pt) -- +(45.000000:8.485281pt and 8.485281pt) -- +(135.000000:8.485281pt and 8.485281pt) -- +(225.000000:8.485281pt and 8.485281pt) -- cycle;
\draw (12.000000, 15.000000) node {$H$};
\end{scope}
\begin{scope}
\draw[fill=white] (12.000000, 30.000000) +(-45.000000:8.485281pt and 8.485281pt) -- +(45.000000:8.485281pt and 8.485281pt) -- +(135.000000:8.485281pt and 8.485281pt) -- +(225.000000:8.485281pt and 8.485281pt) -- cycle;
\clip (12.000000, 30.000000) +(-45.000000:8.485281pt and 8.485281pt) -- +(45.000000:8.485281pt and 8.485281pt) -- +(135.000000:8.485281pt and 8.485281pt) -- +(225.000000:8.485281pt and 8.485281pt) -- cycle;
\draw (12.000000, 30.000000) node {$H$};
\end{scope}
\draw (33.000000,30.000000) -- (33.000000,15.000000);
\filldraw (33.000000, 30.000000) circle(1.500000pt);
\filldraw (33.000000, 15.000000) circle(1.500000pt);
\draw (51.000000,30.000000) -- (51.000000,0.000000);
\filldraw (51.000000, 30.000000) circle(1.500000pt);
\filldraw (51.000000, 0.000000) circle(1.500000pt);
\draw (69.000000,15.000000) -- (69.000000,0.000000);
\filldraw (69.000000, 15.000000) circle(1.500000pt);
\filldraw (69.000000, 0.000000) circle(1.500000pt);
\draw (87.000000,30.000000) -- (87.000000,15.000000);
\begin{scope}
\draw (84.878680, 27.878680) -- (89.121320, 32.121320);
\draw (84.878680, 32.121320) -- (89.121320, 27.878680);
\end{scope}
\begin{scope}
\draw (84.878680, 12.878680) -- (89.121320, 17.121320);
\draw (84.878680, 17.121320) -- (89.121320, 12.878680);
\end{scope}
\draw (105.000000,30.000000) -- (105.000000,0.000000);
\filldraw (105.000000, 30.000000) circle(1.500000pt);
\filldraw (105.000000, 0.000000) circle(1.500000pt);
\begin{scope}
\draw[fill=white] (114.000000, 15.000000) +(-45.000000:8.485281pt and 8.485281pt) -- +(45.000000:8.485281pt and 8.485281pt) -- +(135.000000:8.485281pt and 8.485281pt) -- +(225.000000:8.485281pt and 8.485281pt) -- cycle;
\clip (114.000000, 15.000000) +(-45.000000:8.485281pt and 8.485281pt) -- +(45.000000:8.485281pt and 8.485281pt) -- +(135.000000:8.485281pt and 8.485281pt) -- +(225.000000:8.485281pt and 8.485281pt) -- cycle;
\draw (114.000000, 15.000000) node {$H$};
\end{scope}
\begin{scope}
\draw[fill=white] (138.000000, -0.000000) +(-45.000000:8.485281pt and 8.485281pt) -- +(45.000000:8.485281pt and 8.485281pt) -- +(135.000000:8.485281pt and 8.485281pt) -- +(225.000000:8.485281pt and 8.485281pt) -- cycle;
\clip (138.000000, -0.000000) +(-45.000000:8.485281pt and 8.485281pt) -- +(45.000000:8.485281pt and 8.485281pt) -- +(135.000000:8.485281pt and 8.485281pt) -- +(225.000000:8.485281pt and 8.485281pt) -- cycle;
\draw (138.000000, -0.000000) node {$H$};
\end{scope}
\end{tikzpicture}

	\end{center}
	\caption{Circuit equivalent to those of figure \ref{GHZcircIBM} but with more $2$-qubit gates. \label{eqGHZCirc}}
\end{figure}
Consider the circuit pictured in figure \ref{eqGHZCirc}. Applying the results of section \ref{Optimization} we find that it is equivalent to those of figure \ref{GHZcircIBM}. 
 \begin{figure}[h]
 	\begin{center}
 			\includegraphics[scale=0.3]{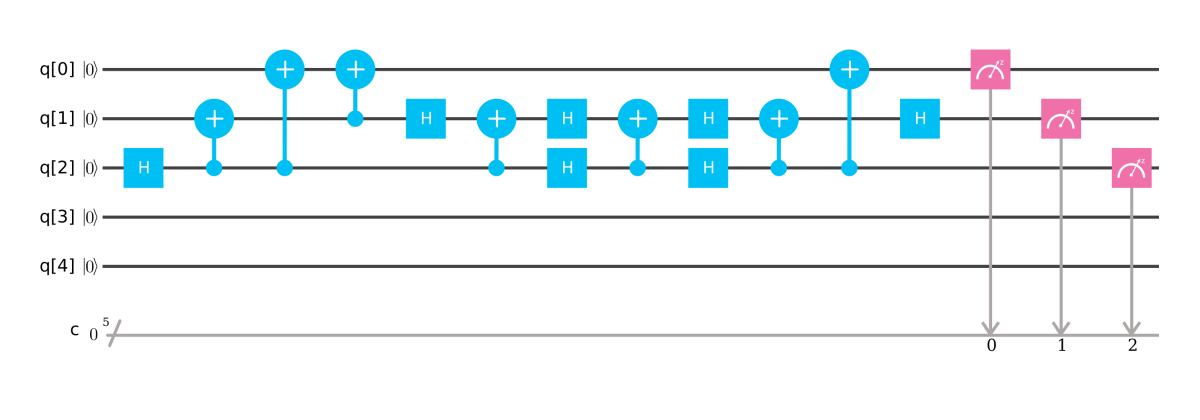}
 	\end{center}
 	\caption{Circuit of \ref{eqGHZCirc} implemented on the IBM Q 5 Tenerife machine. \label{ActualeqGHZCirc}}
 \end{figure}
This circuit has been implemented on the IBM Q 5 Tenerife machine (see figure \ref{ActualeqGHZCirc}). After $1024$ executions, we observe that  reliability is less good than for the circuit of figure \ref{GHZcircIBM}. This is due to the fact that more $2$-qubit gates were used.
\begin{figure}[h]
	\begin{center}
 		\includegraphics[scale=0.5]{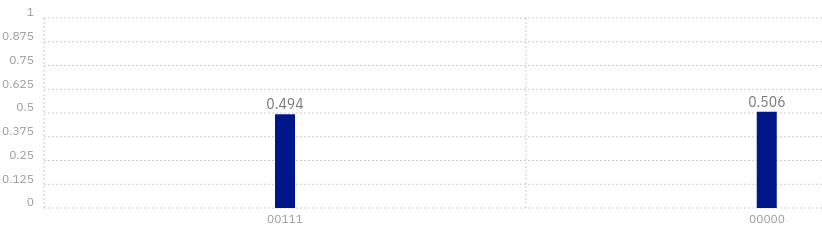}\\
 		\includegraphics[scale=0.5]{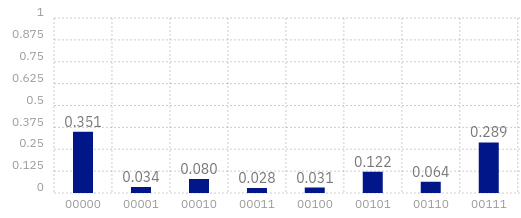}
 	\end{center}
 	\caption{Execution of the circuit of figure \ref{ActualeqGHZCirc}. 
 	The top graphic is a simulation while the  graphic on the bottom is obtained after $1024$ executions on the IBM Q 5 Tenerife  machine. Both graphics have been produced through the IBM website.\label{ActualGHZexpIBM}}
 \end{figure}
\section{Proofs of Theorems \ref{TFirstCox} and \ref{ThPres}\label{ProofTh}}
The proofs are based on two well known facts about group presentations.
\begin{claim}(See eg. \cite{1976Johnson})\label{CPresSemidirect}\\
	Let $G_{1}=\langle \mathcal S_{1}|\mathcal R_{1}\rangle$ and $G_{2}=\langle \mathcal S_{2}|\mathcal R_{2}\rangle$ be two groups given by presentation.
	The semidirect product $G_{1}\rtimes_{\Phi} G_{2}$ is isomorphic to 
	\begin{equation}
		\langle \mathcal S_{1}\cup\mathcal S_{2}|\mathcal R_{1}\cup\mathcal R_{2}\cup \{g_{2}g_{1}g_{2}^{-1}\left(\Phi(g_{2})(g_{1})\right)^{-1}\mid g_{1}\in\mathcal S_{1},g_{2}\in\mathcal S_{2}\}\rangle.
	\end{equation}
\end{claim}
\begin{claim}\label{CSplitR}
	Let $G=\langle \mathcal S|\mathcal R\rangle$. We suppose that $\mathcal S$ splits into two subsets $\mathcal S_{1}$ and $\mathcal S_{2}$ such that $\mathcal S_{2}$ is include to the subgroup of $G$ generted by $\mathcal S_{1}$. Let $\mathcal R_{1}$ denote the set of relations in $\mathcal R$ that involves at least an element of $\mathcal S_{2}$ and $\mathcal R_{2}=\mathcal R\setminus\mathcal R_{1}$. We construct a set $\mathcal R'_{2}$ by replacing in $\mathcal R_{2}$ each occurrence of an element of $\mathcal S_{2}$ by an equivalent product of elements of $\mathcal S_{1}$. Obviously, one obtains that $G$ is isomorphic to $\langle G|\mathcal R_{1}\cup\mathcal R_{2}'\rangle.$
	\end{claim}
	
	\begin{example}\label{EPrescZS3}\rm
		Let us illustrate our purpose by giving a presentation of \cZS$_{3}$. First we remark that $\mathfrak P_{3}$ is isomorphic to $\langle z_{0},z_{1},z_{02}|z_{0}^{2},z_{1}^{2},z_{02}^{2},(z_{0}z_{02})^{2},(z_{0}z_{1})^{2},(z_{0}z_{02})^{2}\rangle$ and $\mathfrak S_{3}$ is isomorhic to $\langle s_{0},s_{1}|s_{0}^{2},s_{1}^{2},(s_{0}s_{1})^{3}\rangle$. One applies Claim \ref{CPresSemidirect} from Theorem \ref{TSemiDirect} and obtains that \cZS$_{3}$ is isomorphic to
		$
		\langle z_{0},z_{1},z_{02},s_{0},s_{1}|\mathcal R\rangle,
		$
		where
		$$\mathcal R=\{z_{0}^{2},z_{1}^{2},z_{02}^{2},(z_{0}z_{02})^{2},(z_{0}z_{1})^{2},(z_{0}z_{02})^{2},(z_{1}z_{02})^{2},
		s_{0}^{2},s_{1}^{2},(s_{0}s_{1})^{3},(z_{0}s_{0})^{2},(z_{1}s_{1})^{2},s_{1}z_{0}s_{1}z_{02},s_{0}z_{1}s_{0}z_{02}\}.$$
		We apply Claim \ref{CSplitR} with $\mathcal S_{1}=\{z_{0},z_{1},s_{0},s_{1}\}$ and $\mathcal S_{2}=\{z_{02}\}$. Indeed we note that $z_{02}=s_{1}z_{0}s_{1}$. Hence $$\mathcal R_{1}=\{z_{0}^{2},z_{1}^{2},(z_{0}z_{1})^{2},s_{0}^{2},s_{1}^{2},(s_{0}s_{1})^{3},(z_{0}s_{0})^{2},(z_{1}s_{1})^{2}\}$$ and
		$$
		\mathcal R_{2}=\{z_{02}^{2},(z_{1}z_{02})^{2},(z_{0}z_{02})^{2},s_{1}z_{0}s_{1}z_{02},s_{0}z_{1}s_{0}z_{02}\},$$
		and so
		$$\mathcal R'_{2}=\{(s_{1}z_{0}s_{1})^{2},(z_{1}s_{1}z_{0}s_{1})^{2},(z_{0}s_{1}z_{0}s_{1})^{2},s_{1}z_{0}s_{1}s_{1}z_{0}s_{1},s_{0}z_{1}s_{0}s_{1}z_{0}s_{1}\}.$$
		Since, $s_{1}^{2}=z_{0}^{2}=1$ we can remove the relation $s_{1}z_{0}s_{1}s_{1}z_{0}s_{1}$ from $\mathcal R'_{2}$. Furthermore, $
		s_{0}z_{1}s_{0}s_{1}z_{0}s_{1}=1$ implies $(z_{1}s_{1}z_{0}s_{1})^{2}=(z_{1}s_{0})^{4}$. Hence, \cZS$_{3}$ is isomorphic to
		$$
		\langle z_{0},z_{1},s_{0},s_{1}|z_{0}^{2},z_{1}^{2},s_{0}^{2},s_{1}^{2},(s_{0}s_{1})^{3},(z_{0}z_{1})^{2},(z_{0}s_{0})^{2},(z_{1}s_{1})^{2},(z_{0}s_{1})^{4},(z_{1}s_{0})^{4},s_{0}s_{1}z_{0}s_{1}s_{0}z_{1}\rangle.
		$$
		Assuming that
	$s_{0}s_{1}z_{0}s_{1}s_{0}z_{1}=(z_{0}s_{0})^{2}=(z_{0}s_{1})^{4}=s_{0}^{2}=s_{1}^{1}=1$, one finds
	$$(z_{1}s_{1})^{2}=s_{0}s_{1}(z_{0}s_{0})^{2}s_{1}s_{0}=1,$$
	$$(z_{0}z_{1})^{2}=s_{0}(z_{0}s_{1})^{4}s_{0}=1,$$
	$$(z_{1}s_{0})^{4}=s_{1}(z_{0}z_{1})^{2}s_{1}=1.$$
	Then we simplify the presentation of  \cZS$_{3}$ as 
	$$\langle z_{0},z_{1},s_{0},s_{1}|z_{0}^{2},z_{1}^{2},s_{0}^{2},s_{1}^{2},(s_{0}s_{1})^{3},(z_{0}s_{0})^{2},(z_{0}s_{1})^{4},s_{0}s_{1}z_{0}s_{1}s_{0}z_{1}\rangle .$$
	\end{example}
Reformulated in terms of presentation Theorem \ref{TFirstCox} reads:
\begin{theo}\label{TFirstCox2}
		Suppose $k\geq 2$. The group \cZS$_{k}$ is isomorphic to the presentation $\langle z_{0},\dots,z_{k-2},s_{0},\dots,s_{k-2}|\mathcal R\rangle$ where $\mathcal R$ is the following set of relations
		\begin{enumerate}
			\item for any $0\leq i\leq k-2$, $z_{i}^{2}=s_{i}^{2}=1$,
			\item for any $0\leq i<j\leq k-2$ such that $j-i>1$, $(s_{i}s_{j})^{2}=1$,
			\item for any $0\leq i\leq k-3$, $(s_{i}s_{i+1})^{3}=1$,
			\item for any $0\leq i<j\leq k-2$, $(z_{i}z_{j})^{2}=1$,
			\item for any $0\leq i,j\leq k-2$ such that $|i-j|\neq 1$, $(z_{i}s_{j})^{2}=1$,
			\item for any $0\leq i,j\leq k-2$ such that $|i-j|=1$, $(z_{i}s_{j})^{4}=1$,
			\item for any $0\leq i\leq k-3$, $s_{i}s_{i+1}z_{i}s_{i+1}s_{i}z_{i+1}=1$.
		\end{enumerate}
			\end{theo}
			{\bf Proof}  If $k=2$ the result is straightforward from the definition. Now we assume $k\geq 3$. We apply Claim \ref{CPresSemidirect} and find a presentation of \cZS$_{k}$ as $\langle \mathcal S|\mathcal R\rangle$ with
			$$ 
			\mathcal S=\mathcal T\cup\mathcal Z
			$$
			and
			$$\mathcal R=\mathcal{RS}\cup \mathcal{RZ}\cup\mathcal{RC},$$
			where $\mathcal T=\{s_{i}\mid 0\leq i\leq k-2\}$, $\mathcal Z=\{z_{\{i,j\}}\mid 0\leq i\leq j-1\leq k-2\}$, $\mathcal{RS}=\{s_{i}^{2}\mid 0\leq i\leq k-3\} \cup \{(s_{i}s_{j})^{2}\mid 0\leq i<j-1\leq k-3\}
			\cup \{(s_{i}s_{i+1})^{3}\mid 0\leq i\leq k-3 \}$, $\mathcal{RZ}=\{z_{\{i,j\}}^{2}\mid 0\leq i<j\leq k-1\}\cup \{(z_{\{i,j\}}z_{\{p,q\}})^{2}\mid 0\leq i\leq j-1\leq k-2,
			0\leq p\leq q-1\leq k-2\}$, and $\mathcal{RC}=\{s_{p}z_{\{i,j\}}s_{p}z_{\{s_{p}(i),s_{p}(j)\}}|0\leq i<j\leq k-1,0\leq p\leq k-2\}. $
			The set $\mathcal{RS}$ means that the subgroup $G_{S}=\langle T|\mathcal {RS}\rangle$ is isomorphic to the symmetric group $\mathfrak S_{k}$. The set $\mathcal{RZ}$ means that the subgroup $G_{Z}=\langle \mathcal Z|\mathcal{RZ}\rangle$ is isomorphic to $\mathbb Z_{2}^{\binom k2}$. The set $\mathcal RC$ encodes the conjugacy of an element of $\mathcal Z$ by an element of $G_{S}$. More precisely, if $\sigma$ is a permutation of $\mathfrak S_{k}$ and $w_{\sigma}$ denotes the image of $\sigma$ in $G_{S}$ then the relations of $\mathcal RS$ and $\mathcal RC$ implies 
			\begin{equation}\label{conjzk}
				w_{\sigma}z_{\{i,j\}}w_{\sigma}^{-1}=z_{\{\sigma(i),\sigma(j)\}}.
			\end{equation}
			For simplicity, in the rest of the proof we set $z_{ij}:=z_{\{i,j\}}$ and  $z_{i}:=z_{ii+1}$. The relation $z_{ij}^{2}=1$ can be recovered from $z_{i}^{2}=1$ and the relations of $\mathcal{RS}\cup \mathcal{RC}$. Indeed it suffices to consider a permutation $\sigma$ sending $\{i,j\}$ to $\{i,i+1\}$ and 
			write $z_{ij}^{2}=(w_{\sigma}^{-1}z_{i}w_{\sigma})^{2}=w_{\sigma}^{-1}z_{i}^{2}w_{\sigma}=1$. In the same way, if $\sigma$ is a permutation sending 
			$\{i,j\}$ to $\{i_{1},i_{1}+1\}$ and $\{p,q\}$ to $\{i_{2},i_{2}+1\}$ for some $0\leq i<j\leq k-1$, $0\leq p<q\leq k-1$, $i\neq p$, $j\neq q$ and $0\leq i_{1},i_{2}\leq k-2$  we have $(z_{ij}z_{pq})^{2}=(w_{\sigma}^{-1}z_{i_{1}}w_{\sigma}w_{\sigma}^{-1}z_{i_{2}}w_{\sigma})^{2}=w_{\sigma}^{-1}(z_{i_{1}}z_{i_{2}})^{2}w_{\sigma}$. So the relation $(z_{ij}z_{pq})^{2}=1$ can be recovered from 
			$(z_{i_{1}}z_{i_{2}})^{2}=1$ and the relations of $\mathcal{RS}\cup \mathcal{RC}$. If ($i=p$ and $j\neq q$) or ($i\neq p$ and $j=q$), there exists no permutation sending $\{i,j\}$ to $\{i_{1},i_{1}+1\}$ and $\{p,q\}$ to $\{i_{2},i_{2}+1\}$ for some $0\leq i_{1},i_{2}\leq k-2$. We introduce the set of relations $\mathcal{RSZ}:=\{(z_{i}s_{j})^{4}|0\leq i,j\leq k-2, |i-j|=1\}$.
			These relations can be recovered from $\mathcal R$ since $(z_{i}s_{i+1})^{4}=(z_{i}z_{ii+2})^{2}$ and $(z_{i}s_{i-1})^{4}=(z_{i}z_{i-1i+1})^{2}$. 
			If ($i=p$ and $j\neq q$) or ($i\neq p$ and $j=q$) then there exists a permutation $\sigma$ sending $\{i,j\}$ to $\{i_{1},i_{1}+1\}$ and $\{p,q\}$ to $\{i_{1},i_{1}+2\}$ for
			 some $0\leq i_{1}\leq k-3$. Hence, $(z_{ij}z_{pq})^{2}=(w_{\sigma}^{-1}z_{i_{1}}z_{i_{1}i_{1}+2}w_{\sigma})^{2}=
			 w_{\sigma}^{-1}(z_{i_{1}}z_{i_{1}i_{1}+2})^{2}w_{\sigma}=w_{\sigma}^{-1}(z_{i_{1}}s_{i_{1}+1})^{4}w_{\sigma}$.\\
			As a conclusion, we deduce that $\langle \mathcal S|\mathcal R\rangle=\langle \mathcal S|\mathcal R'\rangle$ with $\mathcal R'=\mathcal{RS}\cup \mathcal{RZ'}\cup\mathcal{RC}\cup \mathcal{RSZ}$ and $\mathcal{RZ}'=\{z_{i}^{2}\mid 0\leq i\leq k-2\}\cup \{(z_{i}z_{p})^{2}\mid 0\leq i,p\leq k-2\}$.\\
			Now let us remove redundancies in $\mathcal {RC}$ and proves that it can be replaced by $\mathcal {RC'}:=\{(s_{j}z_{i})^{2}\mid |i-j|\neq 1\}\cup \{s_{i}s_{i+1}z_{i}s_{i+1}s_{i}z_{i+1}\mid 0\leq i\leq k-3\}\cup\{s_{j-1}z_{ij}s_{j-1}z_{ij-1}\mid 0\leq i\leq j-2\leq k-3\}$ in the presentation.
			 The first and the last sets of the definition of $\mathcal {RC}'$ are include in $\mathcal {RC}$.
			  Furthermore, in $\langle \mathcal S|\mathcal R'\rangle$, 
			we have $s_{i}s_{i+1}z_{i}s_{i+1}s_{i}z_{i+1}=s_{i}z_{ii+2}s_{i}z_{i+1}\in\mathcal {RC}$. Conversely, we set $\mathcal {R''}:=\mathcal {RS}\cup\mathcal {RZ'}\cup \mathcal {RC'} \cup \mathcal {RSZ}$ and we prove that for any $w\in\mathcal {RC}$, we have $w=1$ in $\langle S|\mathcal {R}''\rangle$. Indeed, we have $\mathcal {RC}\setminus \mathcal {RC'}\subset \{s_{p}z_{ij}s_{p}z_{ij}|p\not\in\{i-1,i,j-1,j\}\cup \{s_{j}z_{ij}s_{j}z_{ij+1}\mid 0\leq i\leq j-1\leq k-3\}
			\cup 
			\{s_{i-1}z_{ij}s_{i-1}z_{i-1j}\mid 1\leq i<j\leq k-1 \}\cup 
			\{s_{i}z_{ij}s_{i}z_{i+1j}\mid 0\leq i\leq j-2\leq k-3\}.$ So we have to consider $4$ cases:
			\begin{enumerate}
			\item Consider the element $s_{j}z_{ij}s_{j}z_{ij+1}$ with $0\leq i\leq j-1\leq k-3$. We have $s_{j}z_{ij}s_{j}z_{ij+1}=(z_{ij+1}s_{j}z_{ij}s_{j})^{-1}=s_{j}(s_{j}z_{ij+1}s_{j}z_{ij})^{-1}s_{j}=s_{j}^{2}=1$ in $\langle \mathcal S\mid\mathcal R''\rangle$.{}
			\item Consider the element $s_{i-1}z_{ij}s_{i-1}z_{i-1j}$ with $1\leq i<j\leq k-1$. If $i=j-1$ then, using the second and third sets of the definition of $\mathcal {RC'}$ one obtains $s_{i-1}z_{ii+1}s_{i-1}z_{i-1i+1}=s_{i-1}z_{i}s_{i-1}s_{i}z_{i-1}s_{i}=s_{i-1}z_{i}z_{i}s_{i-1}=1$ in $\langle \mathcal S\mid\mathcal R''\rangle$. If $i<j-1$ then $s_{i-1}z_{ij}s_{i-1}z_{i-1j}=s_{i-1}s_{j-1}z_{ij-1}s_{j-1}s_{i-1}s_{j-1}z_{i-1j-1}s_{j-1}=
			s_{j-1}s_{i-1}z_{ij-1}s_{i-1}z_{i-1j-1}s_{j-1}$ and, using an induction on $|i-j|$ one finds $s_{j-1}s_{i-1}z_{ij-1}s_{i-1}z_{i-1j-1}s_{j-1}=s_{j-1}^{2}=1$ in $\langle \mathcal S\mid\mathcal R''\rangle$.{}
			\item Consider the element $s_{i}z_{ij}s_{i}z_{i+1j}$  for $0\leq i<j-2\leq k-3$. We have $s_{i}z_{ij}s_{i}z_{i+1j}=s_{i}(s_{i}z_{i+1j}s_{i}z_{ij})^{-1}s_{i}=1$ from the previous case.
				\item Suppose $p\not\in\{i-1,i,j-1,j\}$. We proceed by induction on $|i-j|$. If $j=i+1$ then the result is directly obtains from the first set of the definition of $\mathcal {RC}'$. If $p=i+1=j-2$ then we use the previous cases and obtains
				$
					s_{i+1}z_{ii+3}s_{i+1}z_{ii+3}=s_{i+1}s_{i+2}s_{i+1}z_{i}s_{i+1}s_{i+2}s_{i+1}s_{i+2}s_{i+1}z_{i}s_{i+1}s_{i+2}.
				$
				Hence using the fact that $z_{i}$ and $s_{i+2}$ commute in $\langle \mathcal S\mid\mathcal R''\rangle$ together with the braid relations, one obtains
				$
				s_{i+1}s_{i+2}s_{i+1}z_{i}s_{i+1}s_{i+2}s_{i+1}s_{i+2}s_{i+1}z_{i}s_{i+1}s_{i+2}=
				s_{i+2}s_{i+1}s_{i+2}z_{i}s_{i+2}z_{i}s_{i+1}s_{i+2}=1.
				$
				
				If $p\neq j-2$ then $s_{p}z_{ij}s_{p}z_{ij}=s_{p}s_{j-1}z_{ij-1}s_{j-1}s_{p}s_{j-1}z_{ij-1}s_{j-1}=
				s_{j-1}s_{p}z_{ij-1}s_{p}z_{ij-1}s_{j-1}=1$ using the induction hypothesis. Finally, if $p\neq i+1$ then $s_{p}z_{ij}s_{p}z_{ij}=s_{p}s_{i}z_{i+1j}s_{i}s_{p}s_{i}z_{i+1j}s_{i}=
				s_{i}s_{p}z_{i+1j}s_{p}z_{i+1j}s_{i}=1$ using the induction hypothesis.
			\end{enumerate}
			So we have proved that $\langle S\mid\mathcal R\rangle=\langle S\mid\mathcal R''\rangle$.
			Now we apply Claim \ref{CSplitR} with $\mathcal S_{1}=\{z_{i}\mid 0\leq i\leq k-2\}\cup \{s_{i}\mid 0\leq i\leq k-2\}$ and
			$\mathcal S_{2}=\{z_{ij}\mid 0\leq i\leq j-2\leq k-3\}$. We obtain $\mathcal R''=\mathcal R_{1}\cup\mathcal R_{2}$ with $\mathcal R_{1}=\mathcal {RS}\cup\mathcal {RZ'}\cup\mathcal {RSZ}\cup \{(z_{i}s_{j})^{2}\mid |i-j|\neq 1\}\cup \{s_{i}s_{i+1}z_{i}s_{i+1}s_{i}z_{i+1}\mid 0\leq i\leq k-3\}$ and
			$\mathcal R_{2}=\{s_{j-1}z_{ij}s_{j-1}z_{ij-1}\mid 0\leq i\leq j-2\leq k-3\}$. To obtain $\mathcal R'_{2}$, we substitute each occurrence of $z_{ij}$ in $\mathcal R_{2}$ by $s_{j-1}s_{j-2}\cdots s_{i+1}z_{i}s_{i+1}\cdots s_{j-2}s_{j-1}$. In other words, $$\mathcal R'_{2}=\{s_{j-1}\cdot s_{j-1}s_{j-2}\cdots s_{i+1}z_{i} s_{i+1}\cdots s_{j-1}\cdot s_{j-1} \cdot{}
			s_{j-2}\cdots s_{i+1}z_{i} s_{i+1}\cdots s_{j-2}\mid  0\leq i\leq j-2\leq k-3\}.$$ 
			But each $s_{j-1}\cdot s_{j-1}s_{j-2}\cdots s_{i+1}z_{i} s_{i+1}\cdots s_{j-1}\cdot s_{j-1} \cdot{}
			s_{j-2}\cdots s_{i+1}z_{i} s_{i+1}\cdots s_{j-2}$ reduces to $1$ by using rules of $\mathcal{RS}$ and $\mathcal {RZ}'$. Hence we deduce that $\langle \mathcal S|\mathcal R\rangle=\langle \mathcal S_{1}|\mathcal R_{1}\rangle$, as expected.
			$\Box${}\\ \\
Theorem \ref{ThPres} is restated as
\begin{theo}\label{ThPres2}
	For $k\geq 2$, the group \cZS$_{k}$ is isomorphic to the group $\langle g_{0},g_{1},g_{2},\dots,g_{k-1}|\mathcal R_{k}\rangle$, where 
	 $\mathcal R_{k}$ is the set of the following relations:
	 \begin{enumerate}
		 \item\label{Rk1} For any $0\leq i\leq k-1$, $g_{i}^{2}=1$,
		 \item\label{Rk2} For any $1\leq i<j\leq k-1$ such that $|i-j|>1$, $(g_{i}g_{j})^{2}=1$,
		 \item\label{Rk3} For any $1\leq i\leq k-2$, $(g_{i}g_{i+1})^{3}=1$,
		 \item\label{Rk4} For any $i=1,3,\dots k-1$, $(g_{0}g_{i})^{2}=1$,
		 \item\label{Rk5} $(g_{0}g_{2})^{4}=1$,
		 \item\label{Rk6} $(g_{0}g_{2}g_{3}g_{1}g_{2})^{4}=1$.
	 \end{enumerate}
The explicit isomorphism sends $Z_{0}$ to $g_{0}$ and each $S_{i}$ to $g_{i+1}$.
\end{theo}
{\bf Proof}
		We find several redundancies in the presentation of Theorem \ref{TFirstCox2}. First we compute
		\begin{equation}
			\begin{array}{rcl}
				(z_{i}s_{i-1})^{4}&\displaystyle\mathop=^{(vii),(i)}&(s_{i-1}s_{i}z_{i-1}s_{i})^{4}\\
				&\displaystyle\mathop=^{(iii),(v)}&(s_{i-1}s_{i}z_{i-1}s_{i-1}s_{i}z_{i-1}s_{i-1}s_{i})^{2}\\
				&\displaystyle\mathop=^{(iii),(i)}&s_{i}s_{i-1}(s_{i}z_{i-1})^{4}s_{i-1}s_{i}.
			\end{array}
		\end{equation}
		The overscripted numbers correspond to the rules of Theorem \ref{TFirstCox2} used to obtain each equality. Assuming $(z_{i-1}s_{i})^{4}=1$ and applying the rule $(i)$, we show that $(z_{i}s_{i-1})^{4}=1$. So this relation can be removed from the presentation.\\                
               Now let us consider the relations $\{(z_{i}z_{j})^2=1\mid 0\leq i<j\leq k-2\}$ of point $(iv)$ of Theorem \ref{TFirstCox2}.
                If $j=i+1$ we have :
                  $$\begin{array}{rcl}
                      (z_iz_{i+1})^2&\displaystyle\mathop=^{(vii)}&(z_is_is_{i+1}z_is_{i+1}s_i)^2\\
                                    &\displaystyle\mathop=^{(v)}&(s_iz_is_{i+1}z_is_{i+1}s_i)^2\\
                                    &\displaystyle\mathop=^{(i)}&s_i(z_is_{i+1})^4s_i.
                    \end{array}$$ Assuming $(z_is_{i+1})^4=1$ and applying the rule $(i)$ we have $(z_iz_{i+1})^2=1$
                    so this relation can also be removed from the presentation.

If $j=i+2$ then we have 
                  $$\begin{array}{rcl}
                      (z_iz_{i+2})^2&\displaystyle\mathop=^{(vii)}&(s_{i-1}s_iz_{i-1}s_is_{i-1}z_{i+2})^2\\
                                    &\displaystyle\mathop=^{(v)}&(s_{i-1}s_iz_{i-1}z_{i+2}s_is_{i-1})^2\\
                                    &\displaystyle\mathop=^{(i)}&s_{i-1}s_i(z_{i-1}z_{i+2})^2s_is_{i-1}\\
                                    &\displaystyle\mathop=^{(vii)}&s_{i-1}s_i(z_{i-1}s_{i+1}s_{i+2}z_{i+1}s_{i+2}s_{i+1})^2s_is_{i-1}\\
                                    &\displaystyle\mathop=^{(v)}&s_{i-1}s_i(s_{i+1}s_{i+2}z_{i-1}z_{i+1}s_{i+2}s_{i+1})^2s_is_{i-1}\\
                                    &\displaystyle\mathop=^{(i)}&s_{i-1}s_is_{i+1}s_{i+2}(z_{i-1}z_{i+1})^2s_{i+2}s_{i+1}s_is_{i-1}.
                    \end{array}$$ Hence by induction on $i$, assuming that $(z_{0}z_{2})^2=1$, we show that $(z_iz_{i+2})^2=1$.
                                    
                  If $j>i+2$, then we have 
                  $$\begin{array}{rcl}
                      (z_{i}z_{j})^2&\displaystyle\mathop=^{(vii)}&(z_is_{j-1}s_jz_{j-1}s_js_{j-1})^2\\
                                    &\displaystyle\mathop=^{(v)}&(s_{j-1}s_jz_iz_{j-1}s_js_{j-1})^2\\
                                    &\displaystyle\mathop=^{(i)}&s_{j-1}s_j(z_iz_{j-1})^2s_js_{j-1}\\
                    \end{array}.$$ Hence by induction on $j$, assuming that $(z_{i}z_{i+2})^2=1$, we show that $(z_iz_{j})^2=1$.

                  To summarize we have shown that, assuming points $(vii),(vi),(v),(i)$ of Theorem \ref{TFirstCox2}, all the relations of point $(iv)$ (i.e $\{(z_{i}z_{j})^2=1\mid 0\leq i<j\leq k-2\}$) are redundancies except one : $(z_0z_2)^2=1$.
                 We also notice that  $(z_0z_2)^2=(z_0s_1s_2s_0s_1z_0s_1s_0s_2s_1)^2=(z_0s_1s_2s_0s_1)^4$

		Now we apply Claim \ref{CSplitR} by setting $\mathcal S_{1}=\{z_{0},s_{0},\dots,s_{k-2}\}$ and $\mathcal S_{2}=\{z_{1},\dots,z_{k-2}\}$ since
		\begin{equation}
			z_{i}=(s_{i-1}s_{i})\cdots (s_{0}s_{1})z_{0}(s_1s_{0})\cdots (s_{i}s_{i-1}).
		\end{equation}
		We have $$\mathcal R_{1}=\mathcal {RS}\cup\{z_{0}^{2}=1,(z_{0}s_{1})^{4}=1,(z_0s_1s_2s_0s_1)^4=1 \}\cup \{(z_{0}s_{j})^{2}=1\mid j\neq 1\}$$
		and
		$$\begin{array}{rcl}\mathcal R_{2}&=&\{z_{i}^{2}=1\mid 1\leq i\leq k-2\}\cup \{(z_{i}s_{i+1})^{4}=1\mid 1\leq i\leq k-3\}\\&&\cup \{(z_{i}s_{j})^{2}=1\mid 1\leq i,j\leq k-2, j\not\in\{i-1,i+1\}\\&&\cup\{s_{i}s_{i+1}z_{i}s_{i+1}s_{i}z_{i+1}=1\mid 0\leq i\leq k-3\}\\
                    \end{array}$$
                
		So we have $\mathcal R'_{2}=\mathcal T_{1}\cup\mathcal T_{2}\cup\mathcal T_{3}\cup\mathcal T_{4}$ with
		$$\mathcal T_{1}=\{((s_{i-1}s_{i})\cdots (s_{0}s_{1})z_{0}(s_1s_{0})\cdots (s_{i}s_{i-1}))^{2}=1\mid 1\leq i\leq k-2\}$$
		$$\mathcal T_{2}=\{((s_{i-1}s_{i})\cdots (s_{0}s_{1})z_{0}(s_1s_{0})\cdots (s_{i}s_{i-1})s_{i+1})^{4}=1\mid 1\leq i\leq k-3\}$$
		$$\mathcal T_{3}=
		 \{((s_{i-1}s_{i})\cdots(s_{0}s_{1})z_{0}(s_1s_{0})\cdots (s_{i}s_{i-1})s_{j})^{2}=1\mid 1\leq i,j\leq k-2, j\not\in\{i-1,i+1\}$$
		 $$\begin{array}{l}\mathcal T_{4}=
		 \{s_{i}s_{i+1}(s_{i-1}s_{i})\cdots (s_{0}s_{1})z_{0}(s_1s_{0})\cdots (s_{i}s_{i-1})s_{i+1}s_{i}
		 (s_{i}s_{i+1})\cdots (s_{0}s_{1})z_{0}(s_1s_{0})\cdots (s_{i+1}s_{i})=1\\\mid 0\leq i\leq k-3\}
                     .\end{array}$$
                   
                   Remarking that $\displaystyle((s_{i-1}s_{i})\cdots (s_{0}s_{1})z_{0}(s_1s_{0})\cdots (s_{i}s_{i-1}))^{2}\mathop=^{\mathcal R_{1}}1$ we can remove the relation of $\mathcal T_{1}$ from $\mathcal R'_{2}$.\\

                   In order to remove the relations of $\mathcal T_{3}$ from $\mathcal R'_{2}$ we distinguish three cases :
                   \begin{enumerate}
		
                   \item If $j>i+1$ then$$
		\begin{array}{rcl}\displaystyle
			((s_{i-1}s_{i})\cdots(s_{0}s_{1})z_{0}(s_1s_{0})\cdots (s_{i}s_{i-1})s_{j})^{2}&\displaystyle\mathop=^{\mathcal R_{1}}&((s_{i-1}s_{i})\cdots(s_{0}s_{1})z_{0}(s_1s_{0})\cdots (s_{i}s_{i-1}))^{2}s_{j}^{2}\\
			&\displaystyle\mathop=^{\mathcal R_{1}}&1.
		\end{array}$$
		        
              \item  If $j=i$ then we use the the braid relations and obtain 
		\begin{equation}\label{eqbraid}\displaystyle
			(s_{1}s_{0})\cdots (s_{i}s_{i-1})s_{i}\mathop=^{braid}s_{0}(s_{1}s_{0})\cdots (s_{i}s_{i-1}).
		\end{equation}
		Hence,$$
		\begin{array}{rcl}\displaystyle
			((s_{i-1}s_{i})\cdots(s_{0}s_{1})z_{0}(s_1s_{0})\cdots (s_{i}s_{i-1})s_{i})^{2}&\displaystyle\mathop=^{braid}&((s_{i-1}s_{i})\cdots(s_{0}s_{1})z_{0}s_{0}(s_1s_{0})\cdots (s_{i}s_{i-1}))^{2}\\
                                                                                                       &\displaystyle\mathop=^{\mathcal R_{1}}&(s_{i-1}s_{i})\cdots(s_{0}s_{1})(z_{0}s_{0})^{2}(s_1s_{0})\cdots (s_{i}s_{i-1})\\
			&\displaystyle\mathop=^{\mathcal R_{1}}&1.
		\end{array}$$

            \item    If $j<i-1$ then we have $$\begin{array}{rcl}\displaystyle
                                                                 ((s_{i-1}s_{i})\cdots(s_{0}s_{1})z_{0}(s_1s_{0})\cdots (s_{i}s_{i-1})s_{j})^{2}&\displaystyle\mathop=^{\mathcal R_{1}}&((s_{i-1}s_{i})\cdots(s_{j}s_{j+1})(s_{j-1}s_{j})\cdots(s_{0}s_{1})z_{0}\\
                                                                                                                                                &&\cdot (s_1s_{0})\cdots (s_{j}s_{j-1})(s_{j+1}s_{j})(s_{j+2}s_{j+1})s_{j}\cdots (s_{i}s_{i-1}))^{2}\\                                                                                                                              &\displaystyle\mathop=^{\mathcal R_{1}}&((s_{i-1}s_{i})\cdots(s_{j}s_{j+1})(s_{j-1}s_{j})\cdots(s_{0}s_{1})z_{0}\\&&\cdot(s_1s_{0})\cdots (s_{j}s_{j-1})(s_{j+1}s_{j+2})(s_{j}s_{j+1}s_{j})\cdots (s_{i}s_{i-1}))^{2}\\                                                                                                                              &\displaystyle\mathop=^{\mathcal R_{1}}&((s_{i-1}s_{i})\cdots(s_{j}s_{j+1})(s_{j-1}s_{j})\cdots(s_{0}s_{1})z_{0}\\&&\cdot (s_1s_{0})\cdots (s_{j}s_{j-1})(s_{j+1}s_{j+2})(s_{j+1}s_{j}s_{j+1})\cdots (s_{i}s_{i-1}))^{2}\\
                                               &\displaystyle\mathop=^{\mathcal R_{1}}&((s_{i-1}s_{i})\cdots(s_{j}s_{j+1})(s_{j-1}s_{j})\cdots(s_{0}s_{1})z_{0}\\&&\cdot(s_1s_{0})\cdots (s_{j}s_{j-1})s_{j+2}(s_{j+1}s_{j})(s_{j+2}s_{j+1})\cdots (s_{i}s_{i-1}))^{2}\\
                                                                                                                              &\displaystyle\mathop=^{\mathcal R_{1}}&(s_{i-1}s_{i})\cdots(s_{j}s_{j+1})((s_{j-1}s_{j})\cdots(s_{0}s_{1})z_{0}\\&&\cdot(s_1s_{0})\cdots (s_{j}s_{j-1})s_{j+2})^2(s_{j+1}s_{j})(s_{j+2}s_{j+1})\cdots (s_{i}s_{i-1})\\
                                                                 &\displaystyle\mathop=^{\mathcal R_{1}}&1\ \textrm{(using the first case)}
		\end{array}$$
\end{enumerate}
                
		So we can remove the relations of $\mathcal T_{3}$ from $\mathcal R'_{2}$.\\
		Also, using the relation of the symmetric group, one finds
		\begin{equation}
			(s_{1}s_{0})\cdots (s_{i}s_{i-1})s_{i+1}(s_{i-1}s_{i})\cdots (s_{0}s_{1})\mathop=^{\mathcal R_{1}}s_{i+1}s_{i}\cdots s_{2}s_{1}s_{2}\cdots s_{i}s_{i+1}.
		\end{equation}
		Hence,
		$$
			\begin{array}{rcl}
				((s_{i-1}s_{i})\cdots (s_{0}s_{1})z_{0}(s_1s_{0})\cdots (s_{i}s_{i-1})s_{i+1})^{4}
				&\displaystyle\mathop=^{\mathcal R_{1}}&
				(s_{i-1}s_{i})\cdots (s_{0}s_{1})\\&&\cdot(z_{0}s_{i+1}s_{i}\cdots s_{2}s_{1}s_{2}\cdots s_{i}s_{i+1})^{3}\\&&\cdot z_{0}(s_1s_{0})\cdots (s_{i}s_{i-1})s_{i+1}\\
				&\displaystyle\mathop=^{\mathcal R_{1}}&
				(s_{i-1}s_{i})\cdots (s_{0}s_{1})s_{i+1}s_{i}\cdots s_{2}\\
				&&\cdot(z_{0}s_{1})^{3}z_{0}s_{2}\cdots s_{i}s_{i+1}(s_1s_{0})\cdots (s_{i}s_{i-1})s_{i+1}\\
				&\displaystyle\mathop=^{\mathcal R_{1}}&
				(s_{i-1}s_{i})\cdots (s_{0}s_{1})s_{i+1}s_{i}\cdots s_{2}\\
				&&\cdot(z_{0}s_{1})^{4}s_{2}\cdots s_{i}s_{i+1}(s_{1}s_{0})\cdots (s_{i}s_{i-1})\\
				&\displaystyle\mathop=^{\mathcal R_{1}}&1.
			\end{array}
		$$
		We deduce that we can remove the relation of $\mathcal T_{2}$. Finally the relation 
		$s_{i}s_{i+1}(s_{i-1}s_{i})\cdots (s_{0}s_{1})z_{0}(s_1s_{0})\cdots (s_{i}s_{i-1})s_{i+1}s_{i}
		 (s_{i}s_{i+1})\cdots (s_{0}s_{1})z_{0}(s_1s_{0})\cdots (s_{i+1}s_{i})$ reduces to $1$ using only $s_{i}^{2}=z_{0}^{2}=1$. The relations of $\mathcal T_{4}$ are redundancies.
		 Hence, the group \cZS$_{k}$ is isomorphic to $\langle \mathcal S_{1}|\mathcal R_{1}\rangle$ and we recover the statement of Theorem \ref{ThPres2} by sending $z_{0}$ to $g_{0}$ and each $s_{i}$ to $g_{i+1}$.
$\Box$

\section{Entanglement of $|\GHZ_{k}\rangle$\label{appGHZ}}
In this section, we prove that the state $|\GHZ_{k}\rangle$ is not generically entangled. We start with a result of Miyake \cite{2003Miyake} stating that a state is generically entangled if and only if its hyperdeterminant $\Delta$ does not vanish. The hyperdeterminant is a high degree invariant polynomial impossible to compute in practice but its interpretation in terms of solutions to a system of equations allows us to test its nullity, see e.g. \cite{1992GKL} p445. Let us recall briefly how to process. First we consider $k$ binary variables $\mathbf x^{(i)}=(x^{(i)}_{0},x^{(i)}_{1})$, $i=1..k$. To each state $|\phi\rangle=\sum \alpha_{i_{1}\dots i_{k}}|i_{1}\cdots i_{k}\rangle$, we associate the binary multilinear form
\begin{equation}
	f_{\phi}:=\sum_{0\leq i_{1},\dots,i_{k}\leq 1}\alpha_{i_{1},\dots,i_{k}}x_{i_{1}}^{(1)}\cdots x_{i_{k}}^{(k)}.
\end{equation}
The hyperdeterminant vanishes if and only if the system
\begin{equation}
	\{f_{\phi}=0\}\cup\{{d\over dx^{(j)}_{i}}f_{\phi}=0\mid 0\leq i\leq 1, 1\leq j\leq k\}
\end{equation}
has a non trivial solution $\hat\mathbf x^{(1)}, \dots, \hat\mathbf x^{(k)}$, ie such that there exists $1\leq j\leq k$ with $\hat\mathbf x^{(j)}\neq (0,0)$. For $|\GHZ_{k}\rangle$ the system is
\begin{equation}\label{gensys}\begin{array}{l}
	x_{0}^{(1)}\cdots x_{0}^{(k)}+x_{1}^{(1)}\cdots x_{1}^{(k)}=x_{0}^{(2)}\cdots x_{0}^{(k)}=x_{1}^{(2)}\cdots x_{1}^{(k)}=
	x_{0}^{(1)}x_{0}^{(3)}\cdots x_{0}^{(k)}\\=x_{1}^{(1)}x_{1}^{(3)}\cdots x_{1}^{(k)}=\cdots=
	x_{0}^{(1)}\cdots x_{0}^{(k-1)}=x_{1}^{(1)}\cdots x_{1}^{(k-1)}=0.\end{array}
\end{equation}
We check that for $k>3$, $x_{0}^{(1)}=x_{0}^{(2)}=x_{1}^{(3)}=x_{1}^{(4)}=0$ implies (\ref{gensys}). So for $k>3$, $|\GHZ_{k}\rangle$ is not generically entangled. Remark that, when $k=2,3$, all the solutions of (\ref{gensys}) are trivial and so $|\GHZ_{k}\rangle$ is  generically entangled.
\section{Some covariant polynomials associated to $4$ qubit systems\label{appCov}}
In this section, we shall explain how to compute the polynomials which are used to determine the entanglement type of the systems in section \ref{four}.We shall first recall the definition of the transvection of two multi-binary forms on the binary variables $x^{(1)}=(x^{(1)}_0,x^{(1)}_1), \dots, x^{(p)}=(x^{(p)}_0,x^{(p)}_1)$
\begin{equation}
(f,g)_{i_1,\dots,i_p}={\mathrm tr} \Omega^{i_1}_{x^{(1)}}\dots \Omega_{x^{(p)}}^{i_p}f(x'^{(1)},\dots,x'^{(p)})g(x''^{(1)},\dots,x''^{(p)}),
\end{equation}
where  $\Omega$ is the Cayley operator
\[
\Omega_x=\left|\begin{array}{cc}\partial\over \partial x'_0& \partial\over \partial x''_0
\\ \partial x'_1& \partial\over \partial x''_1\end{array}\right|
\]
and $\rm tr$ sends each variables $x', x''$ on $x$ (erases $'$ and $''$). In \cite{2012HLT}, we give a list of generators of the algebra of covariant polynomials for $4$ qubits systems which are obtained by transvection from the ground form
\[{}
A=\sum_{i,j,k,\ell}\alpha_{i,j,k,l}x_{i}y_{j}z_{k}t_{\ell}.
\]
Here we give formulas for some of the polynomials which are used in the paper.
$\begin{array}{cc}&\\
\begin{array}{|c|c|}
\hline \mbox{Symbol}&\mbox{Transvectant}\\\hline
 B_{2200}&\frac12(A,A)^{0011}\\
B_{2020}&\frac12(A,A)^{0101}\\
B_{2002}&\frac12(A,A)^{0110}\\
B_{0220}&\frac12(A,A)^{1001}\\
B_{0202}&\frac12(A,A)^{1010}\\
B_{0022}&\frac12(A,A)^{1100}\\\hline
\end{array}\nonumber&\begin{array}{|c|c|}
\hline \mbox{Symbol}&\mbox{Transvectant}\\
\hline C^1_{1111}&(A,B_{2200})^{1100}+(A,B_{0022})^{0011}
\\\hline
 C_{3111}&\frac13\left((A,B_{2200})^{0100}+(A,B_{2020})^{0010}+(A,B_{2002})^{0001}\right)\\
 C_{1311}&\frac13\left((A,B_{2200})^{1000}+(A,B_{0220})^{0010}+(A,B_{0202})^{0001}\right)\\
 C_{1131}&\frac13\left((A,B_{2020})^{1000}+(A,B_{0220})^{0100}+(A,B_{0022})^{0001}\right)\\
 C_{1113}&\frac13\left((A,B_{2002})^{1000}+(A,B_{0202})^{0100}+(A,B_{0022})^{0010}\right)\\
 \hline
\end{array}\nonumber \end{array}
$

$\begin{array}{cc}\begin{array}{c}\begin{array}{|c|c|}\hline
\mbox{Symbol}&\mbox{Transvectant}\\\hline
D_{2200}&(A,C_{1111}^1)^{0011}\\
D_{2020}&(A,C^1_{1111})^{0101}\\
D_{2002}&(A,C^1_{1111})^{0110}\\
D_{0220}&(A,C^1_{1111})^{1001}\\
D_{0202}&(A,C_{1111}^1)^{1010}\\
D_{0022}&(A,C_{1111})^{1100}\\\hline
D_{4000}&(A,C_{3111})^{0111}\\
D_{0400}&(A,C_{1311})^{1011}\\
D_{0040}&(A,C_{1131})^{1101}\\
D_{0004}&(A,C_{1113})^{1110}\\\hline\end{array}\\ \\
\begin{array}{|c|c|}\hline
\mbox{Symbol}&\mbox{Transvectant}\\\hline
E^1_{3111}&(A,D_{2200})^{0100}+(A,D_{2020})^{0010}+(A,D_{2002})^{0001}\\
E^1_{1311}&(A,D_{2200})^{1000}+(A,D_{0220})^{0010}+(A,D_{0202})^{0001}\\
E^1_{1131}&(A,D_{2020})^{1000}+(A,D_{0220})^{0100}+(A,D_{0022})^{0001}\\
E^1_{1113}&(A,D_{2002})^{1000}+(A,D_{0202})^{0100}+(A,D_{0022})^{0010}\\\hline
\end{array}\end{array}&
	\begin{array}{|c|c|}\hline
\mbox{Symbol}&\mbox{Transvectant}\\\hline
F_{4200}&(A,E^1_{3111})^{0011}\\
F_{4020}&(A,E^1_{3111})^{0101}\\
F_{4002}&(A,E^1_{3111})^{0110}\\
F_{0420}&(A,E^1_{1311})^{1001}\\
F_{0402}&(A,E^1_{1311})^{1010}\\
F_{0042}&(A,E^1_{1131})^{1100}\\
F_{2400}&(A,E^1_{1311})^{0011}\\
F_{2040}&(A,E^1_{1131})^{0101}\\
F_{2004}&(A,E^1_{1113})^{0110}\\
F_{0240}&(A,E^1_{1131})^{1001}\\
F_{0204}&(A,E^1_{1113})^{1010}\\
F_{0024}&(A,E^1_{1113})^{1100}\\\hline
\end{array}\end{array}\nonumber
$

$\begin{array}{cc}\begin{array}{|c|c|}\hline
\mbox{Symbol}&\mbox{Transvectant}\\\hline
G^1_{3111}&(A,F_{4200})^{1100}\\
G^2_{3111}&(A,F_{4020})^{1010}\\
G^1_{1311}&(A,F_{2400})^{110}\\
G^2_{1311}&(A,F_{0420})^{0110}\\
G^1_{1131}&(A,F_{2040})^{1010}\\
G^2_{1131}&(A,F_{0240})^{0110}\\
G^1_{1113}&(A,F_{2004})^{1001}\\
G^2_{1113}&(A,F_{0204})^{0101}
\\\hline\end{array}&
\begin{array}{|c|c|}\hline
\mbox{Symbol}&\mbox{Transvectant}\\\hline
G_{5111}&(A,F_{4002})^{0001}+(A,F_{4020})^{0010}+(A,F_{4200})^{0100}\\
G_{1511}&(A,F_{0402})^{0001}+(A,F_{0420})^{0010}+(A,F_{2400})^{1000}\\
G_{1151}&(A,F_{0042})^{0001}+(A,F_{0240})^{0100}+(A,F_{2040})^{1000}\\
G_{1115}&(A,F_{0204})^{0100}+(A,F_{0024})^{0010}+(A,F_{2004})^{1000}\\\hline
\end{array}\end{array}\nonumber\nonumber
$

$\begin{array}{cc}
	\begin{array}{|c|c|}\hline
\mbox{Symbol}&\mbox{Transvectant}\\\hline
H_{4200}&(A,G_{5111})^{1011}\\
H_{4020}&(A,G_{5111})^{1101}\\
H_{4002}&(A,G_{5111})^{1110}\\
H_{0420}&(A,G_{1511})^{1101}\\
H_{0402}&(A,G_{1511})^{1110}\\
H_{0042}&(A,G_{1151})^{1110}\\
H_{2400}&(A,G_{1511}^1)^{0111}\\
H_{2040}&(A,G_{1151})^{0111}\\
H_{2004}&(A,G_{1115}^1)^{0111}\\
H_{0240}&(A,G_{1151})^{1011}\\
H_{0204}&(A,G_{1115})^{1011}\\
H_{0024}&(A,G_{1115}^1)^{1101}\\\hline
H_{2220}^1&(A,G_{1311}^1)^{0101}+(A,G_{3111}^1)^{1001}+(A,G_{1131}^1)^{0011}\\
H_{2220}^2&(A,G_{1311}^2)^{0101}+(A,G_{3111}^2)^{1001}+(A,G_{1131}^2)^{0011}\\
H_{2202}^1&(A,G_{1311}^1)^{0110}+(A,G_{3111}^1)^{1010}+(A,G_{1113}^1)^{0011}\\
H_{2022}^1&(A,G_{3111}^1)^{1100}+(A,G_{1131}^1)^{0110}+(A,G_{1113}^1)^{0101}\\
H_{0222}^1&(A,G_{1311}^1)^{1100}+(A,G_{1131}^1)^{1010}+(A,G_{1113}^1)^{1001}\\
\hline\end{array}\nonumber\\ \\
	\begin{array}{|c|c|}\hline
\mbox{Symbol}&\mbox{Transvectant}\\\hline
I_{5111}^1&(A,H_{4020})^{0010}+(A,H_{4200})^{0100}+(A,H_{4002})^{0001}\\
I_{1511}^1&(A,H_{0420})^{0010}+(A,H_{2400})^{1000}+(A,H_{4002})^{0001}\\
I_{1151}^1&(A,H_{0240})^{0100}+(A,H_{2040})^{1000}+(A,H_{0042})^{0001}\\
I_{1115}^1&(A,H_{0204})^{0100}+(A,H_{2004})^{1000}+(A,H_{0024})^{0010}
\\\hline
\end{array}\end{array}\nonumber
$

$\begin{array}{cc}&\\\begin{array}{|c|c|}\hline
\mbox{Symbol}&\mbox{Transvectant}\\\hline
J_{4200}&(A,I_{5111}^{1})^{1011}\\
J_{4020}&(A,I_{5111}^{1})^{1101}\\
J_{4002}&(A,I_{5111}^{1})^{1110}\\
J_{0420}&(A,I_{1511}^{1})^{1101}\\
J_{0402}&(A,I_{1511}^{1})^{1110}\\
J_{0042}&(A,I_{1151}^{1})^{1110}\\
J_{2400}&(A,I_{1511}^{1})^{0111}\\
J_{2040}&(A,I_{1151}^{1})^{0111}\\
J_{2004}&(A,I_{1115}^{1})^{0111}\\
J_{0240}&(A,I_{1151}^{1})^{1011}\\
J_{0204}&(A,I_{1115}^{1})^{1011}\\
J_{0024}&(A,I_{1115}^{1})^{1101}\\\hline
\end{array}&\begin{array}{c}
\begin{array}{|c|c|}\hline
\mbox{Symbol}&\mbox{Transvectant}\\\hline
K_{3311}&=(A,J_{4200})^{1000}-(A,J_{2400})^{0100}\\
K_{3131}&=(A,J_{4020})^{1000}-(A,J_{2040})^{0010}\\
K_{3113}&=(A,J_{4002})^{1000}-(A,J_{2004})^{0001}\\
K_{1331}&=(A,J_{0420})^{0100}-(A,J_{0240})^{0010}\\
K_{1313}&=(A,J_{0402})^{0100}-(A,J_{0204})^{0001}\\
K_{1133}&=(A,J_{0042})^{0010}-(A,J_{0024})^{0001}\\\hline
K_{5111}&=(A,J_{4200})^{0100}-(A,J_{4020})^{0010}+(A,J_{4002})^{0001}\\
K_{1511}&=(A,J_{2400})^{1000}-(A,J_{0420})^{0010}+(A,J_{0402})^{0001}\\
K_{1151}&=(A,J_{2040})^{1000}-(A,J_{0240})^{0100}+(A,J_{0042})^{0001}\\
K_{1115}&=(A,J_{2004})^{1000}-(A,J_{0204})^{0110}+(A,J_{0024})^{0010}\\\hline
\end{array}\\
\begin{array}{|c|c|}\hline
\mbox{Symbol}&\mbox{Transvectant}\\\hline
L_{6000}&=(A,K_{5111})^{0111}\\
L_{0600}&=(A,K_{1511})^{1011}\\
L_{0060}&=(A,K_{1151})^{1101}\\
L_{0006}&=(A,K_{1115})^{1110}\\\hline\end{array}\end{array}\end{array}\nonumber$
We use the following polynomials in order to determine the entanglement level of a system:
\[
\mathcal L=L_{6000}+L_{0600}+L_{0060}+L_{0006}
\]
\[
 \mathcal K_{3}=K_{3311}+K_{3131}+K_{3113}+K_{1331}+K_{1313}+K_{1133},
\]
\[\overline{\mathcal G}=G_{3111}^{1}G_{1311}^{1}G_{1131}^{1}G_{1113}^{1},\ 
\mathcal G=G_{3111}^{2}+G_{1311}^{2}+G_{1131}^{2}+G_{1113}^{2},\]
\[\mathcal H=H^{1}_{2220}+H^{1}_{2202}+H^{1}_{2022}+H^{1}_{0222},\]
\[\mathcal D=D_{4000}+D_{0400}+D_{0040}+D_{0004},\]
and $\mathcal C=(A,B_{2200})^{0110}+(A,B_{2002})^{1001}$.

\end{document}